\documentclass{emulateapj}
\usepackage{graphicx}
\usepackage{bm}
\usepackage{epsfig}
\usepackage{multirow}
\usepackage{color}
\usepackage{subfigure}
\usepackage{float}
\usepackage{rotating}
\usepackage{rotfloat}

\def\hideprivate{\global\long\def\private##1{\iffalse ##1 \fi}}
\hideprivate

\shorttitle{Identifying the EM Counterparts of GW Mergers}
\shortauthors{Nissanke, Kasliwal \& Georgieva}

\begin{document}

\title{Identifying Elusive Electromagnetic Counterparts to Gravitational
  Wave Mergers: \\ an end-to-end simulation}

\author{Samaya Nissanke\altaffilmark{1,2}, Mansi Kasliwal\altaffilmark{3}, Alexandra Georgieva\altaffilmark{1}}

\newcommand{\samaya}[1]{\textcolor{blue}{[#1]}}
\newcommand{\mansi}[1]{\textcolor{red}{[#1]}}
\newcommand{\sashka}[1]{\textcolor{green}{[#1]}}

\newcommand{\coincident}{\rotatebox[origin=c]{315}{$\parallel$}}

\altaffiltext{1}{Theoretical Astrophysics, California Institute of Technology, Pasadena, California 91125}

\altaffiltext{2}{JPL, California Institute of Technology, 4800 Oak
  Grove Drive, Pasadena, California 91109}

\altaffiltext{3}{The Observatories, Carnegie Institution for Science, 813 Santa Barbara St Pasadena, California 91101}

\begin{abstract}
Combined gravitational-wave (GW) and electromagnetic (EM) observations of compact binary mergers
should enable detailed studies of astrophysical processes in the strong-field
gravity regime. This decade, ground-based GW interferometers promise to routinely
detect compact binary mergers. Unfortunately, networks of GW interferometers have poor angular resolution on the sky and
their EM signatures are predicted to be faint. Therefore, a challenging goal will
be to unambiguously pinpoint the EM counterparts to GW mergers. We
perform the first comprehensive end-to-end simulation that focuses
on: i) GW sky localization, distance measures and volume errors with
two compact binary populations and four different GW networks, ii)
subsequent EM detectability by a slew of multiwavelength telescopes and, iii) final identification of the merger counterpart
amidst a sea of possible astrophysical false-positives.  First, we find that double neutron star binary mergers can be detected out to a maximum distance
of 400 Mpc (or 750 Mpc) by three (or five) detector GW networks
respectively. Neutron Star - Black-Hole binary mergers can be detected a
factor of 1.5 further out; their median
to maximum sky localizations are 50--170 deg$^2$ (or
6--65 deg$^2$) for a three (or five) detector GW network. 
Second, by optimizing depth, cadence and sky area, we quantify
relative fractions of optical counterparts that are detectable by a
suite of different aperture-size telescopes across the globe. \private{A global network of telescopes with wide-field cameras should
detect a sizable fraction of optical counterparts to GW
mergers.} Third, we present five case studies to illustrate the
diversity of scenarios in secure identification of the EM
counterpart. We discuss the case of a typical binary, neither beamed nor nearby, and
the challenges associated with identifying an EM counterpart both at low and high Galactic latitudes. 
For the first time, we demonstrate how construction of low-latency GW volumes in conjunction with local universe galaxy catalogs can help
solve the problem of false positives. We conclude with strategies that would best prepare us for successfully identifying the elusive EM counterpart to a GW merger. 
\end{abstract}


\section{Introduction}
\label{sec:intro}


Our understanding of astrophysical processes in strong-field gravity
regimes is limited by the nature of the electromagnetic (EM) force
and current instrument sensitivity. Low rates, short timescales,
high energies and interactions with their environments characterize transient
strong-field gravity events in the Universe. Such events provide us with a
brief window to study the interplay of fundamental physical
processes. In this respect, gravitational wave (GW) astronomy should
allow for the study of such events in the Universe, currently
inaccessible through EM observations. With GW observations alone, we may infer
the physical and geometric properties of individual sources and
determine event rates. Furthermore, the combination of GW and EM
measurements will lead to improved understanding of
astrophysical processes in the strong-field gravity regime, and to
also construct a demographic census of different strong-field gravity
events (e.g. \citealt{bloometal09,kk09,phinney09}). A challenging goal in this respect is to localize and identify
a strong-field gravity event in the Universe jointly by GW and EM
measurements.

Within the next decade, a worldwide network of advanced
versions of ground-based GW interferometers --- LIGO in the US and
possibly India~\citep{Barish:1999,Sigg:2008,LIGOIndia},
Virgo in Italy~\citep{Accadia:2011}, KAGRA in
Japan~\citep{Somiya:2012} --- will become operational within the
frequency range of 10 Hz to a few kHz.  At these frequencies, inspiraling and merging 
compact-object binaries, composed of
neutron stars (NS) and/or stellar-mass black holes (BH), are expected
to be amongst the most numerous and strongest GW-emitting
sources. Mergers of NS-NS and NS-BH binaries, where at least one
object includes initially neutron-rich material and possibly strong magnetic fields, 
are expected to emit in both GWs and EM waves. 

GW detections of compact binary mergers are anticipated to become {\it
  routine} by the end of this decade.  Based on the observed Galactic binary pulsar
distribution and population synthesis results, predicted event rates
for NS-NS binary mergers range from $0.4$ to $400$ per year [with 40 being
the mean value quoted in~\cite{Abadieetal:2010}] detectable by an
advanced GW three-detector network to distances of several
hundred Mpc. Based solely on population synthesis results due to an
absence of observed NS-BH systems, predicted event rates range from
$0.2$ to $200$ per year for NS-$10 M_\odot$ BH binary mergers with
detectable distances $>1$ Gpc.\footnote{The particular value of $10 M_\odot$ is
chosen to be representative of stellar-mass BHs; however, as discussed in \S\,2, it is
unclear whether a NS-$10 M_\odot$ BH binary merger will produce an EM
signature. We use NS-$5 M_\odot$ BH in this paper.}

EM detections of compact binary mergers are still a matter of debate.
There is growing evidence that short hard gamma-ray bursts (SGRBs)
represent the small fraction of NS-NS or NS-BH mergers beamed towards
us (see e.g. \citealt{Nakar:2007,Berger:2011}).  
Joint GW and EM observations can
unequivocally test this hypothesis and illuminate the nature of the central engine and 
collimated outflows. Additionally, extensive theoretical modeling is underway
to predict the EM signature for post-merger ejecta, produced from all
mergers, that is gravitationally unbound to the final remnant BH. For instance, a plausible EM counterpart in the optical or near infrared, referred to
as a kilonova, macronova or mini-supernova, may be powered by weak radioactive decay
arising from any neutron-rich ejecta (e.g., \citealt{LP98,Kulkarni:2005,Metzger:2010,Chawla:2010,Roberts:2011,Piran:2012,Wanajo:2012}).

To observe in both GWs and EM waves, two scenarios exist which depend
critically on the timescale of the EM emission with respect to the NS
binary merger time. They are: i) a GW event is first detected and is
then followed by a slew of multi-wavelength EM telescopes, and ii) an
EM observation is seen prior to, or coincident with, a GW
measurement. Such `GW-triggered' and `EM-
triggered' searches were implemented with the enhanced LIGO and Virgo
interferometers, before they halted for
their upgrades to their advanced versions \citep{Abadie:2012PRD,AbadieEM:2012,
  Abadielowlatency:2012, Abadie:2012GRB}. In this paper, we focus on the first scenario, in which
we first observe the inspiral of a NS binary in GWs, and we then
detect their EM counterparts using multiwavelength EM observatories. 
This scenario leverages the instantaneous all-sky 
visibility of GW detectors in comparison to the narrow field-of-view
(FoV) of EM facilities. Due to the improvement in the
instrument sensitivity and hence GW-detectable distance by an
approximate factor of ten, EM follow-up in
the advanced GW interferometric era presents a new set of challenges
to that faced by the initial versions. 

Here, we view the EM follow-up of a GW event in three
steps: i) localization on the sky with GW measured areas and volumes
using a network of GW interferometers, ii) detectability using
different multiwavelength EM facilities, particularly the optical, and iii) strategies to reduce the number of false-positive signatures that might mimic an EM counterpart of a NS
binary merger within the same GW localization volume. 

Several works over the past few years have begun to explore GW sky
localization. A single GW interferometer has poor directional sensitivity for transient
signals because of its broad antenna function. Localizing any source on the sky depends primarily on triangulating the
GW signal's arrival times at detectors using networks of three or more
GW interferometers. Studies estimate sky localization errors
for unmodeled and modeled GW sources to vary from less than one to a
few hundred square degrees using networks of GW interferometers \citep{Fairhurst:2010, Wen:2010,Nissanke:2011, Klimenko:2011,
  Schutz:2011,Veitch:2012,Fairhurst:2012}.  Previous works assume that
the GW source is fixed at high signal-to-noise ratio (SNR) or distance, and use
analytically-derived Fisher matrix estimates (e.g., \citealt{Fairhurst:2010, Wen:2010,Klimenko:2011,
  Schutz:2011,Veitch:2012,Fairhurst:2012}). However, a significant fraction of expected signals will
be at SNR threshold and degeneracies between parameters in the
predicted GW strain become important. In contrast to earlier work, we compute
explicit GW errors of volumes, distances and sky errors using the full
predicted GW wavestrain for astrophysically distributed NS binary mergers. 

Recent works have also begun to explore EM detectability.
Some papers take on a statistical approach; \cite{Singer:2012}
divide sky localization errors between telescopes and advocate a
coordinated response or \cite{Nuttall:2010} assign a probabilistic
ranking statistic for host galaxies out to 100 Mpc. Some focus on 
a particular wavelength (e.g., X-rays,
\citealt{Kanner:2012}). \cite{Metzger:2012} seek to identify the most 
promising amongst proposed EM counterparts by defining cardinal virtues and discount
follow-up in optical and radio compared to $\gamma$-rays.  There, \cite{Metzger:2012}
assume single numbers for detectable distances\footnote{For clarity, we define here different distance definitions that are
used in the literature. By \emph{average detectable} distance, we refer to the \emph{average} distance that a
single GW interferometer with idealized Gaussian instrument noise
can observe NS binary inspirals averaged over all possible
sky positions and binary orientations. On the other hand, the
\emph{horizon} distance refers to the maximum detectable distance that a
single GW interferometer with idealized Gaussian noise can detect a NS
binary event that is located directly above the interferometer and is
optimally orientated face-on. The \emph{horizon} distance improves on the
\emph{average detectable} value by an approximate geometric factor of
$\sim$ 2.24 (e.g. see~\citealt{Finn:1993}).} 
and sky localization. 
Here, we take on a different approach. We simulate an astrophysical NS
binary population and consider
the full range of distances, localizations and GW networks. We quantitatively divide the pie of binaries 
by beaming angle, Galactic latitude and distance. We then consider the challenges and optimal multi-wavelength strategies 
in each slice.

In this work, we present an end-to-end simulation with the following five steps.\footnote{Discussed in
detail throughout the paper, our results are necessarily limited by the assumptions we make; for instance, we
assume that joint observable GW-EM events are non-spinning NS-NS and
NS -5 M$_{\odot}$ binary systems, and idealized Gaussian GW interferometric noise and observing
conditions at optical telescopes.} First, we construct astrophysically-motivated distributions of NS-NS
and NS-5 M$_{\odot}$ BH binary mergers detectable by different GW networks using different triggering
criteria (\S2). Each binary will have specific geometric properties:
an orientation, a sky position and a luminosity distance. Second, by simulating GW data streams using analytically modeled GW strains, we estimate source
parameters measured by different GW networks. For parameter
estimation, we use MCMC methods developed in
\citealt{Nissanke:2010,Nissanke:2011} (henceforth, N10 and N11 respectively). 
As well as estimating the sources' sky areas, we compute the sky
volume errors (\S3). We summarize the distributions of sky errors,
volumes and distances for NS binary merger populations detected by
different networks and different trigger criteria (\S4). Third, armed
with localizations and distances of each binary in the simulation, we
assess the feasibility of detecting an EM counterpart with a wide suite of current and planned EM facilities.
 We pay close attention to the trade-off between depth and area given finite telescope time (\S5).  
 Fourth, using detailed case studies, we present
the challenges and discuss possible strategies to pinpoint the GW event
amongst the anticipated few to many false-positive transients in
different wavelengths (\S6). In conjunction with a galaxy catalog, we discuss how fractional
reductions in volume error can reduce the overall number
of false-positive transients within a GW-localized event. 
Finally, we conclude with strategies that maximize the success in identifying EM counterparts to GW events
(\S7).


\section{Construction of two NS binary merger catalogs}
\label{sec:pop}


We construct two distinct catalogs with either $4 \times 10^4$ NS-NS or $3 \times
10^4$ NS-$5\,M_{\odot}$ BH binary populations.\footnote{We choose a
  sufficiently large number of binary systems in each catalog such
  that GW networks will detect a sizable number of systems; the
  particular values of $4 \times 10^4$ NS-NS or $3 \times
10^4$ NS-$5\,M_{\odot}$ BH systems are specified somewhat arbitrary.} For every binary in each catalog, we assign physical and geometric source
properties as described below. In this study, physical source parameters are the
individual compact objects' masses $m_1$ and $m_2$ (the spin of the NS and/or BH is
assumed negligible). The geometric source parameters comprise the
luminosity distance $D_L$, the sky position in spherical polar
coordinates (which points from the
center of the Earth to the binary) $\mathbf{n} \equiv (\theta, \phi)$, and the binary's
inclination angle $\cos \iota = \mathbf{\hat L}\cdot\mathbf{\hat n}$,
where $\mathbf{\hat L}$ is the unit vector normal to the binary's
orbital plane. The colatitude $\theta$ and longitude $\phi$ describe ${\bf n}$, and
are related to the declination $\delta$ and right ascension $\alpha$, by $\theta = \pi/2
- \delta$ and $\phi=\alpha-$GAST respectively, where GAST is Greenwich
Apparent Sidereal Time (see N10 for details on the binary's and Earth
coordinate systems used in this work). Let us consider now how we assign specific
source parameters to each binary. 

Regarding the physical source parameters, we assume that each NS has a physical mass of $1.4 \,M_{\odot}$ and each BH has a
physical mass of $5.0 \,M_{\odot}$, and that the objects are non-spinning. In practice, we expect the NS
binary population in the Universe to have continuous NS and/or BH mass
distributions. In the case of NS-BH binaries, instead of the fiducial
$10 \,M_{\odot}$ BH used in standard GW
literature, we choose BHs with a small enough
mass that the NS companion does not plunge directly into the
gravitational potential well of the central BH. Therefore, we can expect some tidal
disruption of the NS to occur and to observe an accompanying EM counterpart. Tidal disruption occurs if the tidal disruption
radius is greater than the BH's innermost circular orbit (ICO); the tidal
disruption radius being a function of the BH's spin, the NS's equation of state
and the binary's mass ratio (see discussions in e.g. \citealt{Taniguchi:2007,Shibata:2008,Shibata:2009,Kyutoku:2011,Foucart:2011,Foucart:2012}). The ICO describes the last stable circular orbit
of the binary system prior to the merger and can be approximated to
a test particle's innermost stable circular orbit (ISCO) radius of $6 \, G
M/c^3$ for a non-spinning BH, where $G$ is the Gravitational Constant,
$c$ is the speed of light and $M$ is the BH's mass
\citep{shapteuk}. In addition, in the actual Universe, we expect BHs in
NS-BH systems to have considerable spin, and tidal disruption may occur
for a NS orbiting prograde around a highly spinning 10 M$_{\odot}$
BH \citep{Foucart:2011}. We choose NS-NS and NS-BH systems with binary separations of
$1.0 \times 10^{-3} R_{\odot}$ and $1.4 \times 10^{-3}  R_{\odot}$ respectively to ensure that they will merge within the system's characteristic gravitational
radiation timescale \citep{shapteuk}.

Regarding geometric source parameters, for each of our two
catalogs, we distribute either $4 \times 10^4$ NS-NS binaries or $3 \times 10^4$ NS-BH binaries out to $z = 0.5$ ($\sim 2.82$
Gpc assuming a $\Lambda$CDM Universe given in \citealt{komatsu09}). Each binary in a catalog
is associated with a random orientation such that $p \, (\cos \iota )
\propto \mathrm{const}$ with $\cos \iota \in [-1,1]$, and a
random sky position such that $p \, (\cos \theta) \propto
\mathrm{const}$  with $\cos \theta \in [-1,1]$  and $p \, (\phi) \propto
\mathrm{const}$ such that $\phi \in [0, 2 \pi]$. 

For those NS binary merger events with distances $< 200$ Mpc, we assume that the
spatial distribution of NS binaries traces host galaxy light. We use a ``Census of the Local Universe" (CLU) with information
compiled from different galaxy catalogs that provide B-band
luminosities (e.g., HyperLEDA, NED, EDD; see \citealt{KasliwalPhD:2011} for details). The probability that a binary
is located in a particular galaxy is weighted by the B-band luminosity of that galaxy in CLU. The size of the 
galaxy is assumed to be three times the size given by the surface
brightness contour at apparent magnitude 25 arcsec$^{-2}$ in CLU. 
B-band luminosity incompleteness is taken into account by dividing the catalog into 10\,Mpc bins and choosing random positions
for galaxies which represent the missing luminosity. 

Finally, for those binaries located with distances $> 200$
Mpc, we assume that the NS binary merger distribution has a constant comoving volume density
in a $\Lambda$CDM Universe \citep{komatsu09}.

In summary, we construct two catalogs of $4 \times 10^4$ NS-NS and $3 \times 10^4$ NS-$5
\,M_{\odot}$ BH binary populations, where each binary is described by its
set of physical and geometric source parameters: $\{ m_1, m_2, D_L, \cos \iota,
\cos \theta, \cos \phi \}$.


\section{GW Detectability and Parameter Estimation - Method}
\label{sec:GWmethod}


With the geometric and physical source parameters in hand for each
binary, we can simulate the predicted GW strain emitted for every
inspiraling NS binary in our two catalogs defined above. With knowledge of anticipated GW
interferometric noise curves and by assuming idealized Gaussian
instrument noise, we can simulate the predicted GW
data-stream measured at a particular GW interferometer. Therefore, by matched filtering a GW detector output with a theoretically-predicted
GW waveform, measurements of GWs will allow us: i) to detect NS-NS and
NS-BH binary inspiral and mergers, and ii) to extract the
physical and geometric properties of the source. We first
review our understanding of GW waveforms, and then introduce the
particular GW waveform that we use. Second, we outline the principles of
matched filtering used when detecting and estimating parameters of the
GW source. Finally, we describe the different GW networks considered
and the three triggering criteria used to
construct different GW-detected NS binary merger populations. Further details can
be found in N10 and N11.


\subsection{GW waveform}
\label{sec:waveform}


Turning to models of GW emission and dynamics, we view the GW waveform for merging compact binaries in terms of three phases; the inspiral, merger and
ringdown. The inspiral phase, describing the loss of energy and
angular momentum of the binary due to GWs, can be modeled accurately using the post-Newtonian (PN) approximation in
General Relativity. The PN approximation is an expansion in $\sim v^2/c^2$, where $v$ is
the characteristic orbital speed for gravitationally-bound systems. The state-of-the-art accuracy for non-spinning
inspiraling binaries is 3.5PN, corresponding to an order of ${\cal
  O} \, (1/c^7)$ in a PN expansion (e.g., \citealt{blanchet06}). At 3.5PN, NSs and/or
BHs are modeled using the `point'-particle (`$\delta$'-function) description, with finite-size
effects being formally negligible upto 5PN. However, several orbits prior to the
merger of the two bodies, the weak-field PN approximation is no longer valid, and we require
computationally-expensive, numerical simulations that model the merger
phase by directly solving Einstein Field Equations (see
e.g. \citealt{pretorius05}). After the two bodies have
merged into a final single BH, perturbation techniques of a Kerr BH
describe the ringdown. 

We make two important assumptions for the GW waveform
used in our work. Firstly, we assume that only the
inspiral phase models the GW signal in this work. This is
because the inspiral phase contributes to
the majority of the signal accumulated in the
frequency-band of advanced interferometers for NS-NS and
several stellar-mass NS-BH systems (see
\citealt{Flanagan:1998}). Typically, the inspiral phase of NS binaries
lasts from a few to tens of minutes in the interferometer's
frequency-band \citep{Cutler:1993}. Secondly, we
neglect the spin of NSs and BHs in our analysis. We expect NSs to have
small spins in
NS-NS and NS-BH systems. In contrast, BHs in NS-BH systems should have
moderate to large spins. In the case of NS-10 M$_{\odot}$ BH (with high spin) binaries, we
expect spin precessional effects to increase the dimensionality of the
parameter space and to modulate the GW waveform
significantly, which in some cases can improve GW sky
localization \citep{vandersluys:2008,Raymond:2009}. Consequently, neglecting the merger phase and assuming non-spinning binary
systems will introduce systematic errors when estimating source
parameters. However, as most GW detections will be at
threshold, we estimate that statistical errors will dominate over GW
waveform systematic errors for the majority of NS binary inspirals. 

The GW inspiral encodes a combination of the NS binary's physical and
geometric properties such as its redshifted masses, its luminosity distance, its
orientation, its source position, as well as the time and phase of
coalescence. Specifically, the predicted GW waveform at a particular
detector $a$ comprises the linear sum of the two GW polarizations
$h_{+}$ and $h_{\times}$ weighted by the two antenna functions
$F_{+}$ and $F_{\times}$, and is given by:
\begin{eqnarray}
h_a &=& D^{ij}_a h_{ij}
\nonumber\\
&\equiv& e^{- 2 \pi i ({\bf n}\cdot{\bf r}_a) f} (F_{a,+}h_+ +
F_{a,\times}h_\times)\;,
\label{eq:measuredwave}
\end{eqnarray}
where $D^{ij}_a$ is the detector's response tensor, and ${\bf r}_a$
denotes the detector's position in spherical polar coordinates from
the Earth's center. The scalar product ${\bf n}\cdot{\bf r}_a$ denotes the time-of-flight
of the signal from the source to the GW interferometer. In our work, we use a GW waveform in the frequency domain,
$h_+(f)$ and $h_{\times} (f)$, where the stationary-phase
approximation assumes that $(df/dt)/f \ll f$ \citep{droz99}. The GW
waveform used is accurate up to 3PN order in its phase (this improves the
accuracy of the waveform used in N10 and N11), and Newtonian in its
amplitude. The 3.5PN GW phase depends on physical source parameters such as
the redshifted chirp mass ${\cal M}_z\, = \, (1+z)  \, {\cal M}_c \, =\, (1+z) \, m_1^{3/5} \, m_2^{3/5}\,/\,(m_1
+ m_2)^{1/5}$, $z$ is the binary's
redshift (henceforth, the notation ${\cal M}_c$ refers to the physical
non-redshifted chirp mass), and $\mu_z = (1+z) \, m_1
m_2\,/\,(m_1 + m_2)$ is the redshifted reduced mass. The 3.5PN phase also
includes $t_c$ and $\Phi_c$, which are integration constants and
define the time and phase of
coalescence respectively. In contrast, the Newtonian-order GW
amplitude is a function of the GW frequency derivative $\dot{f}$ or so-called
`chirp' (itself a function which depends at Newtonian order on the
${\cal M}_z$ and at higher 1PN order on the $\mu_z$), and of the geometric parameters such as the binary's
$\cos \iota $, $D_L$ and
its source position $( \cos \theta, \phi )$. Physical source
parameters that appear in the GW phase and $\dot{f}$ can be determined
to a high accuracy for NS binaries because from thousands to tens of
thousands of inspiral GW cycles could sweep up in the frequency
band of the advanced GW detectors. However, only weak constraints on geometric source
parameters, such as $(\cos \iota, D_L)$, are possible because of
strong degeneracies that exist between parameters appearing in
the GW amplitude. This is the case for the majority of threshold-detected GW events (see
N10 for further discussion). However, for $( \cos \theta, \phi )$, differences in time-of-flight among
detectors in the network dominate over GW amplitude effects when
reconstructing the event's sky position. We terminate our
inspiral waveform abruptly at the ISCO, $f_{\rm ISCO}=(6
\sqrt{6} \pi M_z)^{-1}$, where $M_z=(1+z) (m_1+m_2)$ is the redshifted
total mass of the system. Such an abrupt cut-off of the GW waveform should have
little impact on matched filtering for NS-NS binaries where $f_{\rm ISCO}$ occurs at high
frequencies with poor detector sensitivity (see Figure 1). In contrast, for higher total mass NS-BH binaries, $f_{\rm ISCO}$ lies within
the frequency band with high detector sensitivity and such a
unphysical cut-off will introduce non-negligible systematic
errors. Finally, we assume that calibration measurement errors are
negligible \citep{Lindblom:2009,Vitale:2012}.


\subsection{GW Parameter Estimation}
\label{sec:paramest}


Turning to GW parameter estimation, we summarize MCMC methods
discussed in N10 and N11. Our central quantity of interest
is the posterior density function (PDF) of the distribution of inferred
source parameters, denoted by the vector of parameters
$\boldsymbol{\theta}$, following a GW measurement. The PN inspiral
waveform used in our work depends on the vector $\boldsymbol{\theta}$, which comprises the parameters $\{ {\cal
  M}_z, \mu_z, D_L, \cos \theta, \phi, 
\cos \iota, \psi, t_c, \Phi_c \}$. Following \cite{finn92} and \cite{cf94}, we consider a data-stream $s(t)$ measured at a detector $a$
that comprises the instrument noise $n (t)$ and a GW signal
$h(t,{\boldsymbol{\hat\theta}})$, where $\boldsymbol{\hat\theta}$
describes the source's ``true'' parameters, i.e., $s(t) = n(t)+h(t,{\boldsymbol{\hat\theta}})$. We assume that the noise
at each detector has idealized Gaussian statistics. For a network of
detectors, the PDF of the parameters $\boldsymbol{\theta}$
given some set of observed datastreams $\boldsymbol{s}$ is:
\begin{equation}
\label{eq:postPDF}
p({\boldsymbol \theta} \, | \, {\bf s}) \propto \, p^{(0)}
({\boldsymbol{\theta}}) {\cal L}_{\rm TOT} ({\bf s} \, |
\,{\boldsymbol{\theta}}) \,,
\end{equation}
where ${\cal L}_{\rm TOT} (\bf{s} \, | \,
{\boldsymbol \theta} )$ is the total {\it likelihood function} and $p^{(0)}({\boldsymbol{\theta}})$
is the prior PDF that describes our prior knowledge of the signal's
parameter distribution $\boldsymbol{\theta}$. The likelihood function measures the relative
conditional probability of observing a particular datastream $\bf{s}$
given ${\bf h}$ and ${\bf n}$.  By
assuming that the noise is independent at each interferometer, the
total likelihood function ${\cal L}_{\rm TOT} $ is equivalent to the product of the
individual likelihoods at each detector. The likelihood ${\cal L}_a$ for detector
$a$ is given by \citep{finn92}:
\begin{equation}
\label{eq:Like}
{\cal L}_a \, (s \, | \,{\boldsymbol \theta}) = \, e^{ -
\big( h_a({\boldsymbol \theta}) - s_a \, \big| \, h_a({\boldsymbol
\theta}) - s_a \big)/2 } \,.
\end{equation}
The notation $(g|h)$ describes the noise-weighted
cross-correlation of $g (t)$ with $h (t)$ in the vector space and is defined as
\begin{equation}
(g|h) = 2 \int_0^{\infty} df \frac{\tilde{g}^*(f)\tilde{h}(f) +
\tilde{g}(f)\tilde{h}^*(f)}{S_n(f)} \, ,
\label{eq:innerproduct}
\end{equation}
where $S_n(f)$ denotes the instrument's power spectral density. We
discuss the form of $S_n(f)$ in \S\ref{sec:GWdet}.

Using Eqn.~(\ref{eq:measuredwave}), for an ensemble of detector noise realizations, the {\it expected} SNR
at detector $a$ is given by:
\begin{eqnarray}
\nonumber
\left({S\over N}\right)_{a,\ {\mathrm{exp}}} & =& (h_a | h_a )^{1/2}, \\
\nonumber
& =& \sqrt{\frac{5}{96}} \frac{c}{D_L} \frac{2}{\pi^{2/3}} \left(\frac{G
{\cal M}_z}{c^3}\right)^{5/6} \\ & & \times \, \left[ F_{a \, , \,+}^2 (1 +
\cos^2 \iota)^2 \, + \, 4 F_{a \, , \,\times}^2 (\cos^2 \iota) \, \right]^{1/2}
\nonumber \\ & & \times \, \left[ \int_{f_{\rm low}}^{f_{\rm ISCO}}
\frac{f^{-7/3}}{S_h(f)} df \right]^{1/2} \, ,
\label{eq:snr_ave} 
\end{eqnarray}
where $f_{\rm low}$ is the instrument's low-frequency
cut-off.  Averaging over all possible binary sky positions and
orientations, the {\it expected sky-and-inclination-averaged} SNR is given by (see \citealt{dalaletal}):
\begin{eqnarray}
\nonumber
\left({S\over N}\right)_{a,\ {\rm sky-inc-ave}} & =& \frac{8}{5}
\sqrt{\frac{5}{96}} \frac{c}{D_L} \frac{1}{\pi^{2/3}} \left(\frac{G
{\cal M}_z}{c^3}\right)^{5/6} \\ & & \times \, \left[ \int_{f_{\rm low}}^{f_{\rm ISCO}}
\frac{f^{-7/3}}{S_h(f)} df \right]^{1/2} \, .
\label{eq:snr_ave_sky_orien} 
\end{eqnarray}
For a binary that is directly face-on to an observer ($\cos \iota = \pm 1$), its SNR$_{\mathrm{exp}}$ is
a factor of $\sqrt{5/2} \simeq 1.58$ greater than its
inclination-averaged counterpart SNR$_{{\rm inc-ave}}$ at the
same position and distance, and a factor of $\sqrt{5/4} \simeq 1.12$
greater than its sky-and-inclination-averaged SNR$_{{\rm
    sky-inc-ave}}$ (see N10 and \citealt{dalaletal}). Throughout the
rest of the paper, we use Eqns.~(\ref{eq:snr_ave}) and
(\ref{eq:snr_ave_sky_orien}) for GW-detectability (SNR-based) scaling arguments. We define the \emph{expected network SNR} as the root-sum-square of the expected individual detector SNRs.

To infer the geometric source parameters for each NS binary considered, in particular the
subset $( D_L,\cos \iota, \cos\theta, \phi)$, we map out the full
posterior PDF of all source parameters using MCMC, given an observed
data stream $\mathbf{s}_a$ at a detector. The Metropolis--Hastings MCMC algorithm used is
based on a generic version of CosmoMC, described in \cite{lewis02}. We
assume prior distributions in all source parameters to be flat over the region of sample space where
the binary is detectable at an expected network SNR = 3.5. For each
MCMC simulation used on a single NS binary inspiral, we derive marginalized
parameter measures and rms errors over $( D_L,\cos \iota, \cos\theta, \phi)$ at 68$\%$,
95$\%$, and 99$\%$ confidence regions (henceforth, denoted c.r.).

\subsection{GW networks}
\label{sec:GWdet}


We consider GW networks consisting of combinations of LIGO (which comprises LIGO Hanford and LIGO Livingston), Virgo, LIGO India, and
KAGRA. In the rest of the paper, we use the following notation to
describe different GW networks with $n$ detectors:
\begin{itemize}
\item[-] \emph{Net3} or network 3 is LIGO Hanford, LIGO Livingston and Virgo, 
\item[-] \emph{Net4I}  or network 4I is LIGO Hanford, LIGO Livingston, Virgo and LIGO India,
\item[-] \emph{Net4K} or network 4K is LIGO Hanford, LIGO Livingston, Virgo and KAGRA, and 
\item[-] \emph{Net5} or network 5 is LIGO Hanford, LIGO Livingston, Virgo, LIGO India and KAGRA.
\end{itemize}

Apart from LIGO India, the detector's positions (as measured
from the Earth's center) used in this work are
given in Table 1 of N10. For LIGO India's position and orientation, we use East Longitude $\lambda = 76.7$,
North latitude $\varphi = 14.3$, Orientation $\upsilon =  0$,
$x$-arm tilt $\Omega_x = 0$ and $y$-arm tilt 
$\Omega_y = 0$. 

\begin{figure}
\centering 
\includegraphics[angle=90,width=0.5\textwidth]{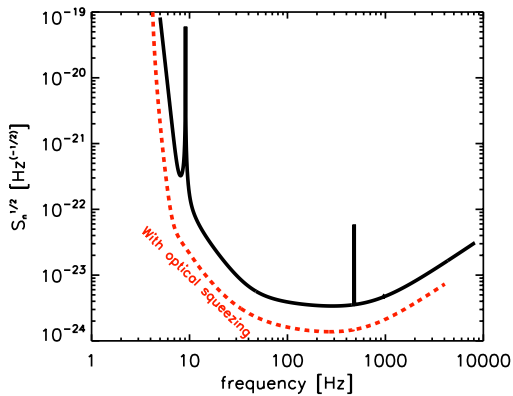}
\caption{Anticipated noise curves for Advanced LIGO (solid black line;
  \citealt{ligo_noise}) and Advanced LIGO with optical squeezed light (dashed red line;
 \citealt{Miao:2012}). In this paper, we assume a low frequency cut-off of 10 Hz. The features at 10 and a few
hundred Hz are various thermal noise resonant modes of mirror suspensions. The modes at a few hundred Hz are suspension fiber resonances.
}
\vspace{0.2in}
\label{fig:noiseoptsq}
\end{figure}

For simplicity, we assume that the noise sensitivity curve for each detector is represented by
the anticipated broadband-tuned sensitivity curve for a single
advanced LIGO detector, shown in Fig.~\ref{fig:noiseoptsq}. We
impose a low-frequency cut off at 10 Hz and frequencies
below 10 Hz are not included in our analysis. In practice, LIGO, Virgo
and KAGRA will have different noise sensitivities in
different frequency bands because of variations in each instrument's
design. We also consider the anticipated sensitivity
curve for LIGO Livingston and LIGO Hanford interferometers using
optically squeezed light \citep{Miao:2012}. To compare different GW
network abilities, for each NS-NS or NS-BH binary, we assign a
unique noise realization to each GW detector, which we keep constant when
adding and subtracting detectors to a network. In addition, we assume
that each GW interferometer operates at an idealized 100\%  of the
time (see \citealt{Schutz:2011} for estimates of different instruments' duty cycles).

\vspace{0.2in}

\subsection{GW triggering criteria}
\label{sec:gwtrigger}


For each binary in the two catalogs, we implement three GW triggering
scenarios that use:

\begin{itemize}

\item a \emph{coincident} trigger criterion (denoted ``{\it a}''): we
  select the binary if at least two GW detectors each have an SNR\,$>$\,6 and 
  if the expected network SNR\,$>$\,12. The estimated number of GW templates
required and desired false-alarm rates determines the choice in particular SNR
threshold values~(\citealt{Owen:1996} and see e.g., N11 and \citealt{Abadie:2012PRD}).

\item a \emph{coherent} trigger criterion (denoted ``{\it b}''): we select the binary if its
GW expected network SNR\,$>$\,8.5
(see e.g., N11 and \citealt{Harry:2011}).

\item an \emph{EM-precursor coherent} trigger criterion: we select the
  binary if its GW expected network SNR\,$>$\,7.5.  Henceforth, to avoid confusion with
the coherent trigger above, we refer to this
case as an \emph{EM-precursor} trigger criterion.  As discussed in N10, we choose to lower the network threshold in the
presence of an already observed EM counterpart because prior
knowledge of the merger time and sky position reduces the number of
searched GW
templates. Models of EM-precursor emission to NS binary mergers include resonant
shattering of NS crusts observable in $\gamma$- and X-rays (e.g. \citealt{Tsang:2012}) and a coherent burst of radio emission
produced by magnetically dominated outflows \citep{Pshirkov:2010,Piro:2012}. 

\end{itemize}

For all scenarios, the interferometers may have significantly non-Gaussian
noise, necessitating an increase in the SNR threshold to take into
account complex detector statistics.

\section{GW-Detected Binary Populations - Results}
\label{sec:results_detbin}


This section presents the \emph{relative} fraction, rates and
distributions of geometric parameters of NS-NS or NS-BH inspirals
detected by different GW networks of interferometers. We define the
term \emph{relative} fraction to be the ratio of GW-detected binary
mergers out of the total number of binaries in each catalog. 

\subsection{Relative fractions of GW-detected events}
\label{sec:fracGW}
For each NS binary merger in our two catalogs, we compute and compare GW SNRs to a defined threshold SNR at a
particular detector or at a network (\S~\ref{sec:gwtrigger}). For
different progenitors, GW networks, and triggering criteria, we thus obtain fractions of
those mergers that are detectable by GWs out of the catalog's total number
of systems. Table~\ref{tab:detbinaries} shows relative fractions of GW-detected merger samples. 

Table~\ref{tab:detbinaries} also indicates the relative fraction of
GW-detected mergers that have their orbital angular momentum vectors oriented towards the Earth such that they could
  show collimated $\gamma$- and X-ray emission. Observations exist for
  two SGRBs indicating a half-jet opening angle $\theta_{\mathrm j}$ of $\sim 7^{\circ}$ \citep{burrowsetal06,Soderberg:2006} and $\sim
3-8^{\circ}$ \citep{Fong:2012}. A handful of other SGRBs exhibit upper
and lower bound jet-break measurements, discussed in
\S\ref{sec:EMreview}. In this work, we define \emph{beamed} binaries to be those binaries whose orbital angular
momentum vector lies within a relatively stringent $\theta_{\mathrm j}$ of 6$^{\circ}$. 

Table~\ref{tab:detbinaries} illustrates several trends
between different samples of GW-detected NS binary
mergers. \cite{Schutz:2011} provides powerful analytically-derived
expressions that show good agreement with our explicit results. First, the
fraction $f_{\mathrm{beamed}}$ of beamed NS binary mergers seen in
GWs, observable from all possible inclination angles $\iota$, is less than those binaries with isotropic orientation (see \citealt{Schutz:2011}
and \citealt{Metzger:2012}). When $\iota \ll \theta_{\mathrm j}$, $f_{\mathrm{beamed}}
\sim 1- \cos \theta_{\mathrm j} \sim \theta_{\mathrm j}^2/2$ for small
$\theta_{\mathrm j}$; in our case, the empirically-derived
range $f_{\mathrm{beamed}} \sim 0.5 - 1 \%$ agrees well with its theoretical
value of $0.7 \%$. Second, the relative fraction of
GW-detected NS-5M$_{\odot}$ BH mergers is greater by a factor from four to five
than GW-detected NS-NS mergers. This follows from Eqn.~(\ref{eq:snr_ave_sky_orien}), where
SNR scales as ${\mathcal M}_c^{5/6}$ and detectable volume thus scales
as ${\mathcal M}_c^{15/6}$. For our NS-NS and NS-5$M_{\odot}$ BH
   inspirals, values for ${\mathcal M}_c$ are 1.21$M_{\odot}$ and 2.22$M_{\odot}$
   respectively. Third, the fractions of GW-detected events where the two LIGO interferometers use squeezed light are a factor from $\sim
  9$ to 10 greater than for those networks where no optical squeezing
  is implemented. Illustrated in Figure~1, the optically-squeezed advanced LIGO noise curve is a
  factor from two to three more sensitive than the standard analog's
  curve. From Eqn.~(\ref{eq:snr_ave_sky_orien}), such an improvement in instrument sensitivity translates to an improvement
  by a factor of $\sim 2^3-3^3$ in detectable volume (because SNR is inversely proportional to $D_L$). Fourth, increasing the number of GW detectors in a network from 3 to 5 increases
  the number of GW-detected mergers by a factor of $\sim$ 2
  or less (see N10). Shown in Eqn.~(\ref{eq:snr_ave_sky_orien}), the network SNR$_{\mathrm{exp}}$
  scales as $\sim \sqrt{n}$, where $n$ is the number of detectors.

\subsection{Estimated relative rates of GW-detected inspirals}
\label{sec:ratesGW}

To convert relative GW-detected fractions into \emph{relative
  GW-detected rate predictions}, we first require
estimates of the astrophysical NS binary merger rate, independent of
GW detection. In the case of NS-NS binaries, different astrophysical merger rates are derived either by extrapolating the
distribution of observed Galactic
binary pulsars or from population synthesis results (see
e.g., \citealt{Phinney:1991} and references in \citealt{Abadieetal:2010}). The
rates range from 0.01 to 10 Mpc$^{-3}$
Myr$^{-1}$, with 1 Mpc$^{-3}$ Myr$^{-1}$ being the mean of the
rate's PDF \citep{Abadieetal:2010}. In contrast, because we have yet to
observe a NS-BH system, all estimates of NS-BH merger rates are based entirely on theoretical population synthesis results. The
rates range from 6 $\times 10^{-4}$ to 1 Mpc$^{-3}$ Myr$^{-1}$, where
0.03 Mpc$^{-3}$ Myr$^{-1}$ is defined as a realistic rate in
\cite{Abadieetal:2010}. Therefore, our estimates for GW-detected merger rates rely on theoretical predictions
of NS binary merger rates that span three orders of magnitude.

Here, we estimate ${\cal R}_{\mathrm{NS-X}}$, NS binary merger rates detected by networks
of GW interferometers, using: 
\begin{equation}
\label{eqn:detrate}
{\cal
  R}_{\mathrm{NS-X}}\, =\, {\cal N}_{\mathrm{NS-X}} \, \times \, f
_{\mathrm{NS-X}} \, \times \, \, V \, \times \frac{1}{k} \,\,,
\end{equation}
where the subscript $X$ denotes a NS or BH, and ${\cal N}_{\mathrm{NS-NS}}$ and ${\cal N}_{\mathrm{NS-BH}}$ are the
astrophysical NS-NS and NS-5 M$_{\odot}$ BH merger rates in Mpc$^{-3}$ yr$^{-1}$ respectively. $V$ is the total volume that
the catalogs encompass (in our case, this corresponds to $\sim 4/3
\times \pi \times (2.82 \, \, \rm{Gpc})^3$) and $f _{\mathrm{NS-X}}$
are the relative fractions of GW-detected NS binary mergers. Table 1 gives values of $f_{\mathrm{NS-X}}$ for different progenitors, GW networks and triggering
criteria. The factor $k \sim 3\sqrt{3}$ applies to all networks with any number of detectors. It incorporates the \cite{Abadieetal:2010} correction for the GW interferometers'
non-stationary and non-Gaussian noise, applied in order to achieve required false-alarm rates. 

\begin{figure}
\centering 
\includegraphics[width=0.45\textwidth]{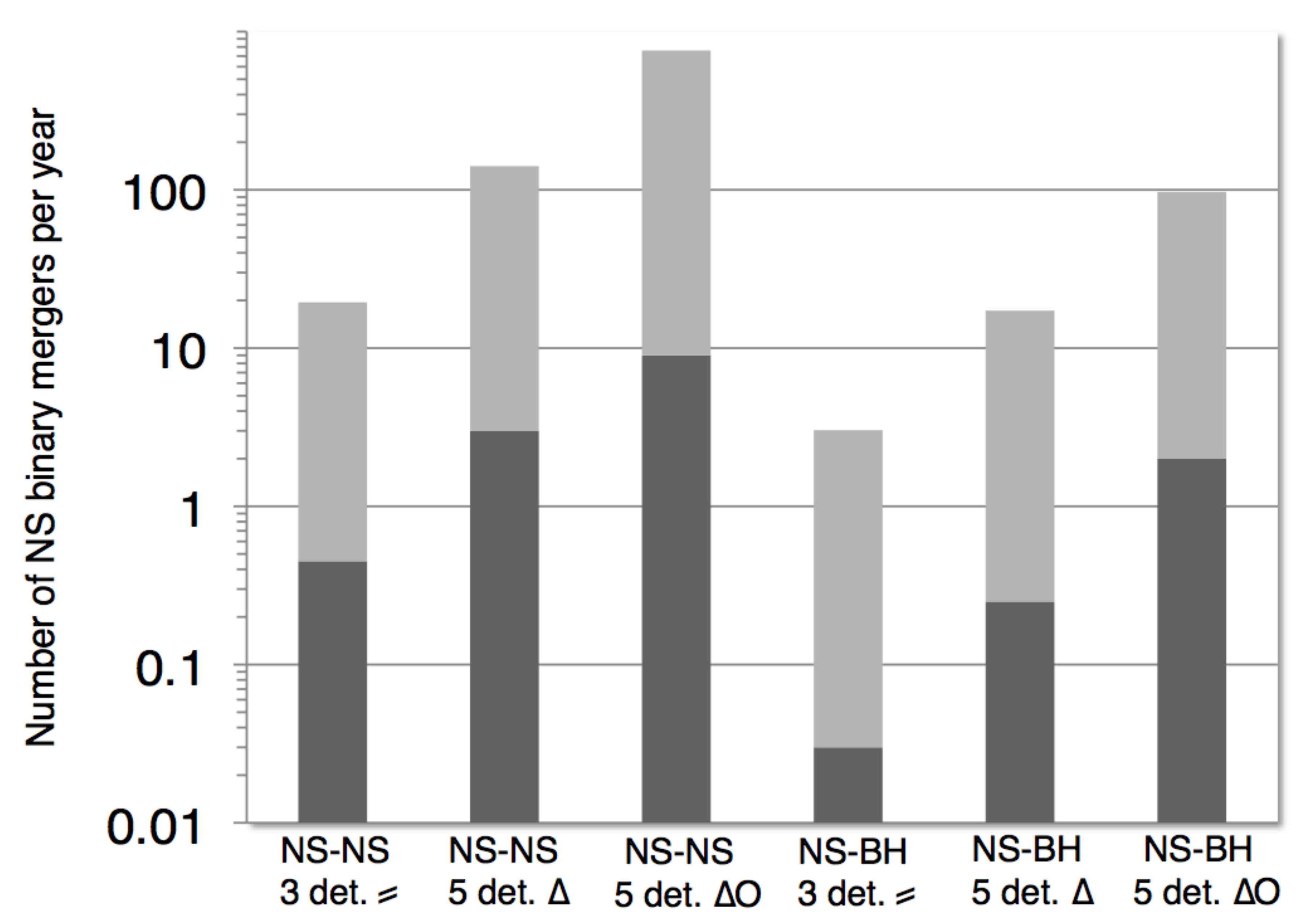}
\vspace{0.2in}
\caption{Relative rate of NS binary mergers detectable by
  different GW networks and triggering criteria. The dark grey shaded regions denote
those binaries that have their orbital angular
  momentum vector lying within a half-jet
  opening angle of 6$^{\circ}$ and the light grey shaded regions denote those
  binaries whose orbital angular momentum have half-jet
  opening angle greater than 6$^{\circ}$. The notation {\bf 3
  det.}\protect\coincident \, refers to \emph{Net3a}, a
coincident-triggered network 3, {\bf 5 det.} $\Delta$  denotes \emph{Net5b}, a coherent-triggered network 5,
and {\bf 5 det.} $\Delta \bigcirc$ represents a \emph{Net5b} with
optical squeezing in the two LIGO interferometers. We use the mean and/or realistic rate of NS binary mergers
quoted in \cite{Abadieetal:2010}.}
\vspace{0.2in}
\label{fig:GWdetrates}
\end{figure}

Figure~\ref{fig:GWdetrates} shows relative rates of GW-detected NS
binary mergers which have either isotropic or beamed emission. We use the mean and/or realistic rate of NS binary mergers
quoted in \cite{Abadieetal:2010}. Given the few orders-of-magnitude uncertainty, we use NS-10 M$_{\odot}$ BH merger rates as
representative for the merger rates of NS-5M$_{\odot}$ BH systems used in this work. For NS-NS mergers, we use a value of
$\sim$ 18 270 for the prefactor $[ \,{\cal N}_{\mathrm{NS-X}} \, \,
\,\frac{V}{k} ]$ in Eqn.~(\ref{eqn:detrate}). For NS-BH mergers, we use a value of $\sim
550$ for the prefactor $[\, {\cal N}_{\mathrm{NS-BH}} \, \,\frac{V}{k}
\, ]$. Let us now discuss several features of Figure~\ref{fig:GWdetrates}. 

First, out of all possible GW networks and triggering schemas, we present
relative NS-NS and NS-BH binary merger rates detected by GWs under
three scenarios:
\begin{itemize}
\item[i)] \emph{Net3a}: {\bf coincident-triggered network 3}
\item[ii)] \emph{Net5b}: {\bf coherent-triggered network 5}
\item[iii)] Net5b with optically-squeezed LIGO.
\end{itemize}
\noindent
Due to the high SNR threshold required at two
detectors or more, Net3a detects the fewest number of NS-NS or NS-BH mergers. It
provides a \emph{lower bound} on the number of detected GW events, indicative of how
the early years of GW measurements might unfold. In
contrast, Net5b detects the largest number of mergers of NS-NS or NS-BH mergers, because the coherent-network SNR scales as
$\sqrt{n}$, where $n=5$ is the maximum number of
detectors. It hence provides an \emph{upper bound} on the number of
GW-detected mergers, suggestive of how a GW network might operate after the first several years of
GW measurements. The third scenario, envisioned later in the timeline
of the development of the GW network, provides a highly optimistic bound for NS-NS and NS-BH
mergers detected using Net5b with optically-squeezed LIGO; we choose to investigate this scenario in a future
study. Summarized in Table 2, we use Net3a and Net5b to indicate
representative bounds for the performance between different GW
networks and triggering schema. Illustrated by Figure~\ref{fig:GWdetrates}, our GW-detected NS binary merger rate estimates show
good agreement with earlier works (e.g.,~\citealt{Abadieetal:2010}). Differences occur because of different triggering criteria invoked and SNR thresholds used. 

Second, the dark grey shaded regions in Figure~\ref{fig:GWdetrates} denote
those binaries that have their orbital angular momentum vector lying
within a relatively stringent $\theta_{\mathrm j} < 6^{\circ}$
and the light grey shaded regions denote those binaries whose orbital
angular momentum have $\theta_{\mathrm j} > 6^{\circ}$. From one to several NS binary mergers 
  per year could have $\theta_{\mathrm j} < 6^{\circ}$ and may exhibit $\gamma$-ray
  collimation associated with SGRBs. Our beamed NS-NS binary merger rates are
  consistent with SGRB rates of 10 Gpc$^{-3}$ yr$^{-1}$, discussed in
  \cite{Metzger:2012}, \cite{Coward:2012}, \cite{Petrillo:2012} and \cite{chenholz:2012}.  

Third, we emphasize that lower and upper bounds to the rate estimates
differ, for instance, for NS-NS mergers by two orders of magnitude below and an order of magnitude above
  from the mean values that we use. Results presented in Figure~\ref{fig:GWdetrates} are instructive in that they illustrate relative GW detectability rates between different GW networks and triggering criteria, but the values given here should be used
    with caution.

\vspace{0.1in}

\begin{center}
\begin{deluxetable*}{llcccc}
\centering
\tabletypesize{\scriptsize}
\tablewidth{17.0cm}
\vspace{0.2in}
\tablecaption{Relative fractions $\times 10^{-4}$ of NS--NS and
  NS--5 M$_{\odot}$ BH Mergers with collimated (denoted `B') and
  isotropic (denoted `I') emission Detectable in GWs Using Three Different Selection Criteria
  with Four GW Networks. The notation `OS' represents optical
  squeezing in the LIGO interferometers. The range given
  represents the 1-$\sigma$ statistical error of our simulation. }
\tablehead{
\colhead{GW Network} && 
\multicolumn{1}{c}{Net3} &
\multicolumn{1}{c}{Net4I} &
\multicolumn{1}{c}{Net4K} &
\multicolumn{1}{c}{Net5}  \\ \\
\colhead{} & & 
\colhead{B$\mid$I} &
\colhead{B$\mid$I} &
\colhead{B$\mid$I} &
\colhead{B$\mid$I} \\
}
\startdata
\hline\\
\multirow{5}{*} {Coincident {\it``a''}}&&&& \\
&NS-NS
&0.3 $\pm$ 0.3$\mid$11 $\pm$ 2&0.3 $\pm$ 0.3$\mid$17 $\pm  $ 2&0.3 $\pm$ 0.3$\mid$17 $\pm
$2&0.3 $\pm$ 0.3$\mid$23 $\pm  $ 2\\
&&&&& \\
&NS-5M$_{\odot}$BH&0.7 $\pm$ 0.5$\mid$50 $\pm$4&1.0 $\pm$ 0.6$\mid$79
$\pm$ 5&1.3 $\pm$ 0.7$\mid$77 $\pm  $ 4&2.3 $\pm$ 0.9$\mid$104 $\pm  $ 6\\
&&&&& \\
\hline \\
\multirow{5}{*}{Coherent {\it``b''}}&&&& \\
&NS-NS&0.8 $\pm$ 0.4$\mid$36 $\pm$ 3&0.8 $\pm$ 0.4$\mid$57 $\pm$ 4&0.8 $\pm$ 0.4 $\mid$59 $\pm  $ 4&1.5 $\pm$ 0.6$\mid$78 $\pm$ 4\\
&&&&& \\
& NS-5M$_{\odot}$BH &2.3 $\pm$ 0.9$\mid$170 $\pm$ 7&3.7 $\pm$
1.1$\mid$251 $\pm$ 9&4.0 $\pm$ 1.2$\mid$243$\pm$ 9&4.7 $\pm$ 1.2$\mid$323 $\pm$ 10\\
&&&&& \\
\hline \\
\multirow{5}{*}{EM precursor}&&&& \\
&NS-NS &0.8 $\pm$ 0.4$\mid$54 $\pm$ 4 &1.0 $\pm$ 0.5$\mid$80 $\pm$4&1.8 $\pm$ 0.6$\mid$81 $\pm$ 4&2.0 $\pm$ 0.7$\mid$113 $\pm$5\\
&&&&& \\
& NS-5M$_{\odot}$BH &3.0 $\pm$ 1.0$\mid$244 $\pm  $ 9&4.0 $\pm$ 1.2$\mid$350 $\pm$11&4.3 $\pm$ 1.2$\mid$350 $\pm$11&6.3 $\pm$ 1.5$\mid$464 $\pm$ 12 \\
&&&&& \\
\hline \hline \\
\multirow{5}{*} {Coincident  {\it``a''} $+$ O.S.}&&&& \\
&NS-NS
&2.0 $\pm$ 0.8$\mid$129 $\pm  $ 7&2.0 $\pm$ 0.8$\mid$140 $\pm  $
7&2.0 $\pm$ 0.8$\mid$140 $\pm  $
7&2.3 $\pm$ 0.9$\mid$152 $\pm  $ 7\\
&&&&& \\
&NS-5M$_{\odot}$BH&6.7 $\pm$ 1.5$\mid$490 $\pm$ 12&8.0 $\pm$ 1.6$\mid$535 $\pm$
13&8.7 $\pm$ 1.7$\mid$534 $\pm$ 13&9.3 $\pm$ 1.7$\mid$579 $\pm  $ 13\\
&&&&& \\
\hline \\
\multirow{5}{*} {Coherent {\it``b''} $+$ O.S.}&&&& \\
&NS-NS
&4.7 $\pm$ 1.2$\mid$364 $\pm  $ 11&4.7 $\pm$ 1.2$\mid$391 $\pm  $ 11&4.7 $\pm$ 1.2$\mid$390 $\pm  $
11 &5.0 $\pm$ 1.3$\mid$418 $\pm  $
12 \\
&&&&& \\
&NS-5M$_{\odot}$BH&27.7$\pm$3.0$\mid$1517 $\pm$ 21&30.0 $\pm$ 3.2$\mid$1643 $\pm$
21&30.7 $\pm$ 3.2$\mid$1640 $\pm$ 21&33.0 $\pm$ 3.3$\mid$1777 $\pm$ 22\\
&&&&& \\
\hline \\
\multirow{5}{*} {EM precursor $+$ O.S.}&&&& \\
&NS-NS
&8.0 $\pm$ 1.6$\mid$517 $\pm  $
13&9.0 $\pm$ 1.7$\mid$565 $\pm  $
13 &9.3 $\pm$ 1.8$\mid$557 $\pm  $
13&10.0$\pm$ 1.8$\mid$610 $\pm  $
14\\
&&&&& \\
&NS-5M$_{\odot}$BH&33.0$\pm$ 3.3$\mid$2062 $\pm $ 23&36.0 $\pm$
3.4$\mid$2248 $\pm$ 24&37.0 $\pm$ 3.5$\mid$2246 $\pm$ 24&39.3 $\pm$
3.6$\mid$2425 $\pm$ 25\\
&&&&& \\
\enddata
\label{tab:detbinaries}
\end{deluxetable*}
\end{center}

\vspace{0.1in}
\begin{deluxetable*}{lcc}
\tablewidth{16cm}
\tablecaption{Representative GW Network Scenarios for Detectable Samples
  of NS-NS and NS-BH mergers. The notation med. and max. refer to the
  median and maximum values of parameter distributions.}
\tablehead{
\colhead{Feature} &
\colhead{Lower Bound Scenario} &
\colhead{Upper Bound Scenario} 
}

\startdata
& & \\
Relative fractions \& rates & Coincident 3 detector: \bf{Net3a} & Coherent 5
detector: \bf{Net5b} \\
&  19 yr$^{-1}$ (NS-NS) $\mid$ 3 yr$^{-1}$ (NS-BH)&
138  yr$^{-1}$ (NS-NS) $\mid$ 17  yr$^{-1}$ (NS-BH) \\
& & \\
Detectable distance& Coincident 3 detector: \bf{Net3a} & Coherent 5
detector: \bf{Net5b} \\
& 220--400 Mpc (med-max; NS-NS) & 390--750 Mpc (med-max; NS-NS)  \\
&350--600 Mpc (med-max; NS-BH) & 650--1250 Mpc (med-max; NS-BH)\\
& & \\
Sky Area Errors & Coherent 3 detector:  \bf{Net3b} &Coincident 5
detector: \bf{Net5a} \\
& 55--180 deg$^2$ (med-max; NS-NS) & 7--120 deg$^2$ (med-max; NS-NS)  \\
& 50--170 deg$^2$ (med-max; NS-BH) & 6--65 deg$^2$ (med-max; NS-BH)
\\
\enddata
\label{tab:casestudies}
\end{deluxetable*}

\subsection{GW Malmquist effect in detected events' distance and
  inclination angle}
\label{sec:malmquist}

GW detection criterion sets implicit prior distributions on geometric parameters of
NS binary mergers (see N10 and \citealt{Schutz:2011}). Defined here as the GW Malmquist effect, our GW detection criterion
preferentially selects for more face-on (or equivalently beamed and more `GW-luminous' binary inspirals); see Eqn.~(\ref{eq:snr_ave}). The GW Malmquist bias is
analogous to the standard Malmquist effect in
observational astronomy, where intrinsically brighter objects are
detected further out. In GWs, beamed ($\cos \iota \rightarrow \pm
1$) binaries have higher SNRs and are intrinsically more luminous in
GWs (Eqn.~(\ref{eq:snr_ave})). 

\begin{figure}
\centering 
\subfigure[NS-NS mergers detected in GWs by Net3a]{\label{fig:cosincDLNSNS}\includegraphics[width=0.45\textwidth]{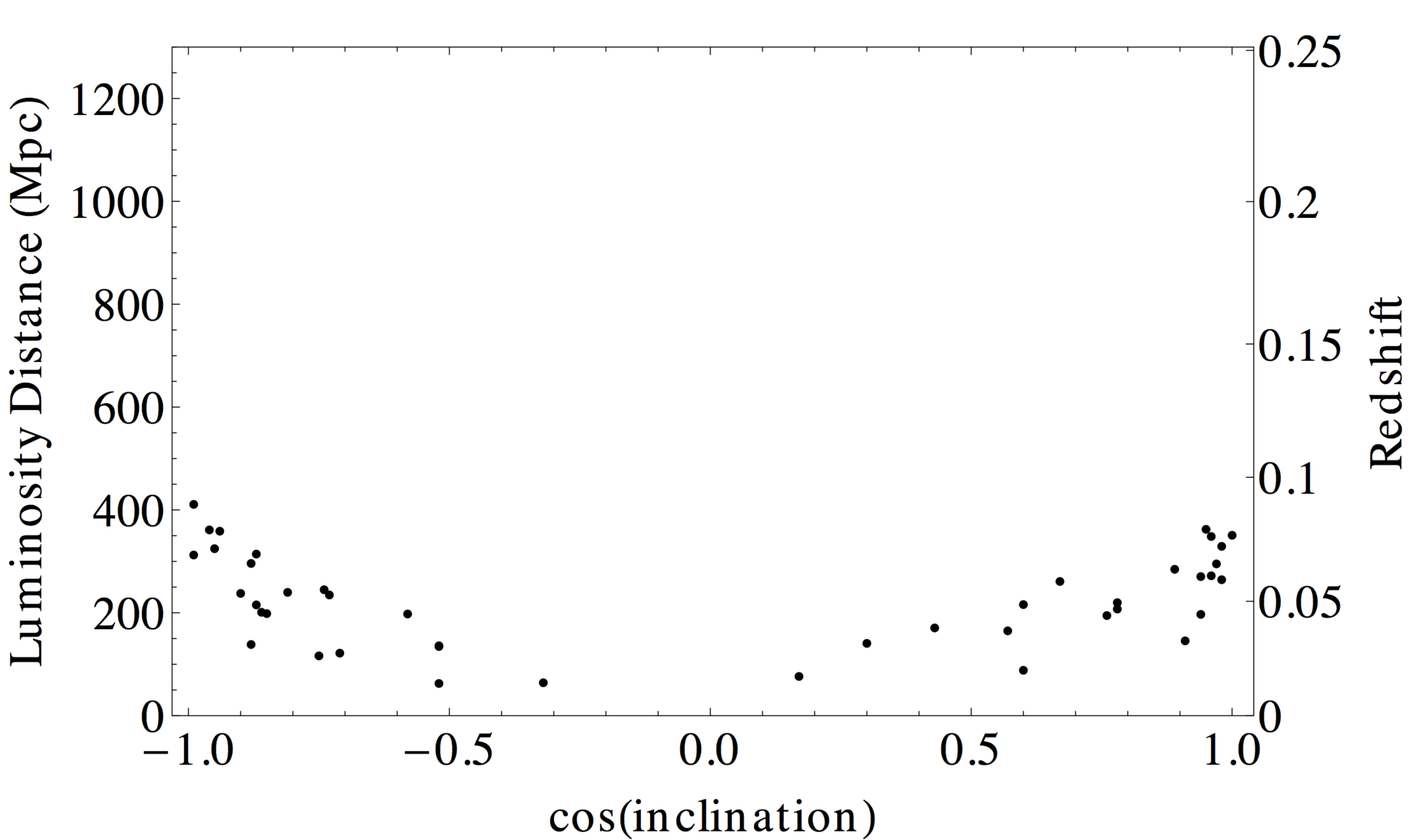}}
\subfigure[NS-NS mergers detected in GWs by Net5b]{\label{fig:cosincDLcoherent}\includegraphics[width=0.45\textwidth]{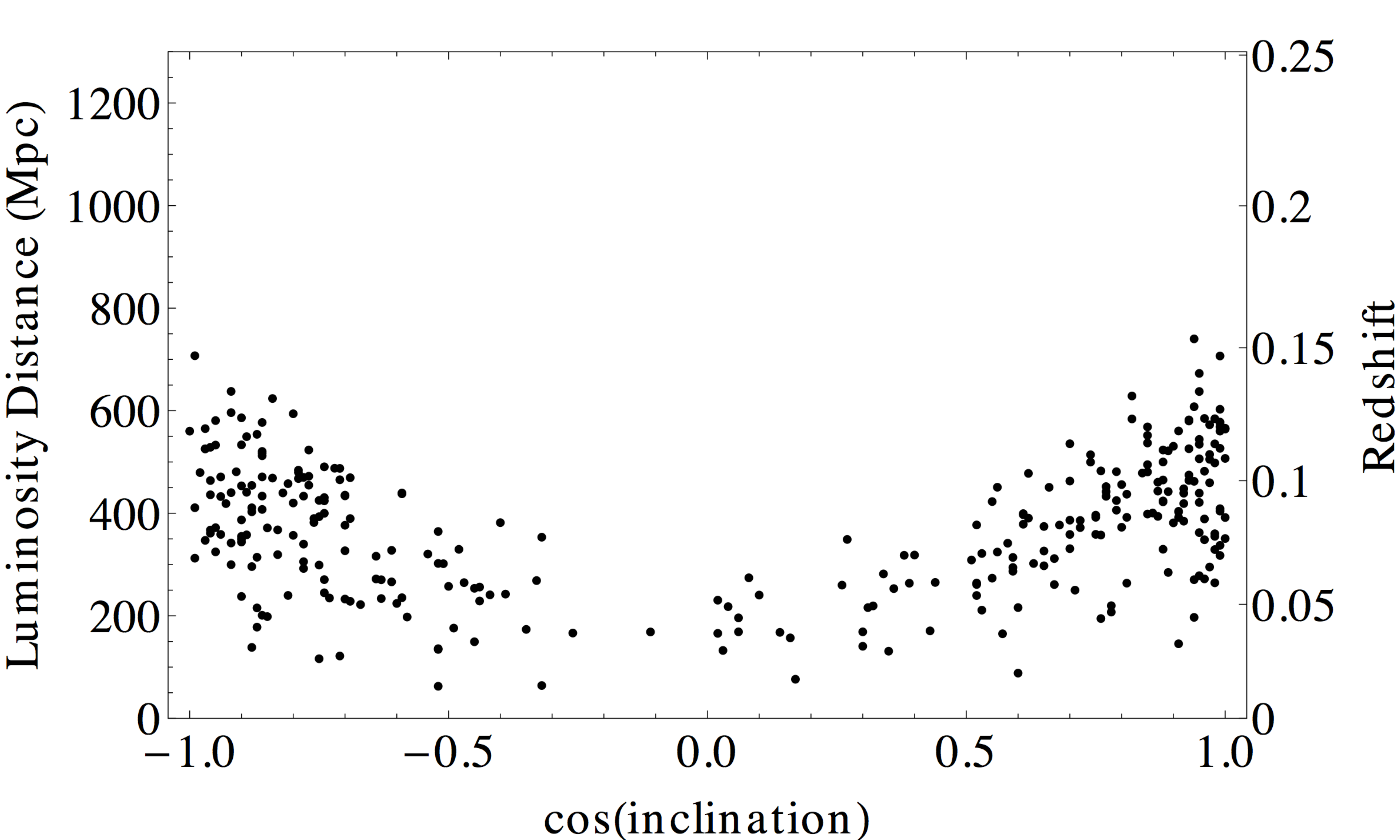}}
\subfigure[NS-BH mergers detected in GWs by Net5b]{\label{fig:cosincDLNSBH}\includegraphics[width=0.45\textwidth]{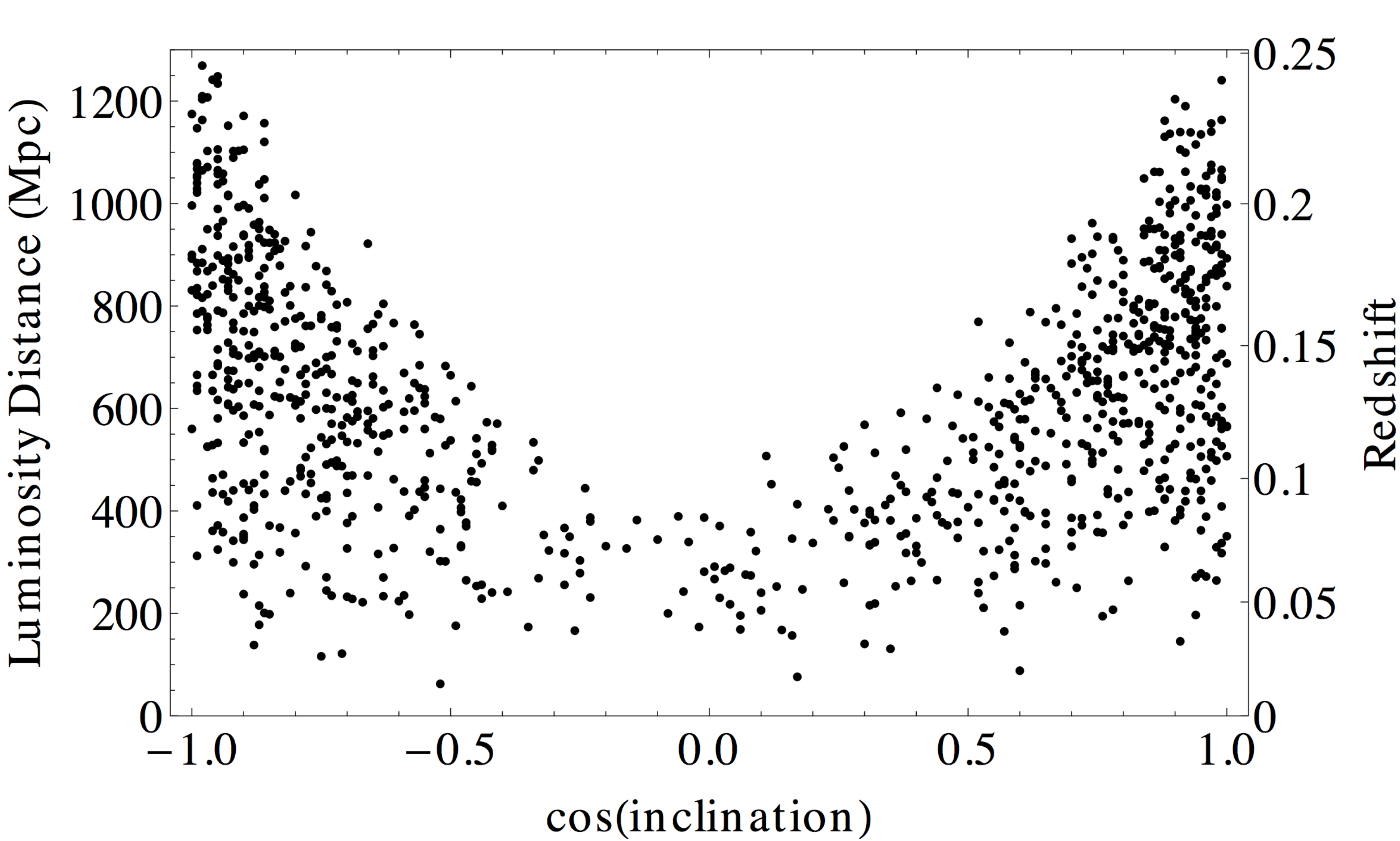}}
\caption{The 2-D marginalized prior distribution in $D_L$ and $\cos
  \iota$ for GW-detectable NS binary
mergers. Each point represents a GW-detectable NS binary merger.  The
top panel shows NS-NS mergers detected by Net3a. The middle panel
shows NS-NS mergers detected by Net5b. The bottom panel shows NS-BH mergers
detected by Net5b. Redshifts are computed
assuming cosmological parameters given in \cite{komatsu09}.
}
\label{fig:cosincDL}
\end{figure}

Figure~\ref{fig:cosincDLNSNS} shows the 2-D distribution for
parameters $\left( D_L \,,  \cos\iota\right)$ of NS-NS mergers
detected in GWs with Net3a. Important for EM follow-up and for
coincident EM and GW observations, we remark on noteworthy features of
the distribution. Figure~\ref{fig:cosincDLNSNS} illustrates the GW Malmquist bias towards detection of beamed binaries, with
$\cos \iota \rightarrow \pm 1$. The distribution exhibits a
characteristic V-shape which is consistent with the analytically-derived PDF of detected values in $\iota$ given in Eqn. (28) of \cite{Schutz:2011}. Unsurprisingly, we detect the
majority of events at threshold and observe a paucity of close-in binaries, detected with distances less than 100
Mpc. The maximum distance for NS binary mergers detectable by Net3a is
$\sim 400$ Mpc. We note that the closest SGRBs with known redshifts
are 080905, 050709, and 050724 at $z = 0.122$ ($\sim$ 560 Mpc), $z = 0.161$ ($\sim$
760 Mpc), and $z = 0.257$  ($\sim$ 1.28 Gpc); see
\cite{Berger:2010}. Therefore, the maximum detectable distance
range of Net3a does not include the distances of the
three closest SGRBs observed so far. Discussed in \cite{Metzger:2012}, the lack of
SGRBs observed within a few hundred Mpc is consistent with the
\emph{Swift} satellite's observational biases: only $\sim$ 1/10 of the
sky is surveyed at a particular epoch and only $\sim$ 1/3
of SGRBs observed by \emph{Swift} have redshifts. Finally, in Figure~\ref{fig:cosincDLcoherent}, we show the 2-D distribution of NS-NS
mergers for the parameters $\left( D_L, \cos\iota \right)$ detected by
Net5b. We note that the maximum detectable
distance increases by a factor of 1.5 compared to Net3a. Moreover, in the case of our NS-5
M$_{\odot}$ BH catalogs, the SNR and hence detectable
distance depends on the chirp mass ${\cal M}_{\rm c}^{5/6}$ (Eqn.~(\ref{eq:snr_ave_sky_orien})). Therefore, the maximum detectable
distance increases to above 1 Gpc in Figure~\ref{fig:cosincDLNSBH}.

\subsection{Cumulative distribution of GW distances and localization
  errors}
\label{sec:cumdistskyloc}

Critical for EM follow-up, we examine cumulative distributions in GW distance and sky
area errors for populations of NS-NS and NS-BH inspirals detected
using different GW detector networks and triggering criteria. For
representative populations of GW-detected NS binary mergers, we randomly
take 200 NS-NS and 200 NS-BH inspirals from their maximum sample
detected in GWs with Net5b. We assume standard advanced LIGO-like
noise curves with no optically squeezed light.

\begin{figure}
\centering 
\subfigure[NS-NS binary mergers]{\label{fig:comparecumlumNSNS}\includegraphics[width=0.45\textwidth]{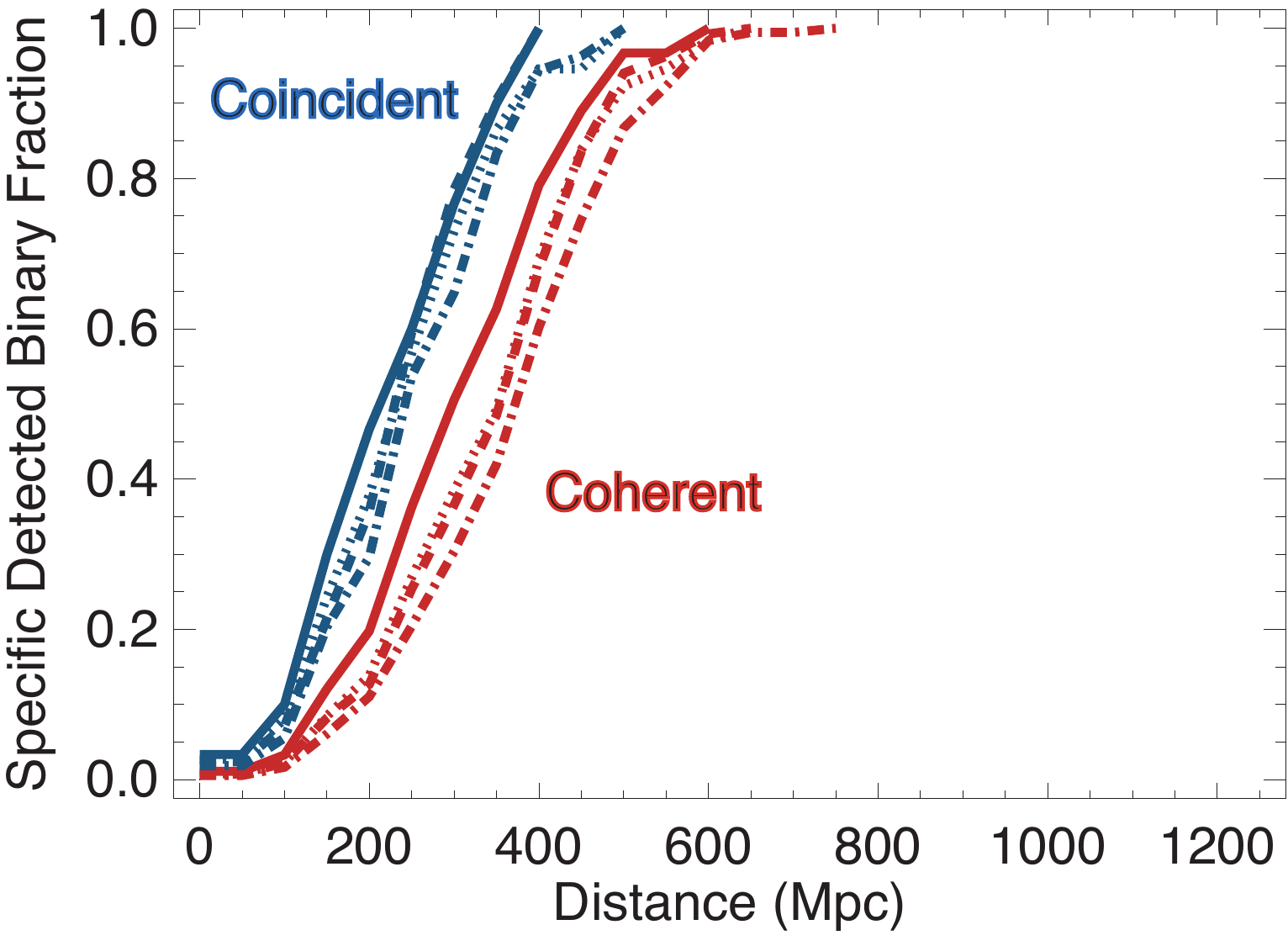}}
\vspace{0.1in}
\subfigure[NS-BH binary mergers]{\label{fig:comparecumlumNSBH}\includegraphics[width=0.45\textwidth]{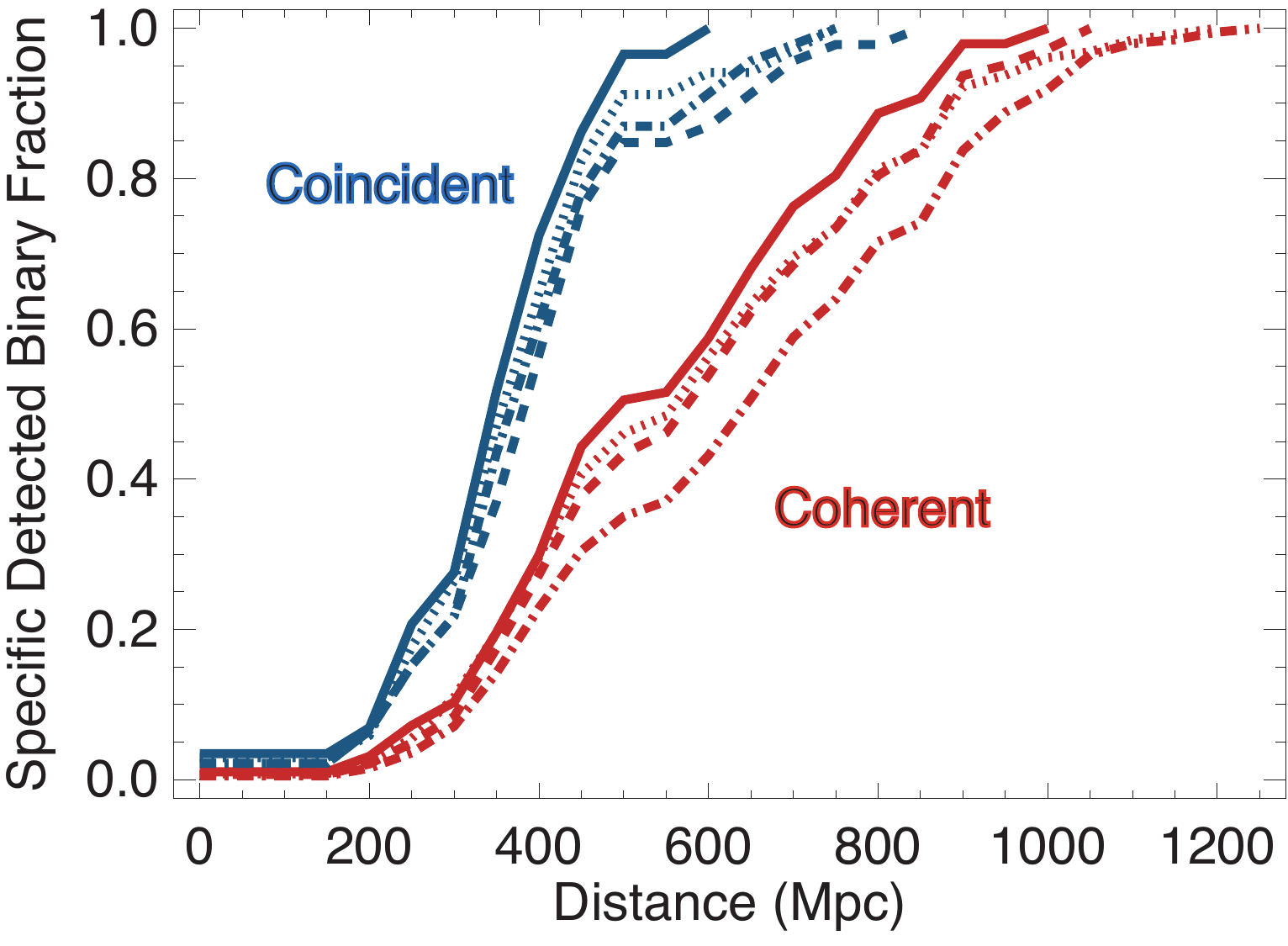}}
\caption{Cumulative luminosity distance (in Mpc)
  distribution of detected sample of NS-NS (top panel) and NS-BH
  (bottom panel) mergers normalized to each specific network and
  trigger criterion. The blue lines denote those NS mergers
  detected using a coincident-trigger criterion; the red represent
  those events detected triggering using the GW network
  coherently. Solid lines represent GW network 3, dotted lines denote
  GW network 4I, dashed are GW network 4K, and dash-dot are GW network 5.
}
\label{fig:comparecumlum}
\end{figure}

Figures~\ref{fig:comparecumlumNSNS} and \ref{fig:comparecumlumNSBH} show
the \emph{specific} distribution in luminosity distance (in
Mpc) of NS-NS and NS-BH mergers detected by different GW networks and triggering
schema. We use the term \emph{specific} because we normalize the cumulative
  distribution to the sample of NS binary mergers that a particular
  triggered GW network can detect. Shown in Table 2, similar to \S\ref{sec:fracGW}, Net3a and Net5b provide representative lower and upper bounds of the GW detectable distance.  For NS-NS mergers, we find
that median detectable distances are 180 Mpc and 370 Mpc with Net3a and Net5b respectively. For NS-BH mergers, we find that the median detectable
distances are 240 Mpc and 660 Mpc with Net3a and Net5b respectively. 

Illustrated in Figures~\ref{fig:comparecumlumNSNS} and
\ref{fig:comparecumlumNSBH}, two distinct distributions for
detectable distance exist depending on whether the GW trigger is coincident {\it versus}
coherent. In contrast, we find that detectable distance ranges depend
only weakly on the number of detectors in a network. From
Figure~\ref{fig:comparecumlumNSBH}, the detectable distance ranges for NS-BH mergers are approximately a
factor of two greater than for NS-NS mergers (see Eqn.~(\ref{eq:snr_ave_sky_orien}), \S\ref{sec:fracGW}
and \S\ref{sec:malmquist}). 

\begin{figure}
\centering 
\subfigure[NS-NS binary mergers]{\label{fig:comparecumareaNSNS}\includegraphics[width=0.45\textwidth]{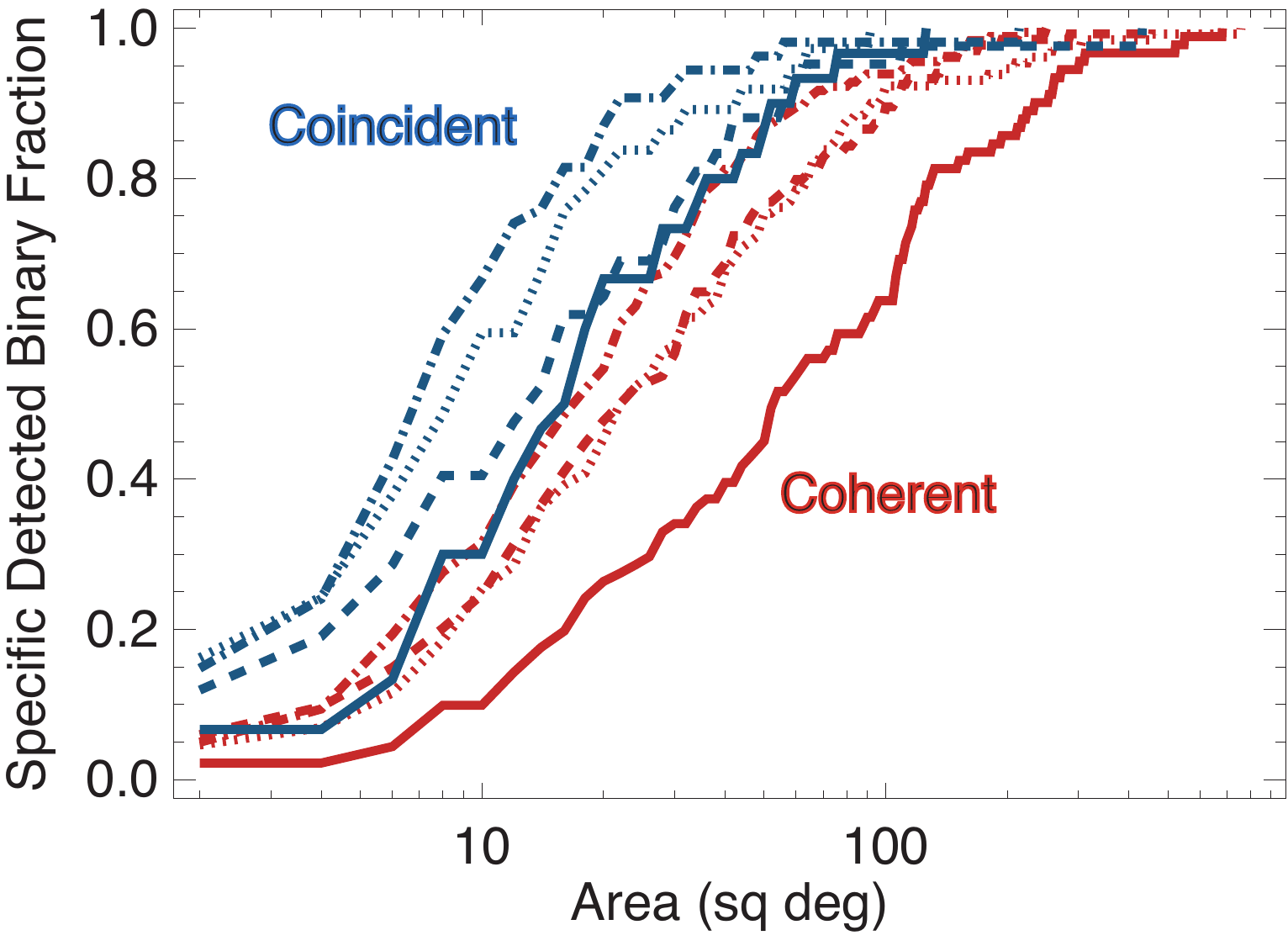}}
\vspace{0.1in}
\subfigure[NS-BH binary mergers]{\label{fig:comparecumareaNSBH}\includegraphics[width=0.45\textwidth]{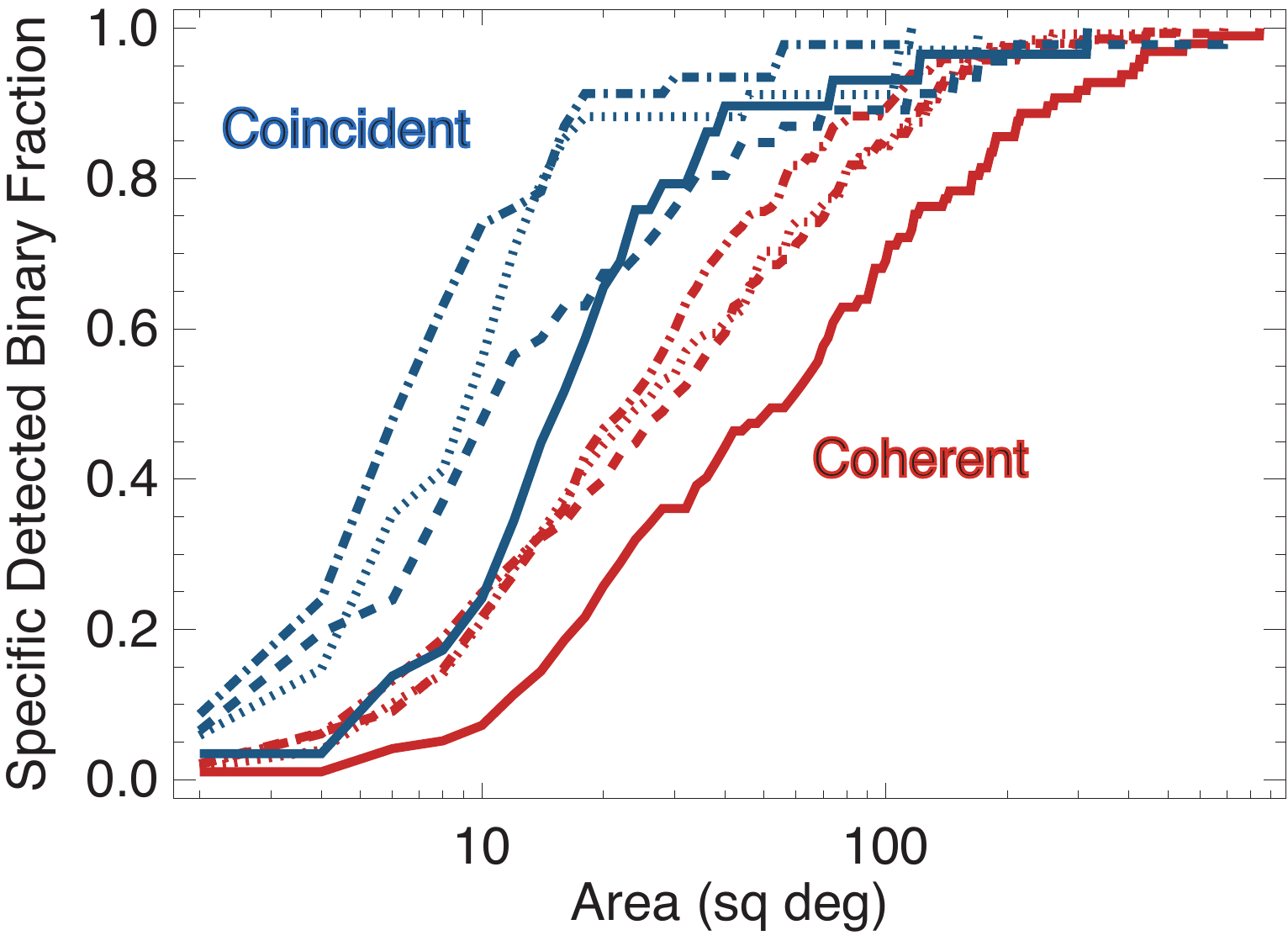}}
\caption{Cumulative sky error (in deg$^2$ at 95 \% c.r.) distributions of detected
  sample of NS-NS mergers normalized to each specific network and
  trigger criterion. The blue lines denote those NS-NS (top panel) and
  NS-BH (bottom panel) mergers
  detected using a coincident-trigger criterion; the red represent
  those events detected triggering using the GW network coherently. Solid lines represent GW network 3, dotted lines denote
  GW network 4I, dashed are GW network 4K, and dash-dot are GW network 5.
}
\label{fig:comparecumarea}
\end{figure}

In addition to distance, EM follow-up detectability
relies on sky area error
ranges; see Figures~\ref{fig:comparecumareaNSNS} and \ref{fig:comparecumareaNSBH} and \S\ref{sec:EMcounterparts}. From Figures
~\ref{fig:comparecumareaNSNS} and \ref{fig:comparecumareaNSBH}\footnote{In contrast, Figures 3 and 4 in N11 show sky error distributions for subsets of NS binary mergers detected by different
GW networks that are normalized to the full detected sample by network
5. In this work, instead of emphasizing the reduced number of
detections, we particularly wish to answer what \% of
NS-binary mergers are detected by a known triggered network to a
certain sky area error.}, we
find that a {\bf coherent-triggered network 3} provides
the largest sky area errors for GW mergers. We refer to this scenario
as \emph{Net3b} and it represents our `lowest-bound' on sky area
errors (Table 2). On the other
hand, a {\bf coincident-triggered network 5}, denoted as \emph{Net5a}, provides the smallest sky area errors and represents
our `upper bound' for sky localization (Table 2). 

From Figures~\ref{fig:comparecumareaNSNS} and \ref{fig:comparecumareaNSBH}, we find that 50\% of NS-NS mergers are detected to within $7
\,\mbox{deg}^2$ with Net5a and to within $60\,\mbox{deg}^2$ with Net3b. For NS-BH mergers, we find that 50\%
of events are detected to within $6 \,\mbox{deg}^2$ with Net5a and
to within $55 \,\mbox{deg}^2$ using Net3b. As expected, similar distributions
in sky area error exists between GW-detected NS-NS and NS-BH merger populations
because most events are detected at
threshold SNR.

We find elliptically-shaped sky errors for the majority of our examined NS
binary mergers (see \S\ref{sec:indbin} for an example); differing GW arrival times at each 
detector dominate sky area reconstruction rather than parameter
degeneracies in the GW waveform's antenna functions (see N11). In a handful of
cases, we find multimodal peaks for especially weak SNR events because
of larger uncertainties in arrival times at detectors.\private{ and so, different values in the GW waveform's amplitude become
indistinguishable from each other.} In addition, we find that sources
located in (or close-by to) the degenerate Net3 plane have relatively poor angular
resolution (see also \citealt{Fairhurst:2010, Wen:2010}, N11,
\citealt{Veitch:2012}). An improvement by a factor of two in the normalized cumulative sky error is seen with network 4I (LIGO+Virgo with
LIGO India) compared to network 4K (LIGO+Virgo with KAGRA) only in the
case of using a coincident trigger. Given that LIGO India is
located further away from the degenerate LIGO-Virgo plane than KAGRA,
such a factor of two improvement in sky area error is expected (e.g.,
\citealt{Schutz:2011}). 

\subsection{GW volume estimates}
\label{sec:volred}

Measurements by GW networks provide us with distance and sky area
errors. With both values in hand, we can construct GW volumes, which aid in identifying the EM
counterparts of NS binary mergers (as \S\ref{sec:vol_pop} describes in
detail).

As a first attempt, we introduce and define below the term \emph{low-latency} GW
volumes. Such volumes in principle can
be computed within a few to tens of minutes of a GW detection (and do not
rely on the full MCMC machinery used in this work) and are hence
critical for EM follow-up. In their final
science run before halting for upgrades to their advanced versions, LIGO
and Virgo sent triggers to EM telescopes within $\sim 30$ minutes of a
possible GW signal being detected \citep{AbadieEM:2012,
  Abadielowlatency:2012, Evans:2012}; most of this time was spent for human-limited
verification checks at each detector site. In the era of advanced
detectors, efforts are underway to reduce the latency timescale to 
less than ten minutes \citep{Singer:2012,Cannon:2012}. 

In this work, we compute low-latency GW volumes by using only marginalized 2-D sky area errors and marginalized 1-D distance
measures (all at 95\% c.r.). As we now discuss, although in this work we derive sky
area and distance errors by marginalizing the full 9-D PDF, we could instead have used $\sim$ minutes timescale
computations of approximate sky
area errors and distance measures. Regarding sky localization
errors for the majority of GW-detected mergers, analytically-derived
formulae (e.g. \citealt{Wen:2010,Fairhurst:2011}), computed on the seconds timescales, allow
for sky reconstruction estimates that are in good agreement with
explicitly-derived 2D sky errors presented in N11. This is because different GW arrival times at each 
detector dominate over amplitude corrections in the GW
waveform. Regarding distance measures, Fisher matrix-based estimates allow for
rough distance measures on a second timescale (see e.g.,
\citealt{ajithetbose09}). In practice, however, measured distances for the majority of threshold events will have significantly larger errors
(by a factor of a several) from their Fisher-matrix derived
counterparts (N10). This is because degeneracies between
the sources' geometric parameters that appear in the GW waveforms'
amplitude inhibit measurement inference for low SNR events (see N10
for a detailed discussion). Therefore, a possible solution when estimating low-latency distance
measures is to use their Fisher matrix-derived errors multiplied
by a factor of three (see N10 and \citealt{DelPozzo:2012}). 

\begin{figure}
\centering
\subfigure[NS-NS binary mergers observed by GW Net3a]{\label{fig:volabs_net1}\includegraphics[width=0.48\textwidth]{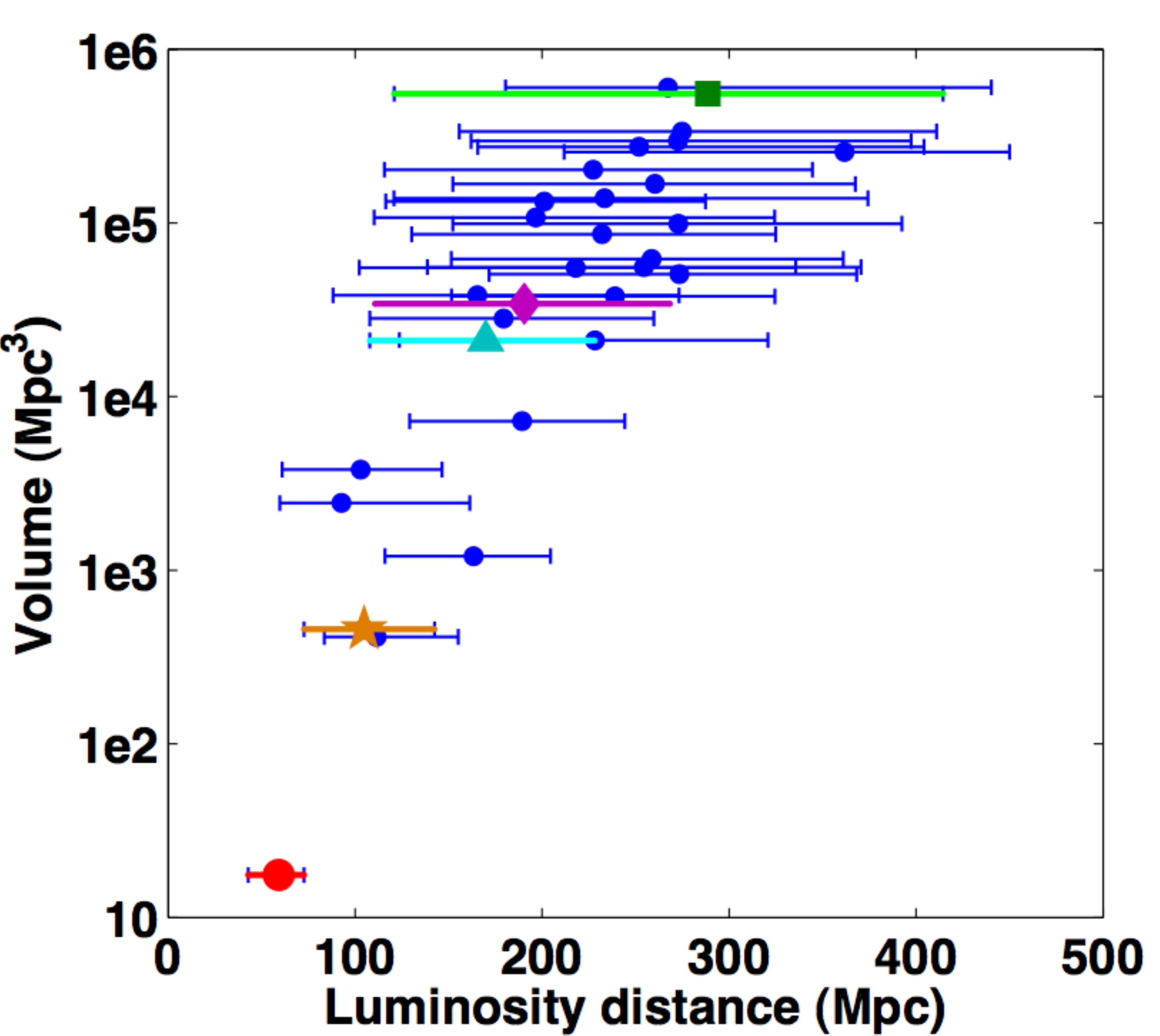}} 
\vspace{0.2in}
\subfigure[NS-BH binary mergers observed by GW Net5b]{\label{fig:volabs_net4}\includegraphics[width=0.48\textwidth]{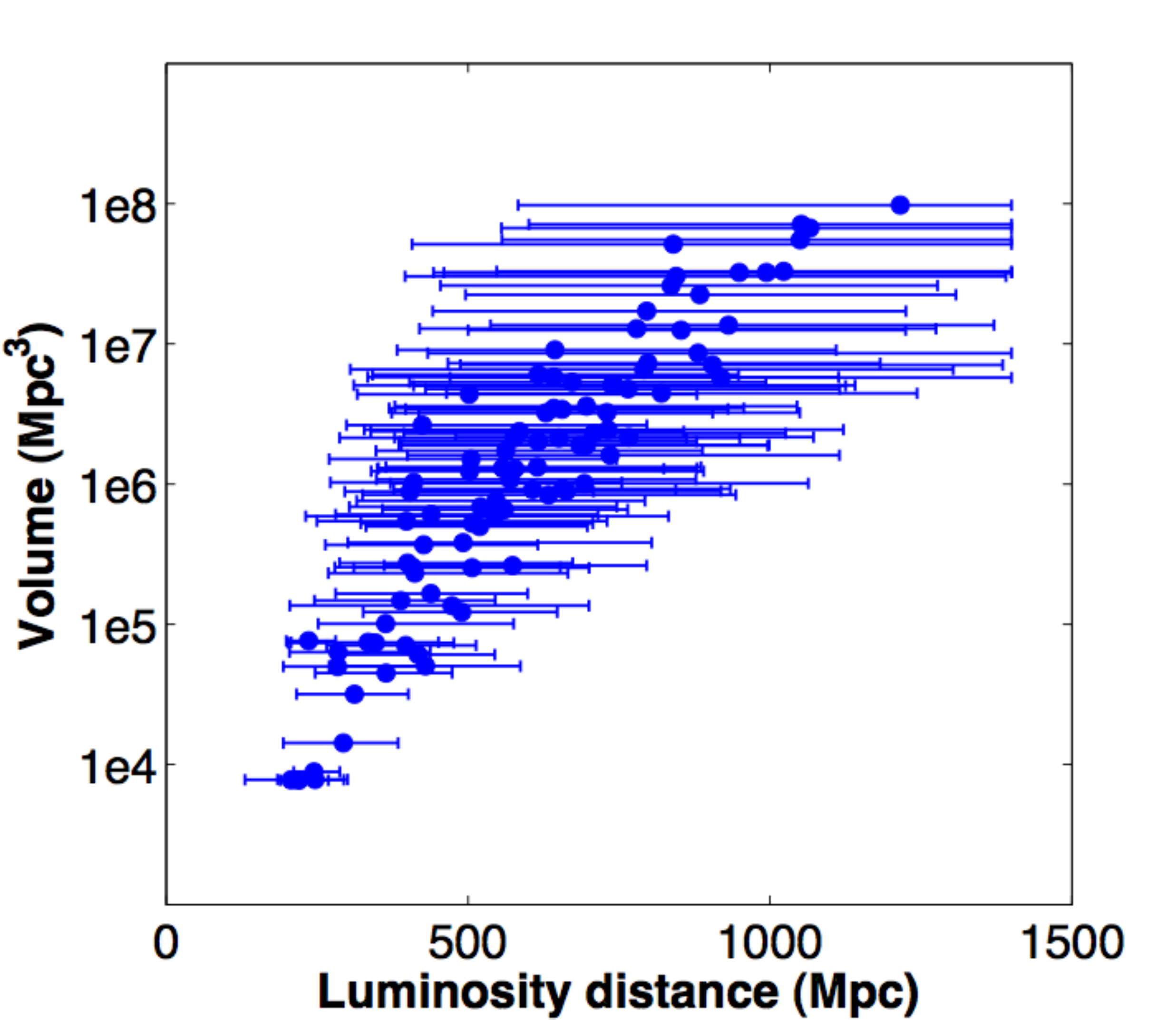}}
\caption{Absolute measures in volume (Mpc$^3$) for a detected sample of
  NS-NS mergers observed by GW Net3a (top panel) and NS-BH mergers
  observed by GW Net5b (bottom panel). Each filled point represents a detected
  NS binary merger at is mean luminosity distance. The horizontal
  error bars represent the upper and lower 1-D marginalized distance
values at 95\% c.r. for each NS binary merger. The different colors
represent different case studies of NS-NS mergers examined in
\S\ref{sec:indbin}: green is Case I (beamed binary), red is Case II (a
nearby binary), light blue is Case III (a merger at low Galactic
latitude), purple is Case IV (a merger at high Galactic latitude) and orange is Case V (a binary in a dense galaxy cluster
environment). Specifically: green square is Case I, red circle is Case
II, light blue triangle is Case III, purple diamond is Case IV, and
orange star is Case V.}
\label{fig: absvol}
\end{figure}

To compute low-latency GW volumes, we use upper, mean and lower distance measures; we define $\mathrm{d}_u$
and $\mathrm{d}_l$ to be the upper and lower 1-D marginalized distance
values at 95\% c.r.. We replace $\mathrm{d}_u$ with $\mathrm{d}_h$ the horizon or maximum detectable distance of a
coincident- or coherent-triggered GW network when
$\mathrm{d}_u > \mathrm{d}_h$. We show absolute volumes in Mpc$^3$ as a function of the
mean distance with their upper and lower distance errors for NS-NS and NS-BH mergers
detected by Net3a and Net5b respectively
(Figures~\ref{fig:volabs_net1} and ~\ref{fig:volabs_net4}). As
expected, the GW measured upper and lower distance ranges are noticeably
smaller for those NS-NS and NS-BH binaries with true
distances less than 200 and 500 Mpc respectively.

\vspace{0.2in}


\section{EM detectability}
\label{sec:EMcounterparts}


In this section, we first review characteristics of suggested EM
counterparts to compact binary mergers. We then discuss EM
detectability by upcoming or current optical and infrared telescopes.

\subsection{Predicted EM counterparts to NS binary mergers}
\label{sec:EMreview}

\emph{Short gamma ray bursts (SGRBs)}: The leading progenitor models
for the majority of observed SGRBs are NS-NS and NS-BH mergers (see e.g. \citealt{Eichler:1989,Paczynski:1991,Narayan:1992}). The
hypothesis has been further supported by around forty SGRB
observations triggered by the \emph{Swift} satellite and followed-up by rapid
multiwavelength observations
(e.g. \citealt{Berger:2005,Fox:2005,Hjorth:2005,Bloom:2006,Berger:2011}). Theoretical
models assume that accretion by a rotationally-supported disk onto a
newly formed BH (or rapidly rotating NS) powers a relativistically
collimated outflow, which results in the observed prompt gamma-ray
emission (e.g., \citealt{Ruffert:1997}, \citealt{Rosswog:2003},
\citealt{Shibata:2008}, \citealt{Rezzolla:2011}). Due to their high
Lorentz factors and energies, the prompt $\gamma$-ray emission is
assumed to be relativistically beamed with initial gamma-ray emission
that lasts for $< 2s$ (hence the use of the name ``short'' when classifying SGRBs; see
\citealt{Nakar:2007} for a review). Jet-break observations in at least two SGRBs suggest collimation of half opening
angles of $\sim 7^{\circ}$ \citep{Soderberg:2006,burrowsetal06} and $3-8^{\circ}$ \citep{Fong:2012} respectively. Upper and lower limits
exist in a few other cases (e.g. \citealt{Fox:2005,Grupe:2006}). As
the relativistic beamed outflow interacts with the
surrounding medium, we expect to observe afterglow signatures in the
X-ray and optical occurring at longer timescales from minutes to days. Observations of EM
afterglows suggest energies of $E \lesssim 10^{51}$ erg
and circumburst densities of $n \lesssim 0.1$ cm$^{-3}$ \citep{Berger:2005,Soderberg:2006}. Afterglow
model predictions as a function of $E$ and $n$ are given in \cite{vanEerten:2011,vanEerten:2012}.

\emph{R-process radioactivity transients--kilonova}:  Initially proposed
by \cite{LP98}, rapid (r)-process radioactivity-powered transients
are weak supernovae-like events. In this
paper, we refer to such transients as `kilonovae', so-called because
their predicted peak luminosities are estimated to be a factor $\sim 10^3$ greater than standard
novae \citep{Kulkarni:2005,Metzger:2010,Roberts:2011}. A central premise of the model is that NS mergers
produce ejecta from either dynamically-ejected tidal tails, or accretion disk outflows driven by early neutrino winds
or late thermonuclear driven winds
(e.g. \citealt{Kulkarni:2005,Metzger:2008,Metzger:2009,Dessart:2009}). The
ejecta is gravitationally-unbound and does not fall back onto the newly-formed
BH or rapidly rotating NS. Numerical relativity and SPH
simulations predict two orders of magnitude difference in the mass of
the ejecta (0.001-0.1M$_{\odot}$) and a factor of a few in the
ejecta's velocity (0.1-0.3 $c$); see e.g., \cite{Rosswog:1999,Rantsiou:2008,Foucart:2011,Piran:2012,East:2012a,East:2012}. Following the expansion of neutron-rich material from nuclear
densities, r-process nucleosynthesis produces heavier unstable radioactive elements which
subsequently beta-decay and fission back to stability on longer
timescales. We expect the material to act as a heat source, and the
subsequent emission to radiate isotropically. Based on highly
uncertain opacities, light curves and color evolutions using radiative
transfer models suggest that the emission peaks either in the optical or near infra-red. In the optical, the emission
could peak with luminosities of $10^{41}-10^{42} \, \mathrm{erg \, s}^{-1}$ which
decay on the half to
five day timescales \citep{Metzger:2010}. Peak absolute magnitudes $M_R$ range from $-$14 to
$-$17\,mag and depend on the assumed ejecta mass, velocity, opacity calculations and nuclear
reactions (\citealt{Metzger:2012}). On the other hand, preliminary work estimate peak absolute
magnitudes $M_H =-15.5$\,mag  in the near-infrared ($\sim 1.7 \mu$m) assuming
an ejecta mass of 0.01$M_{\odot}$ at $0.1 c$ with timescales varying from several to
tens of days timescale (Kasen 2012). Efforts are currently underway to
predict the spectroscopic Doppler-broadened signature of kilonovae. 

\emph{Radio counterparts}: 
There are three predicted radio counterparts: i). we expect
observable non-thermal radio emission from beamed
ultra-relativistic ejecta of SGRBs
\citep{Berger:2005,Soderberg:2006,Chandra:2012}, ii). we could
observe a coherent radio burst emitted from a magnetically-driven, relativistic plasma outflow prior to the NS
merger \citep{Hansen:2001,Pshirkov:2010}, and iii). recent work
suggests incoherent radio signatures from blast waves produced by the interaction of
quasi-spherical, sub- or mildy-relativistic ejecta with the
interstellar medium (\citealt{Nakar:2011}). \cite{Nakar:2011} estimate that radio flares may peak at 1.4
GHz emission for weeks out to redshifts of 0.1 ($\sim$ 450 Mpc), and
can be detectable at milliJansky levels. If the outflows are
sub-relativistic, flares may be detectable on the years timescale at
150 Mpc at closer distances with current and near-future surveys. 

\subsection{Differences between EM counterparts}
We highlight four features that distinguish the proposed EM counterparts and help define search strategies.
First, the counterparts exhibit either beamed or isotropic emission. SGRBs have collimated jet emission and only accompany
a very small fraction of NS-NS and NS-BH mergers (Table 1). On the
other hand, kilonovae and radio remnants have predicted isotropic
emission and accompany \emph{all} NS-NS and NS-BH mergers.
Second, there is a wide disparity in timescales for fast and slow counterparts. 
SGRBs last for seconds (and their afterglows decay as a power law in time), kilonovae for hours to days and radio transients for months to years. 
Third, the rate of false positives is considerably different across EM wavelengths. The precise timing of the SGRB overcomes the challenge of poor sky localization
of GW events. The quietness of the transient radio sky is a boon to
the small number of spatially coincident
false positives \citep{Frail:2012a}. The dynamic optical sky results in tens to hundreds of false positives that would be spatially and
temporally coincident with GW events and search strategies are necessary to separate the wheat from the chaff.
Fourth, discovery and follow-up of SGRBs is now a mature field. For the handful of mergers beamed towards us, 
we have rehearsed what needs to be done.  On the other hand, off-axis and orphan SGRB
afterglows, kilonovae and radio transients are an uncharted
territory.  Both observational and theoretical progress is ongoing
in leaps and bounds as we prepare for GW detectors to come online. Theoretical models continue to
become more sophisticated with their predictions. Observationally, synoptic surveys in the optical 
are already uncovering entirely new classes of fainter and rarer transients. A suite of new radio facilities and radio transient searches are also
coming online.  For instance, wide-field low frequency (say, $<$\,1\,GHz) radio detectors (e.g.  LOFAR, MWA, EVLA)
should be sensitive to pre-merger coherent emission and
coincident timing can be used to connect them to GW detections. 
Relatively higher sensitivity and higher frequency radio detectors (e.g. EVLA, ASKAP, Apertif) are well-suited to searching for radio relic emission
months to years after the GW detection.

\subsection{Detectability of EM counterparts}
\label{sec:results_detxem}
Next, we discuss the detectability of isotropic EM counterparts by
optical and infrared telescopes. As quantified in 
\S4.4, the localization and distance horizon distribution are dependent on the 
number of detectors and the threshold criterion in the GW network. We consider here the extreme cases for GW maximum detectable
distances (Net3a and Net5b), and for sky error areas (Net3b and Net5a).


\subsubsection{Optical Facilities}
Astronomers have a diverse arsenal of optical telescopes worldwide. We limit the discussion 
here to only telescopes with cameras larger than 1 deg$^2$ (given the large localization areas) and 
apertures larger than 1m (given the faintness of predicted
counterpart). We consider current or scheduled-to-be operational
telescopes. We divide telescopes into three categories: 1m-3m class telescopes, 4m-7m class telescopes,
8m-10m class telescopes. Table~\ref{tab:opttelescopes} provides a
summary of sensitivity and FoV of each telescope and camera system.

\begin{deluxetable*}{lcccccc}
\tabletypesize{\scriptsize}
\tablecaption{Optical Telescopes}
\tablehead{
\colhead{Telescope} & 
\colhead{Aperture} &
\colhead{Field of View} &
\colhead{Exposure} & 
\colhead{Overhead} &
\colhead{Sensitivity}  &
\colhead{Reference} \\
\hline 
&
\colhead{ (m)} &  
\colhead{ (deg$^2$)} &
\colhead{ (sec) }  &
\colhead{ (Readout) } &
\colhead{(5$\sigma$ mag in R band)}
}
\startdata
Zwicky Transient Facility & 1.2 & 35  &  60 &15  & 20.6 & \tablenotemark{a} \\
La Silla Quest                       & 1.0 & 9.4 (80\%) & 60  &30&20.5  &  \tablenotemark{b} \\
Catalina Real-Time Transient Survey  & 0.7 & 8.0 & 30 & 18 & 19 &\tablenotemark{c} \\
Palomar Transient Factory & 1.2 & 7.1 & 60 & 40 & 20.6 &\tablenotemark{d} \\
Pan-STARRS 1                            & 1.8 & 7.0 & 60  & 3 & 22.0 &   \tablenotemark{e} \\
Skymapper                            & 1.35 & 5.62 &110  &20  & 21.5 &\tablenotemark{f}  \\
\hline
CTIO-Dark Energy Camera & 4.0 & 3.0 & 50 & 17  & 23.7 & \tablenotemark{g}   \\  
WIYN-One Degree Imager  & 3.5 & 1.0 & 60 & 30 & 23 &  \tablenotemark{h} \\
CFHT-Megacam                   & 3.6 & 0.9 & 60 & 40  & 23 &\tablenotemark{i}   \\
\hline 
Large Synoptic Survey Telescope & 8.4 (6.7) & 9.6 & 15 & 2 & 24.5 &  \tablenotemark{j}  \\
Subaru-HyperSuprimeCam & 8.2 & 1.77 & 30 & 20 & 24.5 & \tablenotemark{k}  \\
\hline 
\enddata
\tablenotetext{a}{\citealt{k12}}
\tablenotetext{b}{\citealt{hrb+11}}
\tablenotetext{c}{\citealt{ddm+09}}
\tablenotetext{d}{\citealt{lkd+09}} 
\tablenotetext{e}{see http://pan-starrs.ifa.hawaii.edu} 
\tablenotetext{f}{see http://rsaa.anu.edu.au/observatories/siding-spring-observatory/telescopes/skymapper/skymapper-instrument}
\tablenotetext{g}{\citealt{bkk+12}}
\tablenotetext{h}{see http://www.wiyn.org/ODI/Observe/wiynodioverview.html}
\tablenotetext{i}{see http://www.cfht.hawaii.edu/Instruments/Imaging/Megacam/generalinformation.html}
\tablenotetext{j}{\citealt{aaa+09}}
\tablenotetext{k}{see http://www.naoj.org/Projects/HSC/index.html and http://www.naoj.org/cgi-bin/img\_etc.cgi}

\label{tab:opttelescopes}
\end{deluxetable*}
\vspace{0.2in}

Theoretical predictions of optical EM counterparts span orders of magnitudes in both predicted luminosity and 
predicted timescale (\S~\ref{sec:EMreview}). To evaluate the relative merits of follow-up with
different telescope facilities and to begin to define a search strategy, we need to 
make a conservative assumption on the nature of the counterpart.  Hence, for 
the discussion below, we first assume that the optical counterpart of a NS-NS merger will be brighter 
than M$_{\mathrm R}$ = $-$14\,mag \, for at least two hours (later, we relax this assumption to $-$11\,mag). 
Let us say a particular telescope
takes three images at a separation of one hour, and each of these images has a 5-$\sigma$
depth of M$_{\mathrm R}$ = $-$14\,mag. Then, the counterpart will be discovered with high SNR (e.g., 12.5-$\sigma$ if 
the transient peaked at M$_{\mathrm R}$ = $-$15\,mag) in the first image,
at least at 5-$\sigma$ in the second image and possibly below a 5-$\sigma$ threshold in the third image. This is our ``minimum" criterion for a secure detection.  If the counterpart 
is either more luminous or evolves at a slower rate, it will only
improve the security of our detection. We require a \emph{minimum of
  two detections} to securely
distinguish the optical counterpart
from moving objects in our solar system (asteroids) and artifacts. 

Telescope time is a zero sum game. A telescope with a given FoV of camera and a given aperture has to perform a three-way tradeoff between 
depth, cadence (how frequently the same field is observed), and area covered. Here, we assume that each telescope takes at least three 
sets of images separated by one hour. In this one hour, to attain the
M$_{\mathrm R}$ = $-$14\,mag sensitivity
for the most number of events, the telescope will either integrate longer on a given
field to see events further away or map a larger fraction of the localization area. 
If a telescope has a large aperture and small field camera, it will spend the one
hour taking short exposures on a larger fraction of the localization area.
If a telescope has a small aperture but a large field camera, it will spend the
one hour stacking images to maximize integration time. It is precisely this choice
that determines how many optical counterparts are detectable by a given telescope ({\it modulo}
idealized observing conditions as we shall discuss). 

\begin{figure}
\centering
\subfigure[NS-NS mergers, GW Net5b: M$_R$ = -14]{\label{fig:detEM_net4}\includegraphics[width=0.45\textwidth]{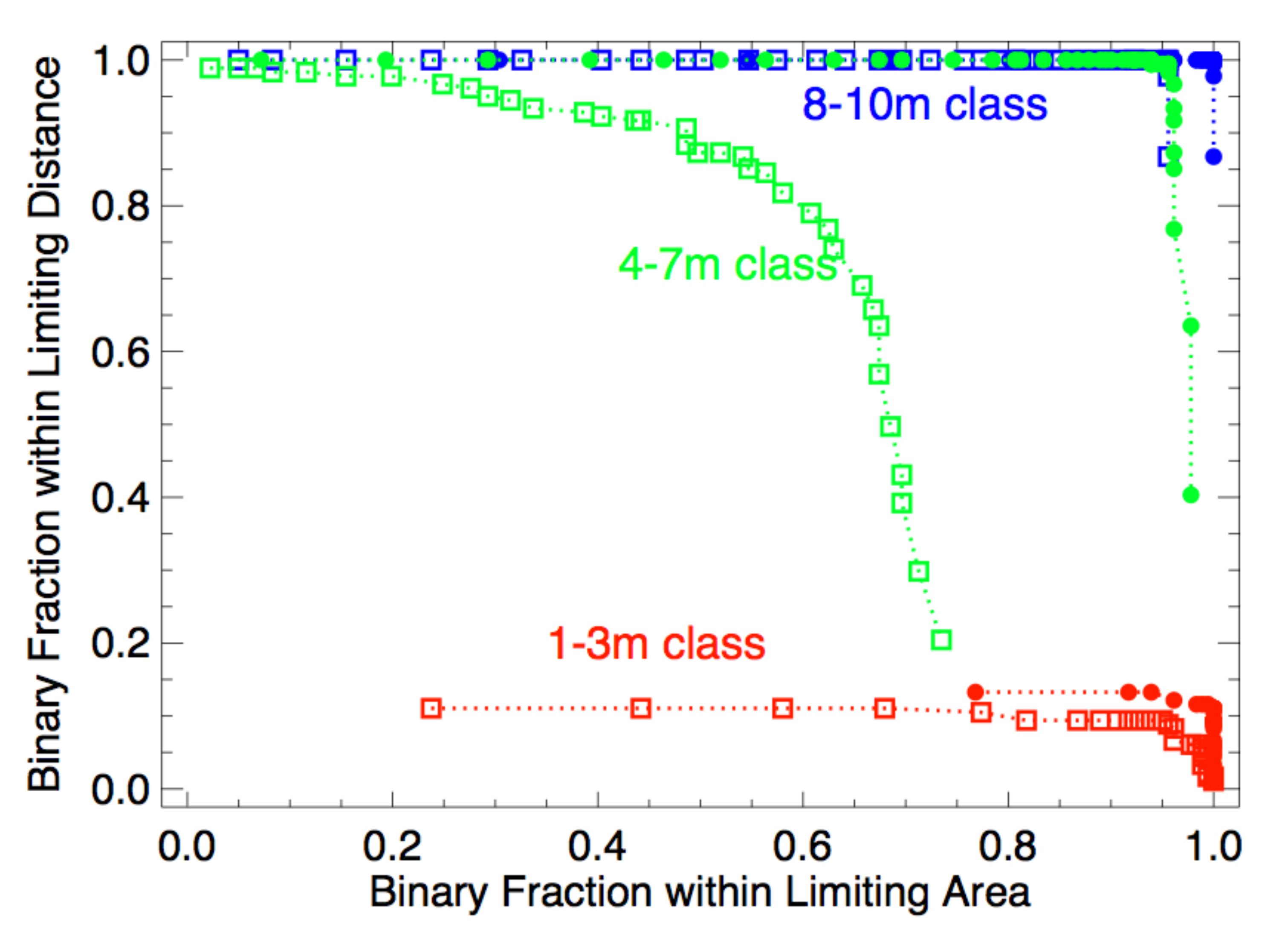}} 
\subfigure[NS-NS mergers, GW Net5b: M$_R$ = -11]{\label{fig:detEM_net4_min11}\includegraphics[width=0.45\textwidth]{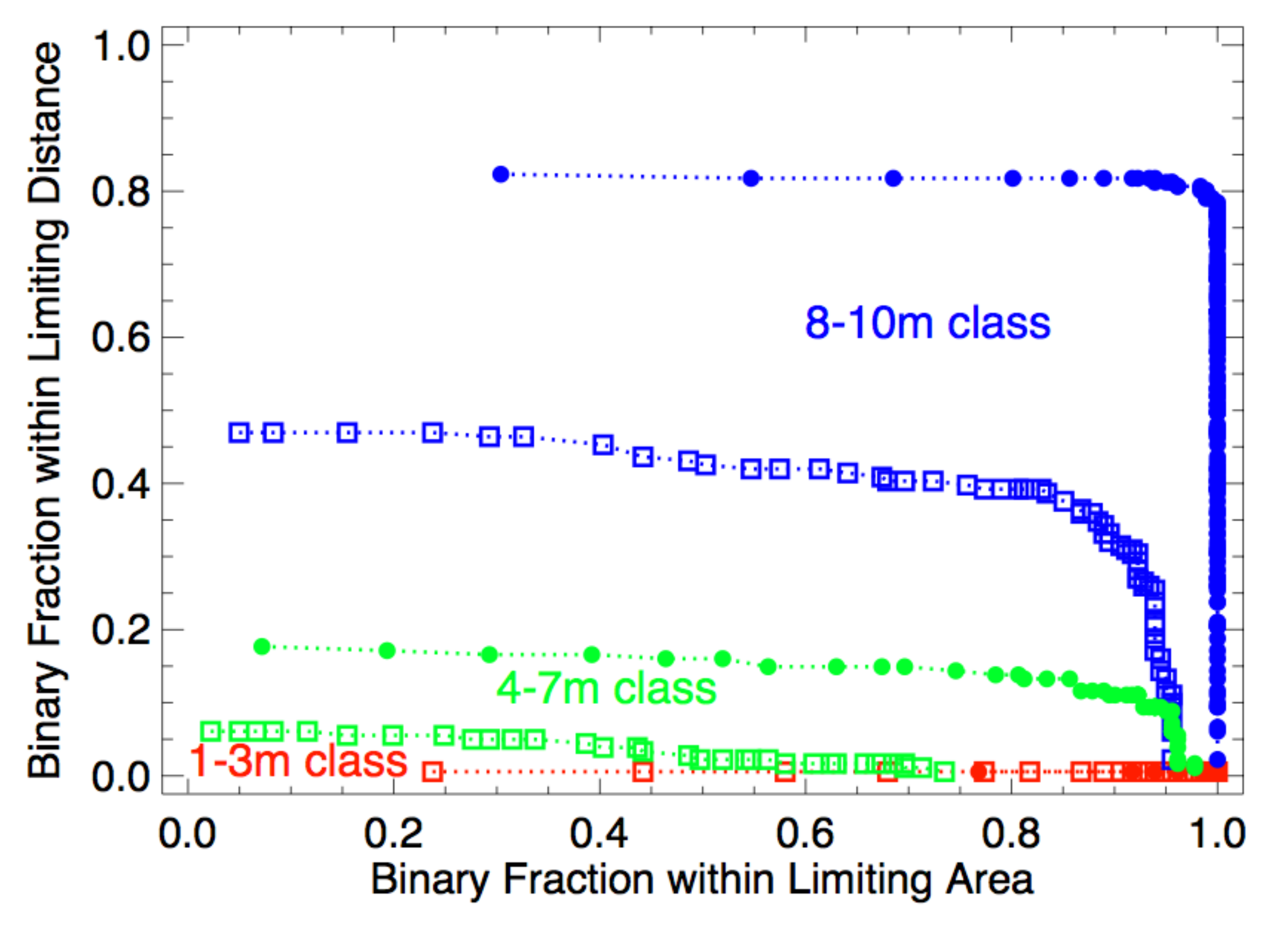}}
\caption{The top panel illustrates the depth versus area trade-off for optical telescopes when searching 
for NS-NS mergers detected by a GW Net5b, where the isotropic optical
counterpart is assumed to be brighter than M$_{\mathrm R}$$=-$14 for
at least two hours. Bottom panel illustrates the depth versus area trade-off for optical telescopes when searching 
for NS-NS mergers detected by a GW Net5b and where the optical
EM counterpart is assumed to be brighter than M$_{\mathrm R}=-$11 for at least two hours. 
The different colors represent different telescope
apertures: Red is 1-3m class, Green is 4-7m class, Blue is 8-10m class telescopes. Open square is a small 
FoV camera and filled circle is a large FoV camera in that aperture class. Specifically: red square is PTF, red circle is ZTF, 
green square is CFHT, green circle is DECAM, blue square is HSC and Blue circle is LSST.}
\label{fig:detEM}
\end{figure}

We quantify the implications of this tradeoff on the number of detected
NS-NS mergers in Figure~\ref{fig:detEM_net4} for a GW Net5b.  

\private{We quantify the implications of this tradeoff on the number of detected
NS-NS mergers in Figure~\ref{fig:detEM_net1} for a GW Net5a and in Figure~\ref{fig:detEM_net4} for a GW Net5b.  }

For example, let us consider the role of CFHT in GW Net5b (green open squares in Figure~\ref{fig:detEM_net4}).  
In 100\,s (60s exposure + 40s readout), CFHT can take a 0.9 deg$^2$
image with a depth of 23\, apparent mag. In one hour, CFHT can take 36 exposures, hence there are 36 possibilities for the tradeoff. We discuss
the first and last point on the curve of green squares. If CFHT spent the entire hour integrating 
on only one field, it would achieve a depth of 24.9\, apparent mag and detect
binaries with distances less than 615\,Mpc (99\%) but localization
areas less than 0.9\,deg$^2$ (2\%). Instead, if CFHT spent the entire hour covering the large localization area and only spent one minute per field, it would achieve
a poorer depth of 23\, apparent mag and detect binaries only out to less than 250\,Mpc (20\%) but localization
areas less than $\sim$ 32\,deg$^2$ (73\%).  

Next, let us take the case of 8m-class Subaru's HyperSuprimeCam (HSC, blue open squares). By spending
only five exposures on a given field, the depth of HSC can cover 100\% of distances of
detected NS binary mergers. However, it's smaller FoV camera limits the total area covered in one hour to 127\,deg$^2$ i.e. 
96\% of mergers. LSST has the same depth but a larger camera, it can
detect 100\% of mergers.

Let us next study the case of 1m-class ZTF (red filled circles in
Figure~\ref{fig:detEM_net4}) in a GW Net5b. In less than five pointings, with its superior 35 deg$^2$ camera, ZTF can cover 100\% of 
all localization areas. But its small aperture limits sensitivity to
apparent 22.6\,apparent mag or 210\,Mpc
i.e. 10\% of the mergers. 

Now, we consider the implications of the isotropic optical counterpart being much less
luminous, for instance $M _{\mathrm R}$ = $-$11\,mag
(Figure~\ref{fig:detEM_net4_min11} and Table~\ref{tab:detEM}). The percentage of detectable counterparts goes down from 100\%
to 82\% for LSST, from 96\% to 42\% for HSC and 97\% to 16\% for DES. 

\begin{figure}
\centering
\subfigure[GW Net3a]{\label{fig:FracdetEMconsnet1}\includegraphics[width=0.45\textwidth]{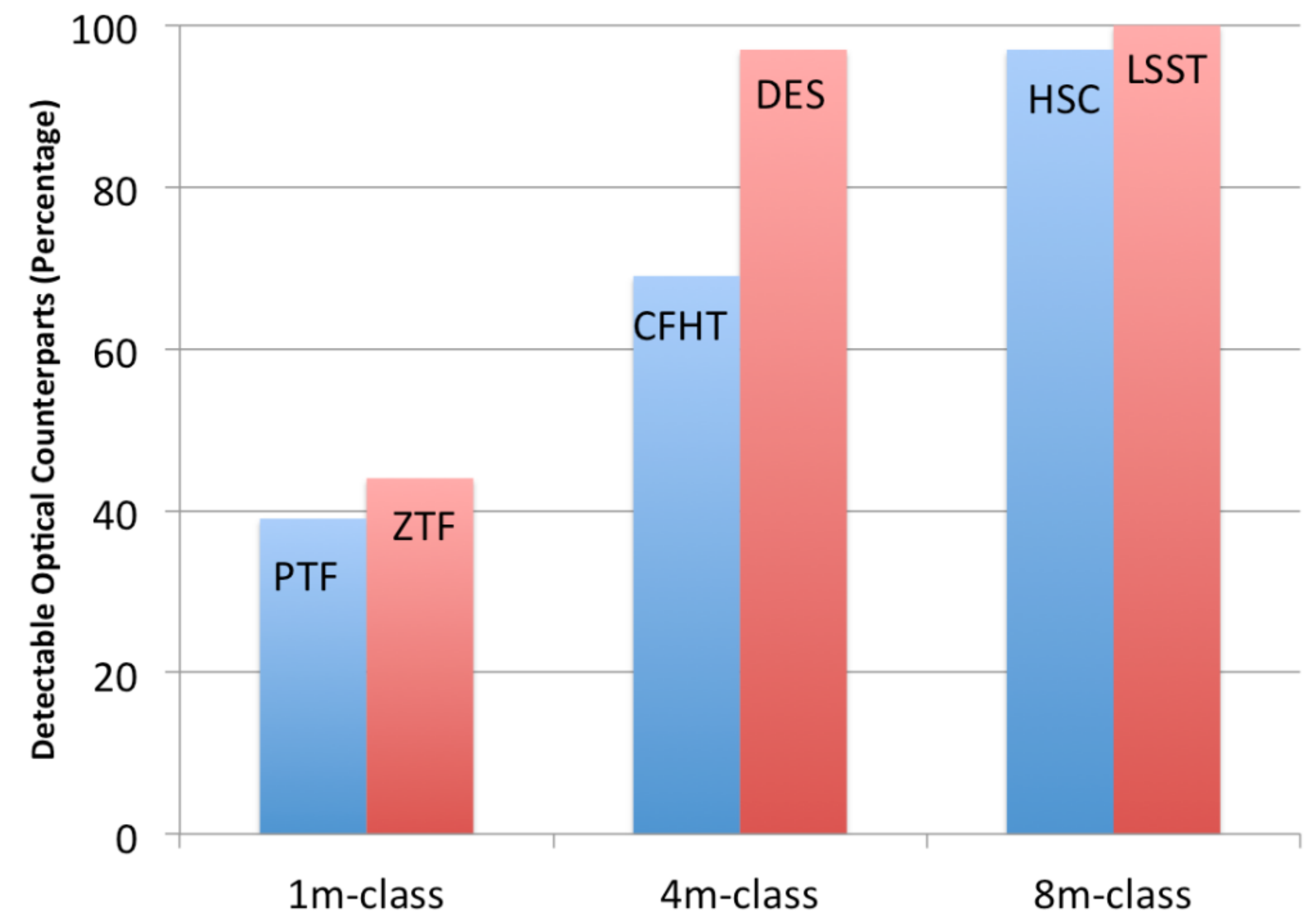}} 
\subfigure[GW Net5b]{\label{fig:FracdetEMcohnet4}\includegraphics[width=0.45\textwidth]{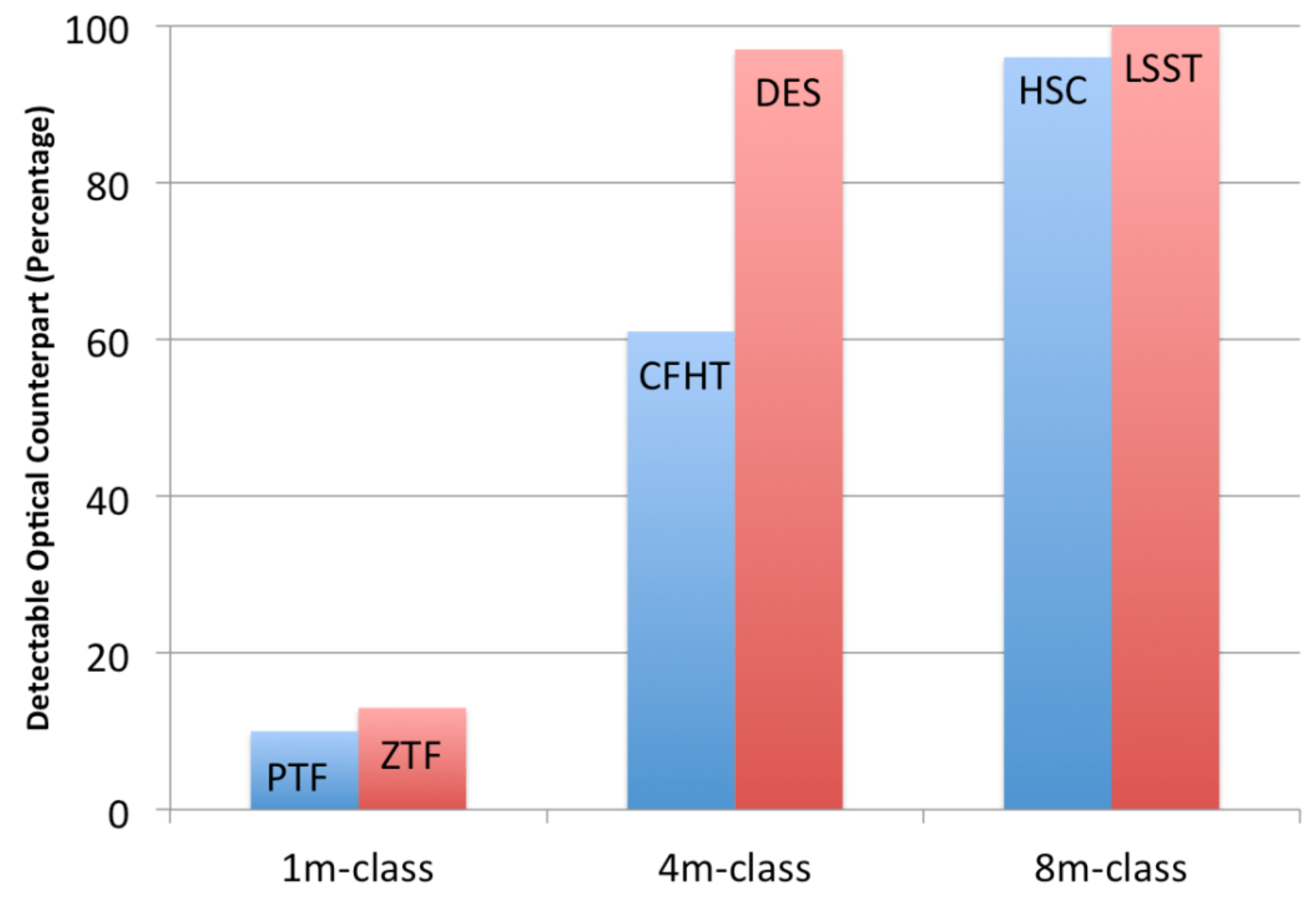}}
\caption{Relative fraction of detectable isotropic optical counterparts
  to NS-NS mergers for GW Net3a and Net5b. Note
that even the small aperture wide field telescopes are sensitive to a
significant fraction, and DES and HSC are almost as sensitive as LSST.}
\label{fig:FracdetEM}
\end{figure}

Assuming the most optimal strategy is chosen for each merger in this simulation,
we can compute the fraction of detectable optical counterparts by each telescope
(Table~\ref{tab:detEM}). As exemplars, the smallest and largest FoV camera in each telescope aperture-class is chosen. 
Initially, when there is a GW three-detector network, binaries would be detected closer in and the localizations would
be poorer. The smaller telescopes with larger FoVs will play an
important role (Figure~\ref{fig:FracdetEMconsnet1}). 
In the era of a GW five-detector network, once localization is
improved and maximal detectable distance pushed further back,
the larger telescopes will be essential (Figure~\ref{fig:FracdetEMcohnet4}).

Finally, we consider the case of NS-BH mergers. Given that NS-BH mergers will
on average be detected a factor of two further away, but have
predicted optical counterparts 1.5\, mag brighter, we get similar detectability fractions as NS-NS mergers (Table~\ref{tab:detEM}).

We emphasize that the detectable fractions presented in Table~\ref{tab:detEM} are relative and subject
to two caveats. First, there would be tiling inefficiency and edge effects 
due to the irregular shapes of GW localization and the rectangular/circular
fields of view of the EM cameras. Second, all optical telescopes in this discussion 
are subject to certain reality checks --- they cannot observe too close to the sun or too close to the moon, 
if it is cloudy or raining, or if the target is in the quadrant of sky not accessible from a given location. 
Typically, these factors amount to
$\frac{1}{2}\times\frac{2}{3}\times\frac{3}{4} = \frac{1}{4}$ of the
targets being visible at a given telescope on a given day respectively. 

We conclude that a network of telescopes at different longitudes, latitudes and
mountain-tops would maximize the odds of follow-up.  Hence, the numbers 
presented here should only be interpreted as illustrative of the relative detectability by different
telescopes.


\begin{table*}
\caption{Relative percentages of isotropic, optical counterparts of GW-detected mergers detectable by
  different telescopes as a function of GW network, triggering
  criterion and peak optical luminosity. Incorporation of realistic
  observing conditions (moon, sun, weather, latitude etc.) reduces efficiency by $\sim 1/4$}
\centering
\begin{tabular}{lllllll}
\hline 
\hline
                   &  PTF & ZTF & CFHT & DES & HSC & LSST \\
\hline
NS-NS merger \& M$_{\mathrm R} < -$14\,mag \\
GW Net3a& 39 & 44 & 69 & 97 & 97 & 100 \\
GW Net5a& 34 & 41 & 95 & 98 & 98 & 100 \\ 
GW Net3b& 18 & 22 & 34 & 82 & 79 & 100 \\
GW Net5b& 10 & 13 & 61 & 97 & 96 & 100 \\
\hline
NS-NS merger \& M$_{\mathrm R} < -$11\,mag \\
GW Net3a& 0 & 0 & 19 & 39 & 86 & 100  \\
GW Net5a& 0 & 0 & 18 & 48 & 91 & 100 \\
GW Net3b& 0 & 0 & 8 & 16 & 45 & 93 \\
GW Net5b & 0 & 0 & 4 & 16 & 42 & 82 \\
\hline
\hline 
NS-BH merger \& M$_{\mathrm R} < -$15.5\,mag \\
GW Net3a& 66 & 79 & 79 & 97 & 97 & 100  \\
GW Net5a& 63 & 72 & 93 & 100 & 93 & 100 \\
GW Net3b&  24 & 39 & 36 & 78 & 76 & 100 \\
GW Net5b & 22 & 28 & 56 & 96 & 94 & 100 \\
\hline
NS-BH merger \& M$_{\mathrm R} < -$12.5\,mag  \\
GW Net3a& 3 & 3 & 17 & 79 & 93 & 100 \\
GW Net5a&  2 & 2 & 15 & 78 & 98 & 100 \\
GW Net3b & 1 & 1& 5 & 33 & 47 & 98 \\
GW Net5b& 1 & 1 & 5 & 30 & 53 & 88  \\
\hline
\hline 
\end{tabular}
\label{tab:detEM}
\end{table*}


\subsubsection{Infrared Facilities}
Recent theoretical calculations of kilonovae opacities suggests that a significant fraction of
the luminosity may be emitted in the redder bands beyond 1 $\mu$\,m \citep{Kasen:2012}. 
Unfortunately, our current suite of near-infrared facilities is not as wide-field as the optical with no camera
larger than a square degree. 

Currently, the two widest field infrared facilities are the 0.594 deg$^2$ VIRCAM on the 4.1m VISTA 
telescope and the 0.19 deg$^2$ WFCAM on the 3.8m UKIRT telescope. Fortunately, efforts are underway to build a 6.5m 
SASIR telescope with a 0.2--1 deg$^2$ camera \citep{SASIR:2012}. Moreover, unlike VIRCAM and WFCAM, SASIR is expected to have a contiguous focal plane and
simultaneously image in YJHK-bands. 

Efforts are also underway to build two wide-field infrared satellites ---
WFIRST \citep{Green:2012} and Euclid \citep{Euclid:2012}. A set of three WFIRST fields is
expected to cover 2 deg$^2$ and eighteen minutes per field will give a
5-$\sigma$ depth of H $\sim$ 25.9\, mag for imaging and H $\sim$ 23.5\, mag
for low-resolution prism spectroscopy. Euclid would need four fields
to cover 2 deg$^2$ and achieve depths of H $\sim$ 25.6\,mag and H
$\sim$ 21.4\,mag in grism mode in the same time. Both missions are constrained to
observe $\sim$ $90^{\circ}$ away from the Sun \citep{Hirata:2012}.

We remind the reader that the median-maximum localization of Net 3b is 55--180 deg$^{2}$ and Net 5a is 7--120 deg$^{2}$
(Table~2). Hence, the infrared follow-up will require an extremely large number of pointings to tile the area and will be limited 
to the best localized binaries.

\section{Identifying EM counterparts}
\label{sec:results_em}


Detection of candidate EM counterparts is only the first step. The most pressing question for EM telescopes
looking at vast sky areas will be whether the transient objects are
true GW emitters or false-positive signals mimicking an EM
counterpart. For instance, the optical sky is so dynamic that there will be hundreds of foreground
and background false positives associated with any
detection. Foreground signals are, for example, M-dwarf flares, CVs and other
stellar variables in the Milky Way. The foreground rates, therefore,
depend strongly on the Galactic latitude and have a wide range of
amplitudes and timescales. Background signals are supernovae (SNe) and AGNs at higher redshift than the GW detectable distance
horizon for NS binary mergers of $\sim 200$ Mpc - 1 Gpc. Thanks to 
systematic optical synoptic surveys, rate estimates of different classes of SNe occuring in
a range of galaxy hosts now exist; for instance,  core-collapse SNe rate is 7.1 ($\pm \, 0.1) \times $
    10$^{-5}$ Mpc$^{-3}$ yr$^{-1}$ (see e.g., \citealt{Leaman:2011,Li:2011b,Li:2011a}).
\private{; for instance,  core-collapse SNe rate is 7.1 ($\pm \, 0.1) \times $
    10$^{-5}$ Mpc$^{-3}$ yr$^{-1}$, and thermonuclear SNe rate is 3.0 ($\pm \, 0.4) \times $
    10$^{-5} $ Mpc$^{-3}$ yr$^{-1}$ }

Hence, panchromatic follow-up (especially optical spectroscopy) is  
critical to unambiguously associate the counterpart with the GW
signal. Given predicted optical lightcurve evolutions, the
timescale for spectroscopic follow-up should be within the hours to day timescale. There are a large number of telescopes in the 3--5m class range which
can easily take low resolution spectra of transients brighter than
21 apparent mag. However, optical counterparts will likely be in
regime where the transient is fainter than 22 apparent mag and a $>$ 6m-class telescope will be needed for spectroscopy.
The list of such telescopes is rather small: the twin Magellan 6.5 telescopes, the MMT 6.5m telescope, the twin Gemini 8m telescopes, the four VLT 8m telescopes, 
the HET 9.2m telescope, the SALT 9.2m telescope and the twin Keck 10m telescopes. Efforts are underway to build even larger 20m-30m class telescopes: GMT, TMT and ELT.  
Spatial coincidence with a nearby Galaxy will distill the large number of counterpart candidates
to a small number that can be promptly followed up spectroscopically \citep{kk09}.

To illustrate the diversity of follow-up scenarios, we
consider below five case studies of NS-NS mergers.
In each case, we discuss optimal strategies for identifying the
EM counterpart of the NS binary merger. Finally, we discuss how we can
leverage volume information to aid EM follow-up strategies for a
\emph{population} of NS binary mergers.

\subsection{Individual binaries}
\label{sec:indbin}


We first examine sky localization and volume errors for one beamed
NS-NS binary merger at 391 Mpc, and four NS-NS
mergers that have distances less than 200 Mpc and lie within
the CLU catalog used in this work. We choose the five NS-NS
mergers described below because their geometric properties or sky locations
represent useful bounds that illustrate the challenges for any EM
follow-up. The five case studies comprise NS-NS mergers with: i) its orbital angular momentum vector
face-on towards the Earth, ii) a close-by event, iii)  a source position at low Galactic
latitude, iv) a source position at high
Galactic latitude, and v) a source position in a dense galaxy cluster environment.

\subsubsection{Case Study I: Beamed binary merger at 391\,Mpc}
\begin{figure}
\centering
\includegraphics[width=0.45\textwidth]{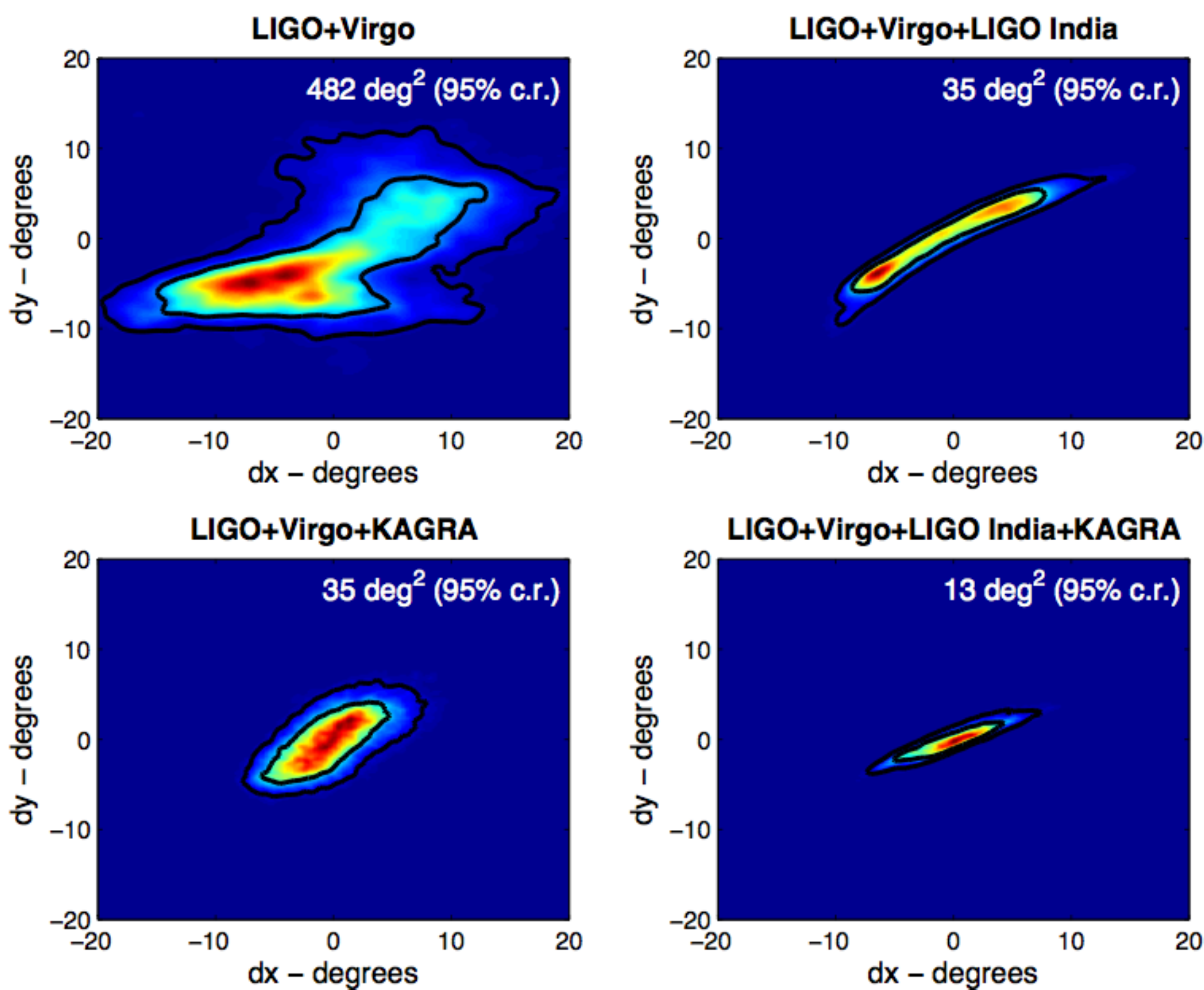}
\caption{Sky localization errors at 95\% c.r. using
  different GW networks for a
  NS-NS binary merger located at $\sim$ 390 Mpc whose orbital angular
  momentum is directly face-on to the Earth.}
\label{fig: binary9}
\end{figure}

We consider the case of a binary merger beamed towards us.
Given the Malmquist bias (\S4.3), these binaries are at threshold
and are thus, on average, located further away. Out of 200 randomly
sampled mergers detected with a GW Net3a, the distances of beamed NS-NS mergers are 391 Mpc, 506 Mpc, 560 Mpc and 564 Mpc.
Illustrated by Figure~\ref{fig: binary9}, using GW networks 3 and 5, the
localization for the closest of these binaries is 483 deg$^2$ (95\% c.r.)
and 13 deg$^2$ (95\% c.r.) respectively. Current $\gamma$- and
X-ray satellites are easily sensitive to SGRBs at these distances (the furthest detected Swift SGRB is 090426 at a redshift of 2.68 or 22 Gpc).  
Advantageously, these satellites have large instantaneous FOVs. Moreover, given the precise
  timing of the gamma ray burst, false-positive signals are not a
  concern \citep{Kanner:2012}. If a precise position (e.g. with XRT onboard the
  Swift satellite or with MIRAX-HXI) is available, prompt follow-up to look for 
the radio and optical afterglows (which will be much brighter than a kilonova signal but decay as a power law in time) will be tractable. 

Unequivocally, wide-field $\gamma$- and X-ray satellites (e.g., Fermi,
Swift, Lobster, MAXI, MIRAX-HXI) are currently the most promising wavelengths to search for EM counterparts of beamed NS-NS and NS-BH mergers. 
However, as Table 1 shows, the \emph{beamed} NS-NS mergers are a very small fraction
($\sim$1.5\% to $\sim$3\%) of the total GW-detected population. Hence,
coincident GW and EM observations of beamed NS binary mergers will be rare. 

\subsubsection{Case study II: Close in binary at 70\,Mpc}

Nearby ($<100$ Mpc) NS binary mergers provide excellent
laboratories in which to study strong-field gravity astrophysical processes
using joint GW and EM observations. Such a ``golden
binary" should result in high SNR detections in GWs and
multiwavelength EM waves, enabling excellent
characterization of the physical properties of the progenitor and
post-merger remnants. 

\begin{figure}
\centering
\includegraphics[width=0.45\textwidth]{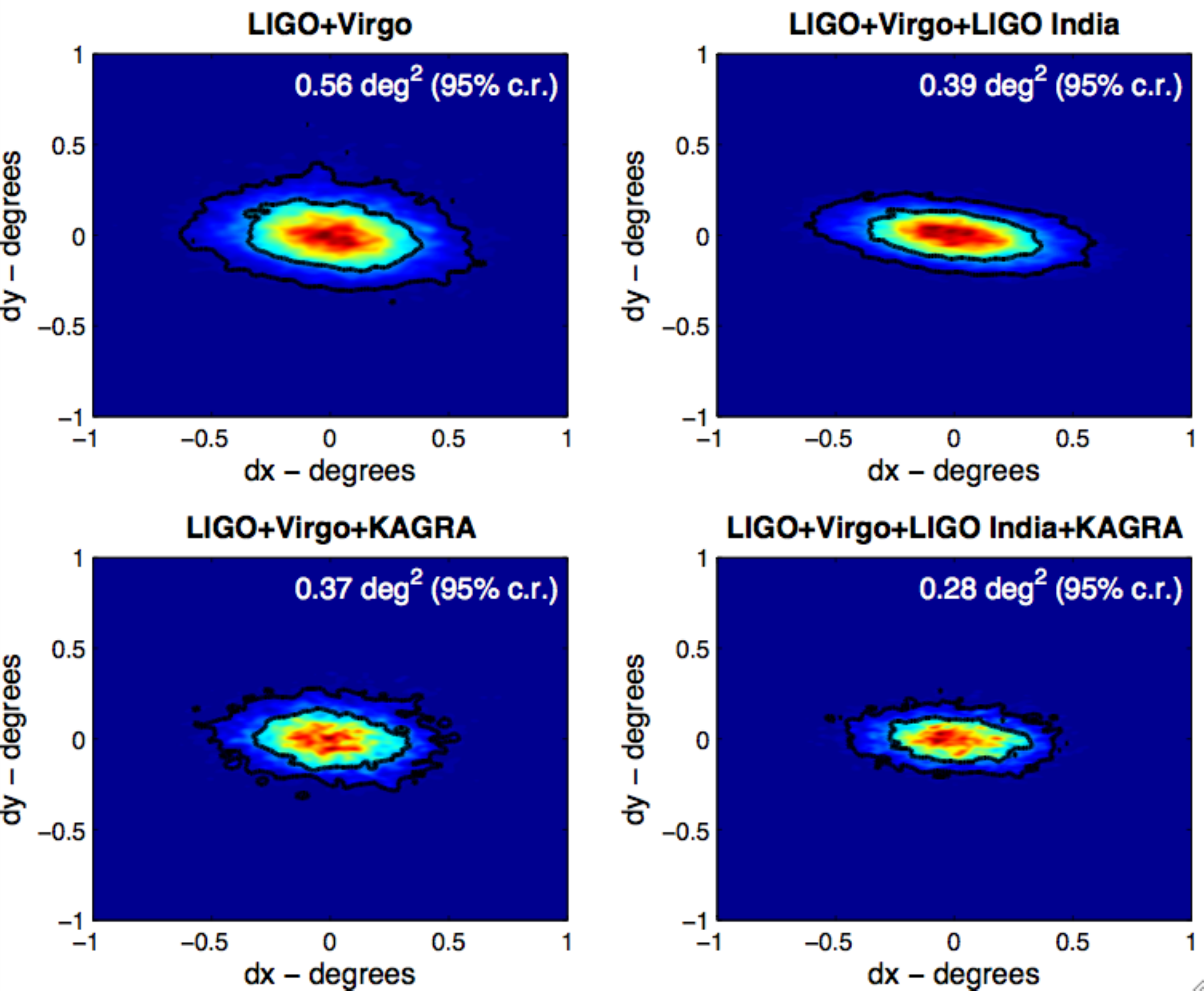}
\caption{Sky localization errors at 95\% c.r. for a
  NS-NS merger located at 69 Mpc with an inclination angle = 150$^{\circ}$ observed
  by different GW networks. The expected SNRs at LIGO Hanford, LIGO
  Livingston, Virgo, LIGO India and KAGRA are 35, 47, 26, 24, 36 respectively.}
\label{fig:binary6}
\end{figure}

\begin{figure}
\centering
\subfigure[GW network 3]{\label{fig:bin6net1}\includegraphics[width=0.3\textwidth]{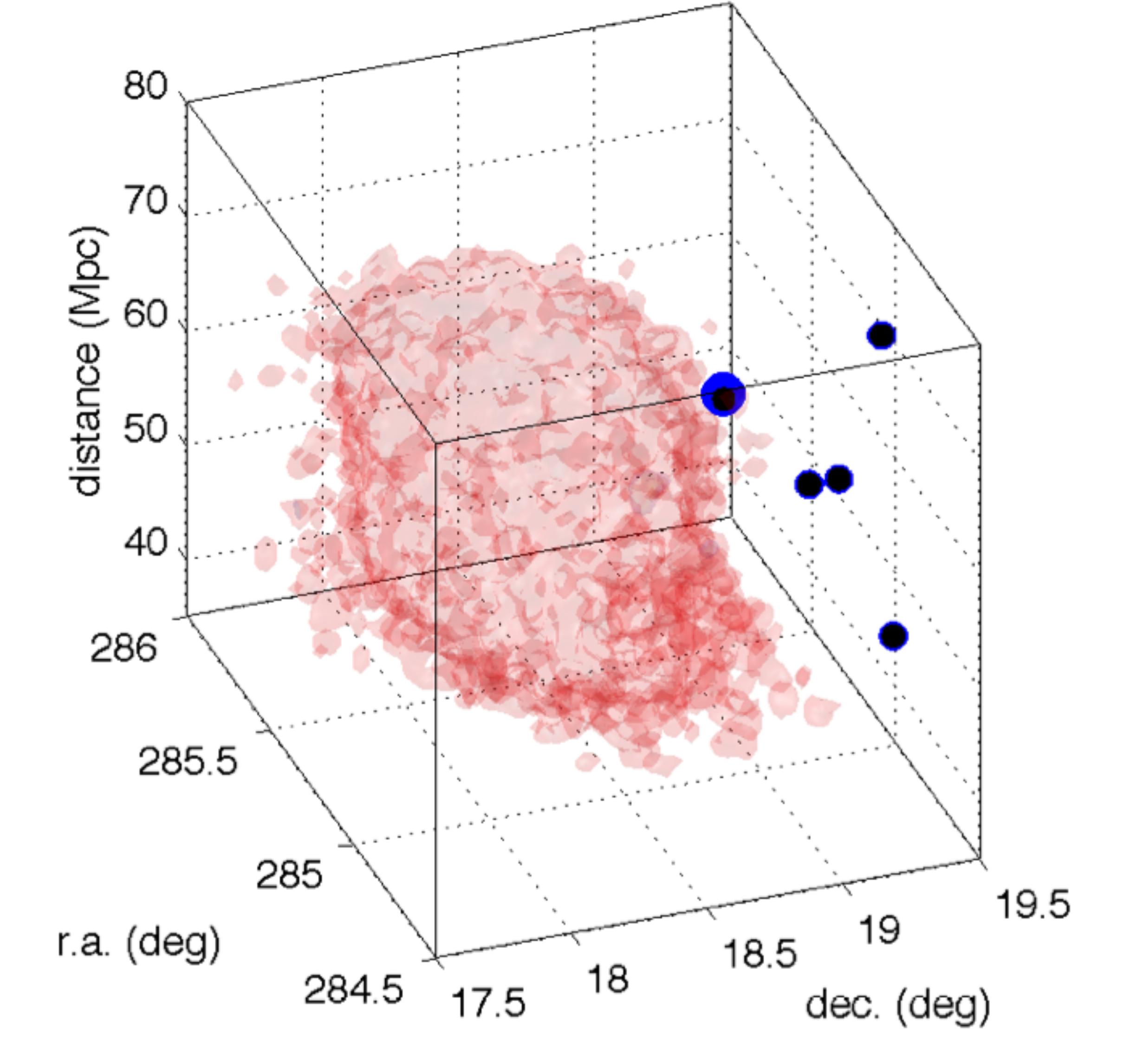}} 
\subfigure[GW network 5]{\label{fig:bin6net4}\includegraphics[width=0.3\textwidth]{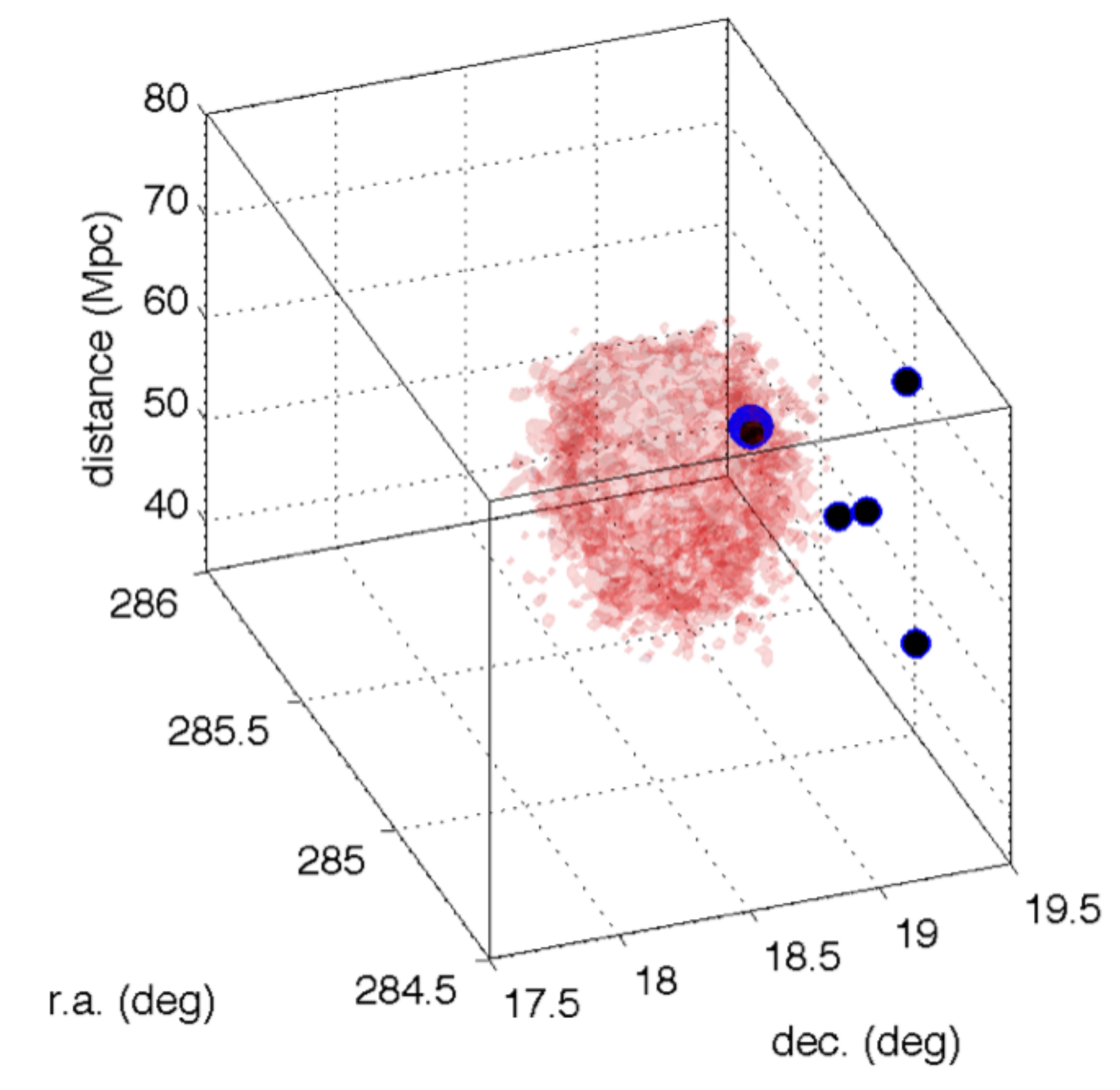}}
\caption{Volume errors at 95\% c.r. for a NS-NS merger located at 69 Mpc with inclination angle = 150 $^{\circ}$ observed
  by GW networks 3 and5. The blue circle marks the merger's true position
  and the black circles denote the $5$ galaxies within the cube's
  volume. Note that the above circles are not to scale. }
\label{fig: binary6vol}
\end{figure}

Let us consider a simulated NS-NS merger located at 69 Mpc. A GW
network 3 measures the source position to 0.6 deg$^2$. A GW network 5
reduces this sky area error by a factor of a couple to 0.3 deg$^2$
(Figure~\ref{fig:binary6}). Using GW network 1, the distance range is from 43 to 73 Mpc at 95\% c.r.
With relatively small localization errors and distance
measures, the number of astrophysical false-positive events that require classification will be nearly zero. In addition, assuming that the NS
binary merger occurred near or within a galaxy, cross-correlating GW localization errors with galaxy catalogs, such as the Census of the Local
Universe (CLU), leaves us with only a handful of candidates for galaxy hosts. As Figures~\ref{fig:bin6net1} and
\ref{fig:bin6net4} show, we expect to see five galaxies within an error
cube of 2 deg$\times$2.5 deg$\times$55 Mpc. The CLU is currently 65\% complete within this distance bin and
efforts are underway to make this catalog more complete. 
With GW network 5, Fig.~\ref{fig:bin6net4} shows that we should be able to identify uniquely
one host galaxy using full 3D-marginalized volumes computed at 95\%
c.r. (this is also the case using the low-latency GW volumes described
in \S\ref{sec:volred}). 

Therefore, for such a well-localized nearby ``golden" binary (with distances $<100$ Mpc), we can undertake
pointed host galaxy follow-up and are no longer limited to large FoV cameras. 
Moreover, we can observe deeper and use a faster cadence than for an
average faint binary. 
Intensive panchromatic follow-up with a wide array of facilities in the optical, infrared, radio
and X-ray would ensure that we leave no stone unturned in studying the EM 
counterpart to a golden binary. We note
that only 10 \% of NS-NS mergers seen by GW Net3a are golden and will have true
distances less than 100 Mpc. Using GW Net5b, we note
that only 1.5\% of binaries will be golden, but they will have high
SNR detections in GWs and EM waves.

\subsubsection{Case Study III: Low Galactic Latitude and at 125 Mpc}

Let us consider the case of a binary with inclination angle 150$^{\circ}$, distance of 125\,Mpc and located very low on the Galactic plane (latitude of -0.11$^{\circ}$, longitude of 63.4$^{\circ}$). 
Using GW network 3, the sky localization of the merger is 1.8 deg$^2$
at 95$\%$ c.r. GW network 5 reduces the sky area error to 1.3 deg$^2$ .

Without any distance information, we would need to search for the
isotropic M$_R$=$-$14 optical counterpart signal out to 450\,Mpc, i.e. 24.3 apparent mag. The number of background supernovae active
in this area at this time would be $\sim$ 120 and the number of foreground M-dwarf flares active would
be $\sim$ 76. Unfortunately, at such a low Galactic latitude, galaxy catalogs are most incomplete and cannot be effectively
used to reduce false positives. 

\begin{figure}
\centering
\subfigure[GW network 3]{\label{bin1net1}\includegraphics[width=0.3\textwidth]{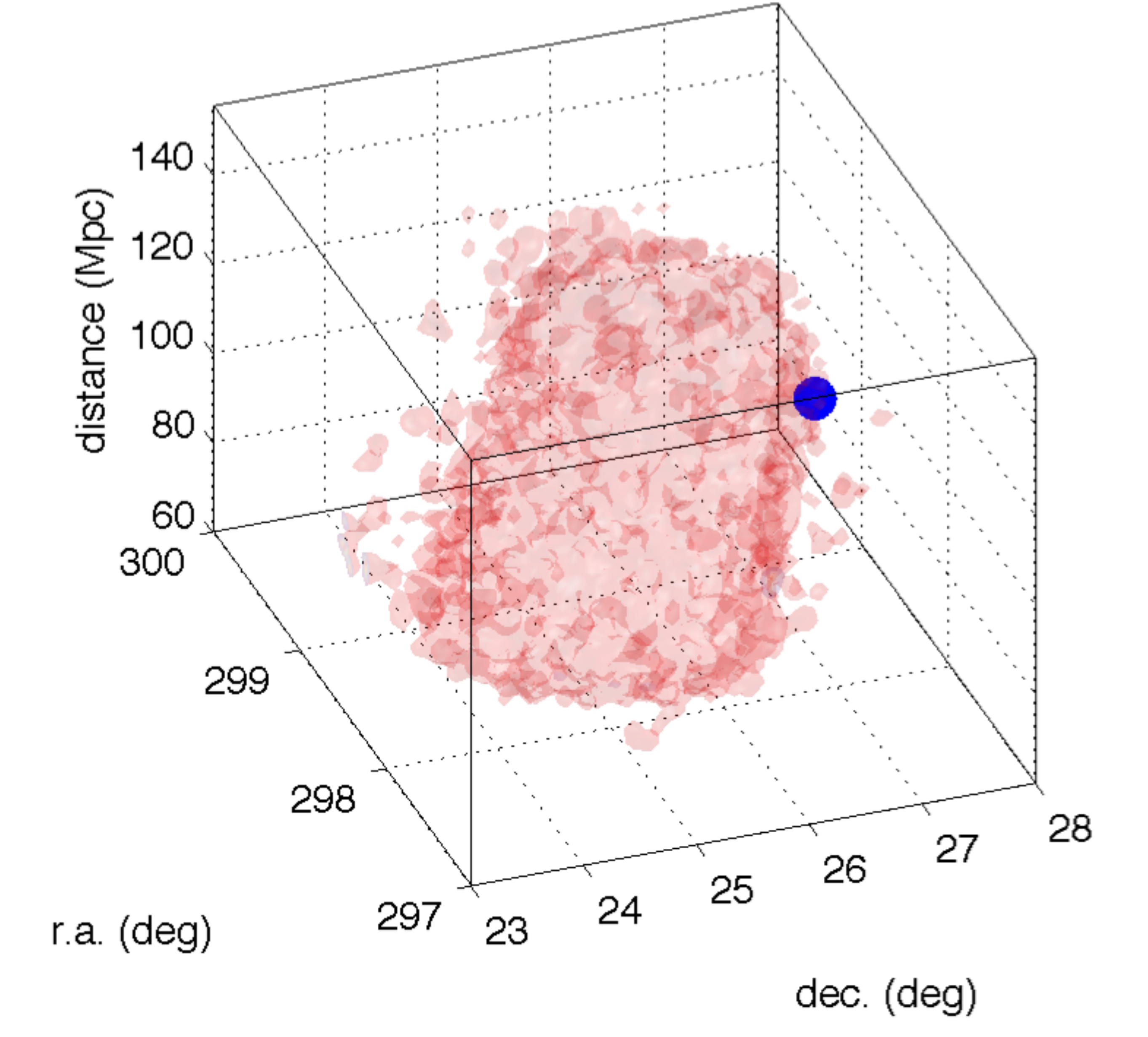}} 
\subfigure[GW network 5]{\label{bin1net4}\includegraphics[width=0.3\textwidth]{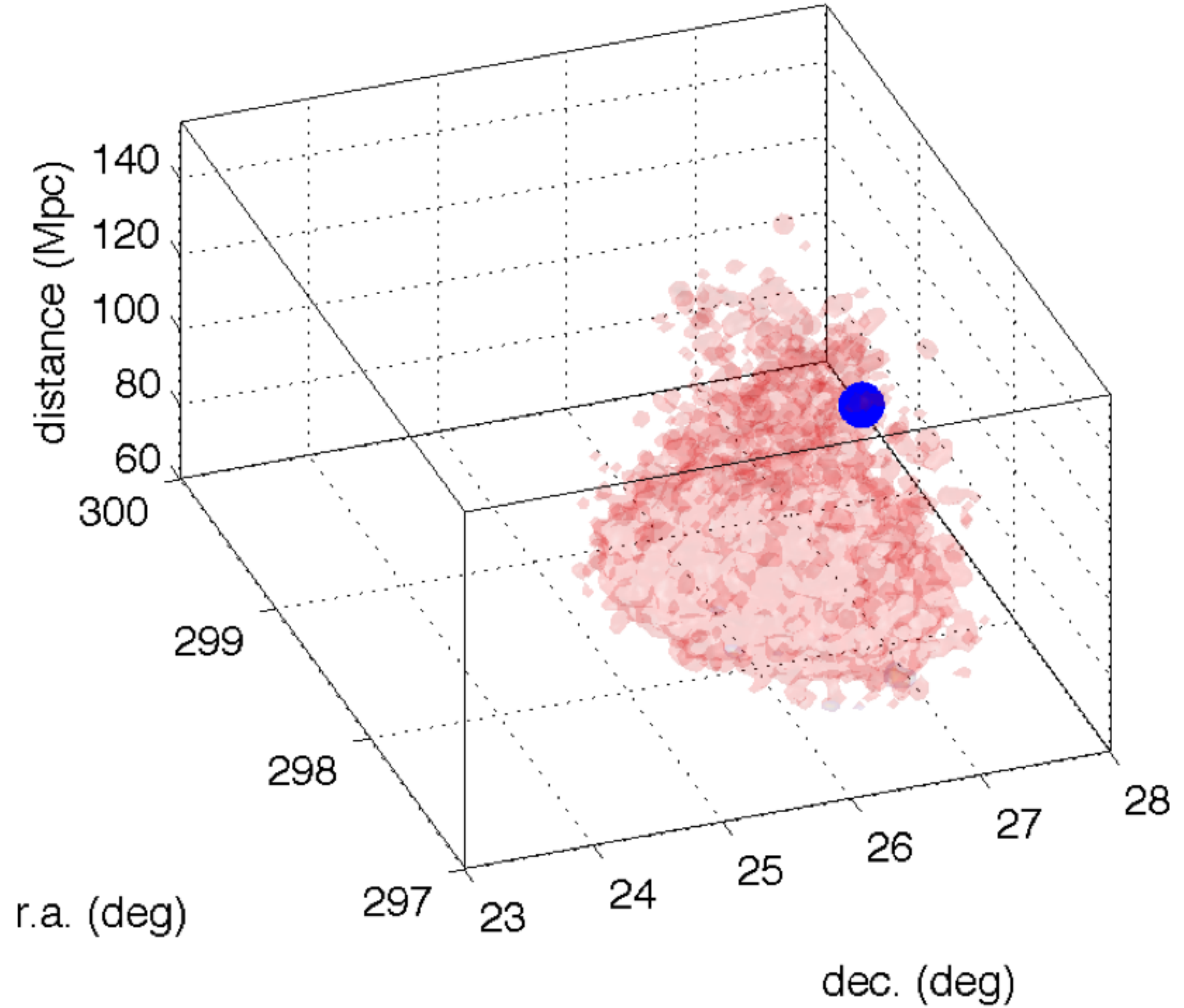}}
\caption{Volume errors at 95\% c.r. for a NS-NS merger
  located at 125 Mpc with inclination angle = 150$^{\circ}$ observed
  by GW networks 3 and 5. The blue circle (not to scale) marks the binary's true position. Due to the low Galactic
  latitude, there are no known galaxies within the volume shown. Using
  network 3, the GW distance measure ranges from 72 Mpc to 142 Mpc
  (95$\%$ c.r.); the reduction in volume is 97$\%$.
}
\label{fig:lowgal}
\end{figure}

However, the derived GW low-latency localization volume can help
reduce false positives. Using GW network 3, the distance measure 
in this volume ranges from 72 Mpc to 142 Mpc at 95$\%$
c.r. (Figure~\ref{fig:lowgal}). The reduction in volume
from the maximum detectable distance of 450\,Mpc to 142\,Mpc is 97\%!
Instead of searching down to 24.3\,mag, we only need to search to 21.8\,mag. This reduces the background 
and foreground false positives to $<$ 10. 

Another effective strategy to deal with foreground false positives is to use a quiescent star catalog (in the optical or infrared) 
that is about two magnitudes deeper than the search depth \citep{Stubbs:2008}. Several
synoptic surveys (e.g., SDSS, PTF, PanStarrs, Skymapper in the optical; VISTA, UKIDSS, WISE in the infrared) are underway to give us
such a catalog. Moreover, some of these surveys will also provide a multi-year historic baseline for variable sources. 

In practice, the limiting factor for follow-up of such a merger would be
the crowding and large line of sight extinction; $\sim$ 8 apparent mag in $R$-band
and $1$ apparent mag in $K$-band. Near-infrared follow-up would be
much easier than optical follow-up. The percentage of GW detected mergers with a Galactic 
latitude less than 5$^{\circ}$ is $\sim$ 9\% and Galactic latitude
less than 10$^{\circ}$ is $\sim$ 18\%.

\subsubsection{Case Study IV: High Galactic Latitude and at 139 Mpc}

The previous three case studies represent a subset of binaries. Let us now consider a canonical binary,
which is neither beamed, nor very close-by, nor too low on the Galactic plane. This binary has an
inclination angle of 64$^{\circ}$, a distance of 139 Mpc and a Galactic latitude of $-$66$^{\circ}$. 
We expect no EM counterparts at $\gamma$- and X-ray wavelengths. Using
GW network 3, GW measurements can localize
the event to 19.5 deg$^2$ on the sky (using GW network 5, the localization error improves to 8 deg$^2$). 

Without any GW distance constraints, we would need to search for an
optical counterpart brighter than M$_{\mathrm R}$ = $-$14 to a horizon
distance of 450\,Mpc i.e. 24.3 apparent mag. Therefore, in a 19.5
deg$^2$ error circle, the extragalactic false-positive number will be
$\sim$ 1300. The Galactic false-positive number will be $\sim$ 100. 

One strategy is to identify the EM counterpart from a false positive
based on the light curve signature. An outburst due to a NS binary merger would be a one-time occurrence.
If we have a good historic light curve of the candidates, foreground false positives and AGN would show previous eruptions.  
Unfortunately, at a depth of 24 \,apparent mag, this may not be the case for most of the sky until LSST is operating for a few years. 
We can also use the theoretical prediction that optical counterparts,
such as kilonovae, evolve faster than supernovae in the same field. Unfortunately, if
we wait too long to obtain multiple epochs, the EM counterpart may fade to a level where it is too faint for spectroscopic follow-up.
We could also use theoretical predictions that kilonovae may be redder than supernovae in the same field. Unfortunately,
given the depth needed and large localization areas, there may not be enough time to obtain data in multiple filters. 

To reduce the number of false positives, a simple approach is to assume that the
merger is spatially coincident with or nearby a galaxy within the
distance reach of the GW network. Galaxies occupy a very small area on the sky. There are only 228 known galaxies in the error circle within 450\,Mpc.
 Allowing a large radius including 50\,kpc around each galaxy, the total area is 60 arcmin$^2$ and the reduction in false
 positives is a factor of 1200!  Allowing an even larger radius including 100\,kpc around each galaxy, the total area is
 240 arcmin$^2$ and the reduction in false positives is still a factor of 300. Unfortunately, the current galaxy catalog is
 grossly incomplete and it is imperative that efforts be made to complete it.
 
 Furthermore, we can leverage our localization volume constraint to further reduce the number of relevant host galaxies.
Using GW network 3, the merger's distance is measured to be between 108\,Mpc and 228\,Mpc,
the volume is smaller than between 0\,Mpc and 450\,Mpc by 88\%.  A smaller volume would correspond to a smaller number of
galaxies. In this case, 73 galaxies are known to lie in the localization volume.  Using GW network 5, the merger's distance is
measured to be between 117\,Mpc and 230\,Mpc and currently, $\sim$ 37 galaxies are known in this volume.  This
list of galaxies is incomplete by a factor of two. If we had a complete catalog of galaxies, and for cases where the number of galaxies was less than few tens, 
we would even consider targeting galaxies individually. We would not
require a large FoV camera, paving the way for using infrared, radio and X-ray facilities.

\begin{figure}
\centering
\subfigure[GW network 3]{\label{bin10net1}\includegraphics[width=0.3\textwidth]{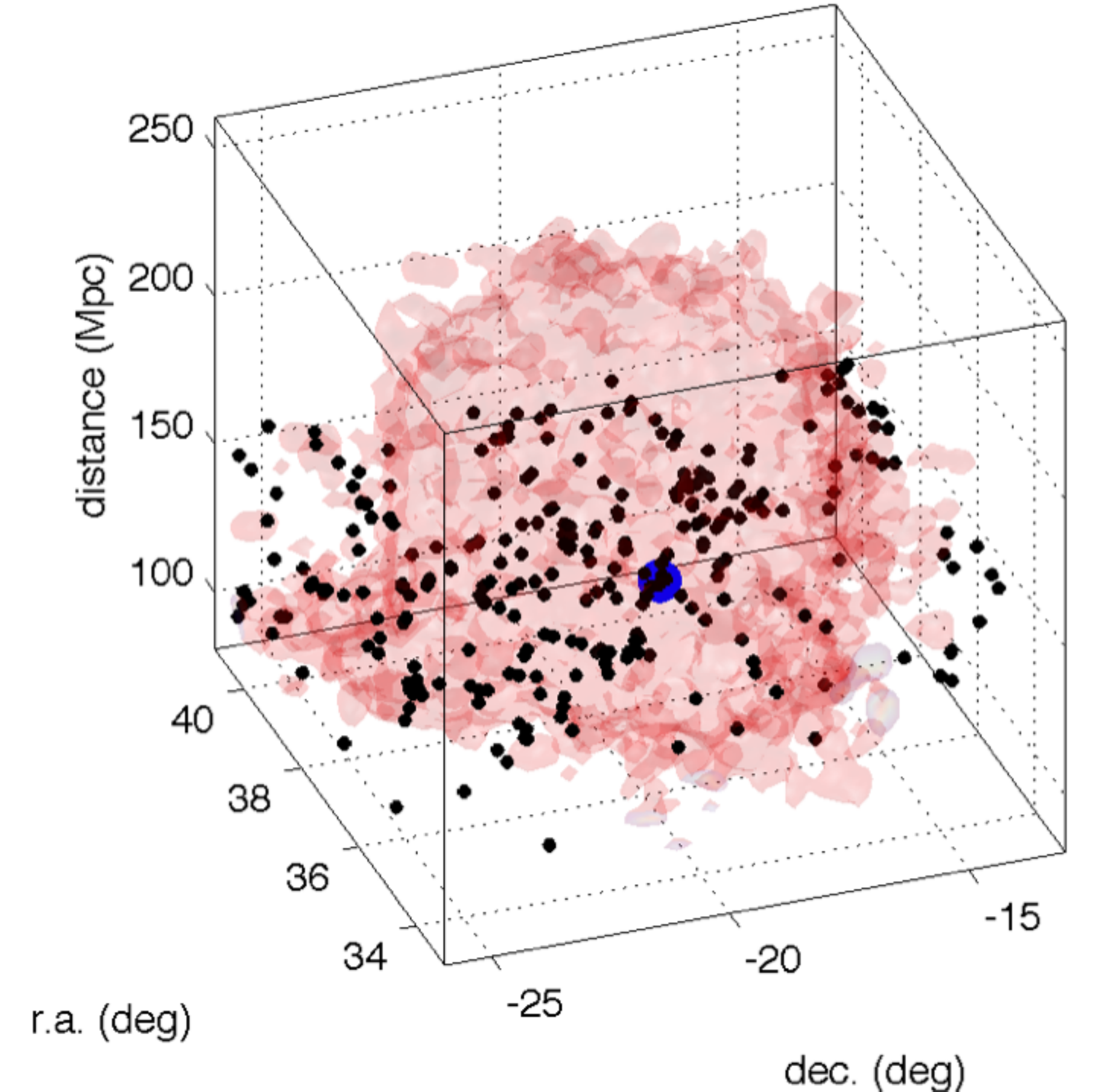}} 
\subfigure[GW network 5]{\label{bin10net4}\includegraphics[width=0.3\textwidth]{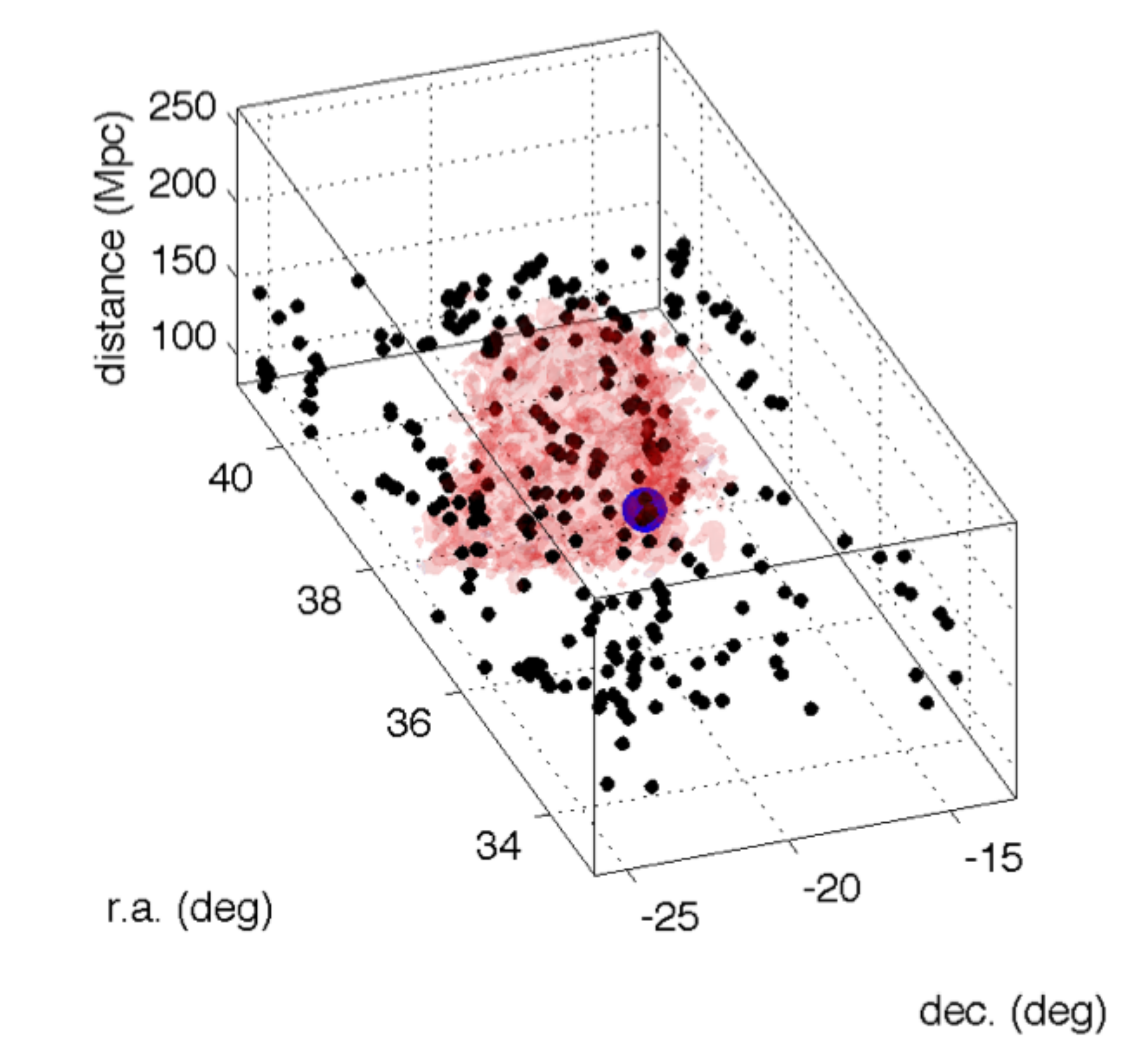}}
\caption{Volume errors at 95\%
  confidence interval for a NS binary merger located at 139 Mpc with inclination
  angle = 64$^{\circ}$ observed by GW networks 3 and 5. The blue circle marks the binary's true position
  and the black circles denote the $279$ galaxies within the cube's
  volume. Note that the circles above are not to scale.
}
\end{figure}

\subsubsection{Case Study V: Dense galaxy cluster environment at high
  Galactic latitude at 115 Mpc}

For our final case study, we choose a simulated NS-NS binary merger
event that occurs in an extremely dense galaxy cluster (ABELL 4038). Here,
the use of localization volumes to target individual galaxies is not
feasible as the number of galaxies is too large.  However, using a galaxy catalog to prioritize
follow-up can still reduce false positives by orders of magnitude. 

This specific binary has a distance of 115 Mpc and Galactic latitude and longitude
of $-$75.8$^{\circ}$ and 25.7$^{\circ}$ respectively. With GW network 3, its sky
localization can be measured to 18.8 deg$^2$ and its distance measure
ranges from 110 Mpc to 269 Mpc. The upper measure of the distance (at
95\% c.r.) is greater than the maximum distance of 200 Mpc used in CLU. Assuming only a horizon distance of
200 Mpc and using the CLU catalog, we find 780 galaxies comprising an area of 1070 arcmin$^2$ within
such an area of the sky. If we then include our lower measure of the distance (at
95\% c.r.), we find 410 galaxies comprising an area of 560 arcmin$^2$. Increasing the network from
three to five interferometers, its sky localization improves to 14.4 deg$^2$ and its distance measure ranges from 97 Mpc to 155 Mpc. Using CLU, 
we then find 390 galaxies comprising an area of 530 arcmin$^2$ within
such an area of the sky. 

In such a dense galactic environment, targeting hundreds of galaxies individually is not feasible.
However, the reduction of false positives is still significant. Specifically, a snapshot of 18.8 deg$^2$ out to 450\,Mpc, 
would give 1250 background supernovae. Restricting the search to candidates spatially coincident with nearby galaxies
(i.e. 1070 arcmin$^2$) reduces the false positives to 20. Further imposing the volume constraint of 269\,Mpc,
reduces the false positives to 4 events. 

\begin{figure}
\centering
\subfigure[GW network 3]{\label{bin6net1}\includegraphics[width=0.3\textwidth]{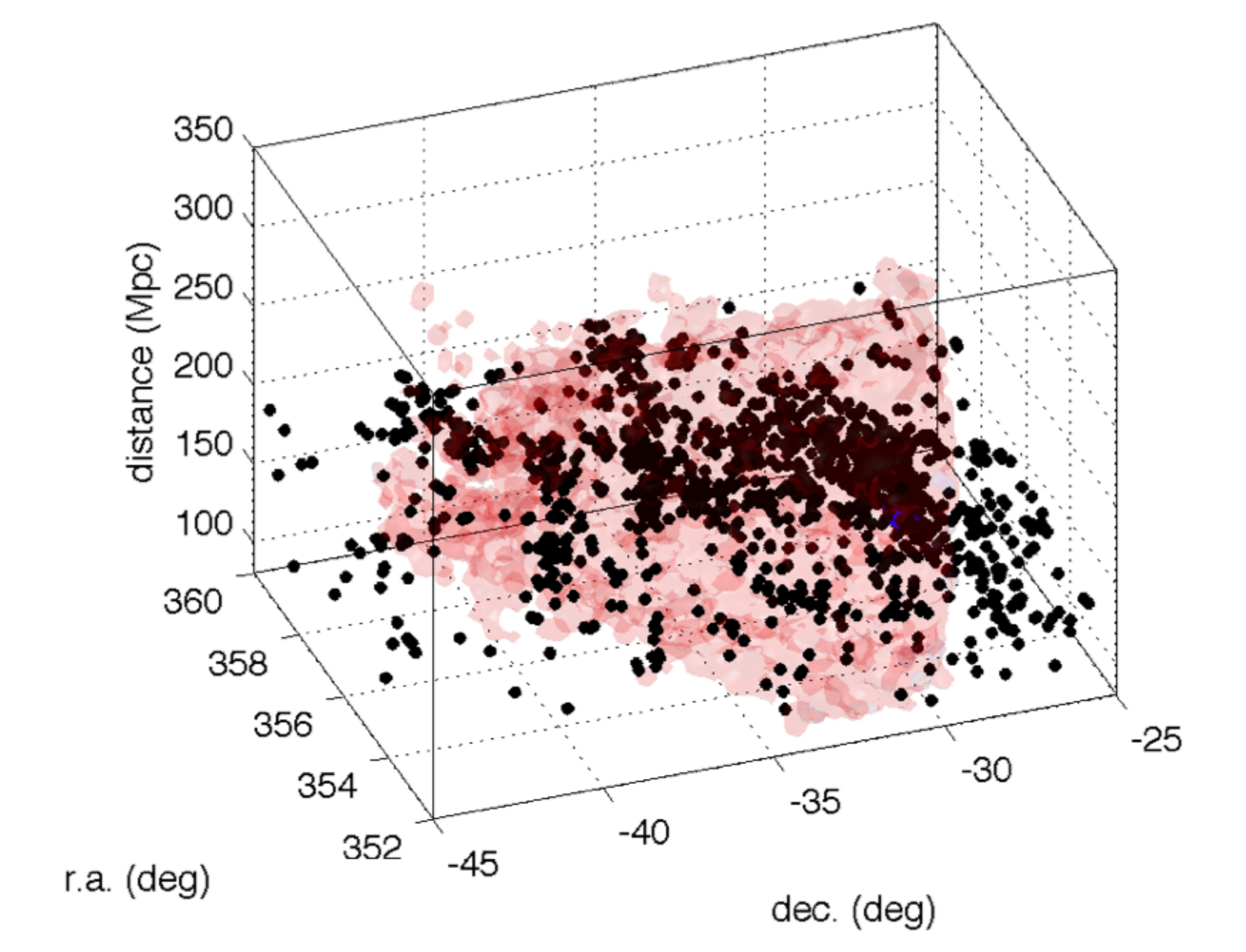}} 
\subfigure[GW network 5]{\label{bin6net4}\includegraphics[width=0.3\textwidth]{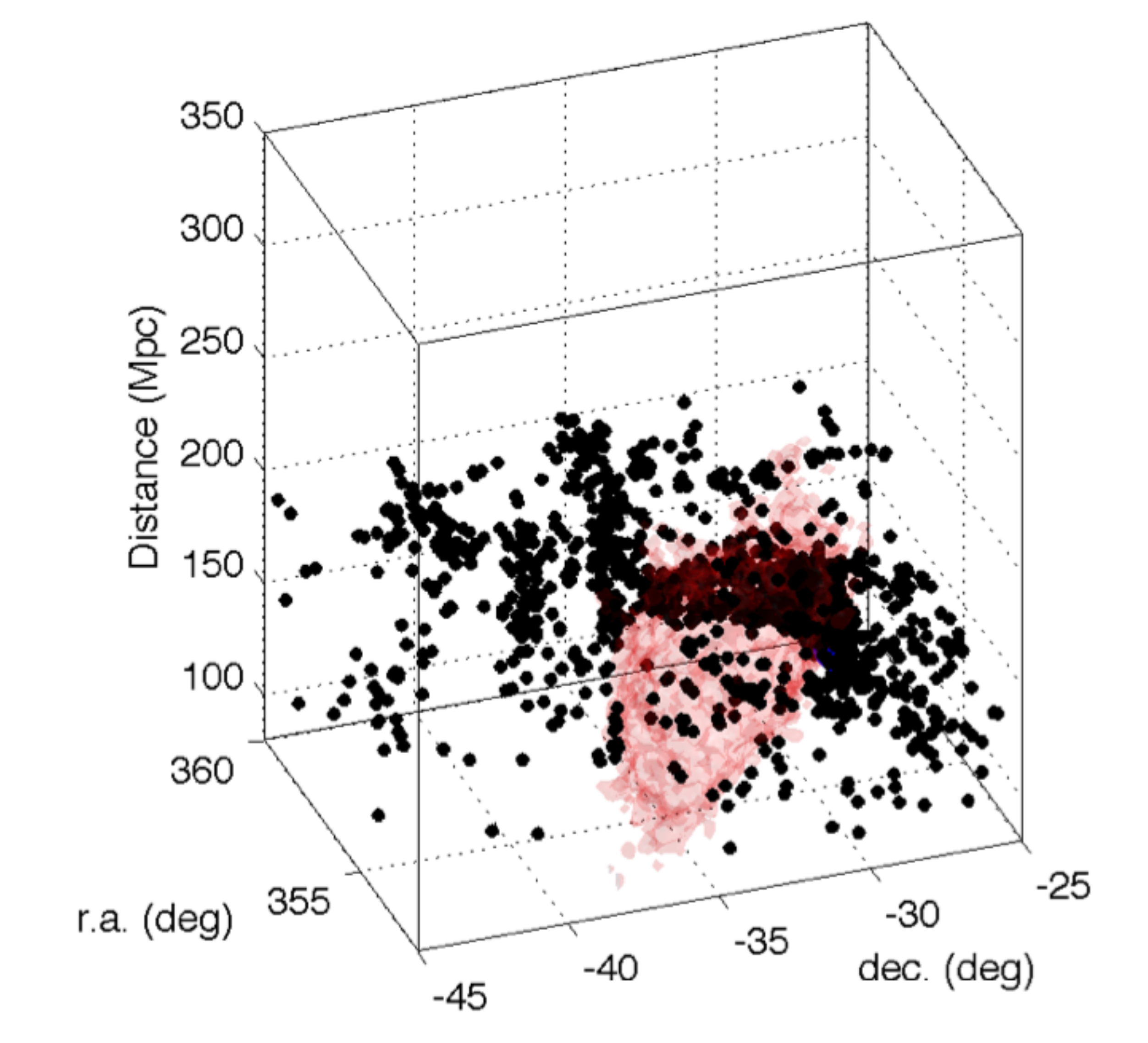}}
\caption{Volume errors at 95\% confidence interval for a NS-NS binary
  merger located at 115 Mpc with inclination angle = 87$^{\circ}$ observed
  by GW networks 3 and 5. The blue circle marks the binary's true position
  and the black circles denote the $1350$ galaxies within the cube's
  volume. Note that the circles above are not to scale.
}
\end{figure}

\subsection{Astrophysical populations}
\label{sec:vol_pop}

As discussed in \S\ref{sec:indbin}, the combination of GW distance
information, GW sky localization and galaxy catalogs could
help reduce the number of galaxies that are possible hosts of the NS
binary merger event and the number of false-positives. The use of spatial coincidence with galaxies in turn could have a substantial effect in
reducing the number of false-positive transients that need to be
considered as possible NS-binary merger events
\citep{kk09}. Alternatively, if we can limit the number of host
galaxy candidates to a few as in the case of a golden nearby GW merger,
targeted follow-up opportunities for individual galaxies becomes a possibility.

\begin{figure}
\centering
\subfigure[NS-NS binary mergers observed by GW Net3a]{\label{volred_net1}\includegraphics[width=0.4\textwidth]{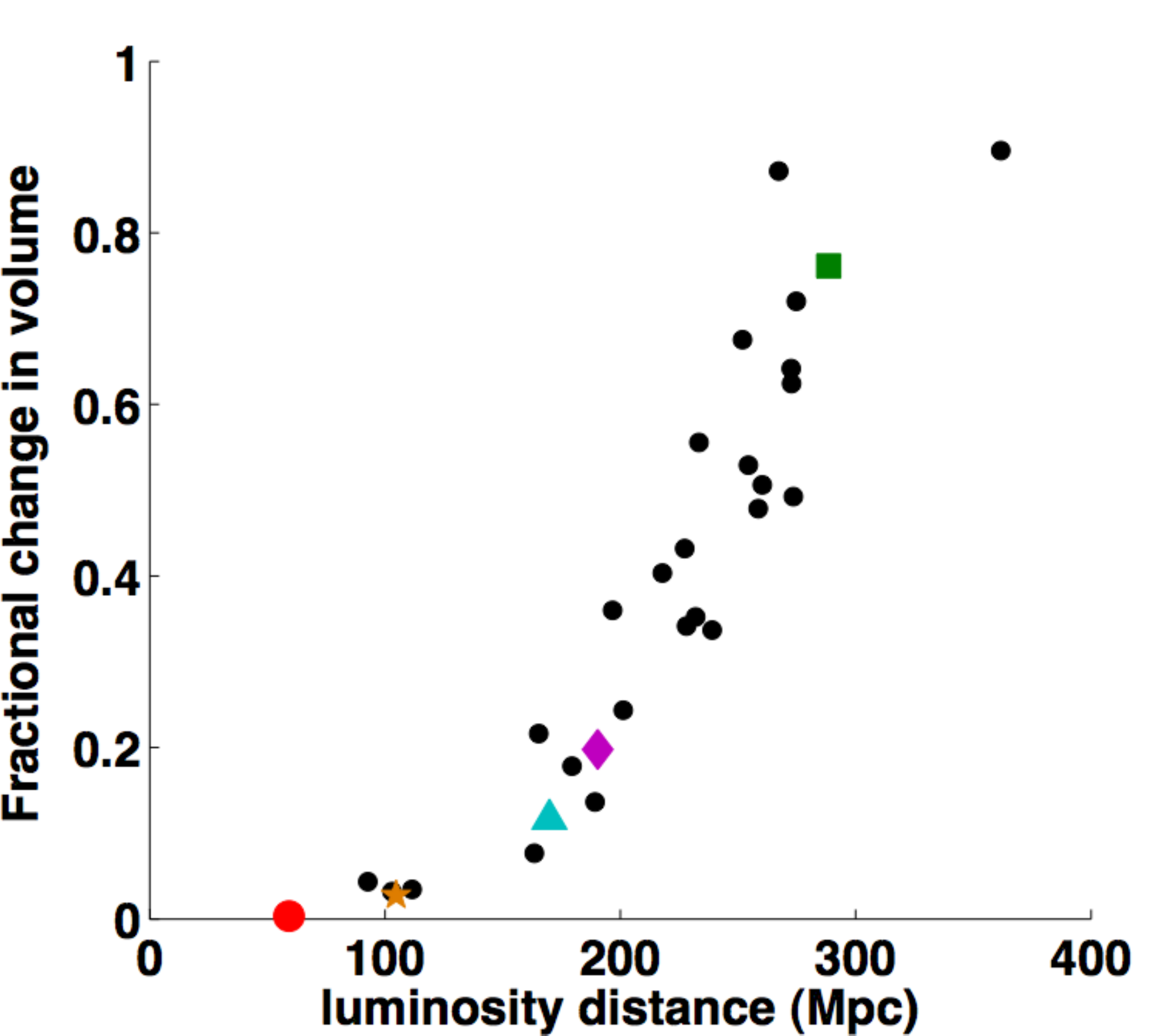}} 
\vspace{0.2in}
\subfigure[NS-BH binary mergers observed by GW Net5b]{\label{volred_net4}\includegraphics[width=0.4\textwidth]{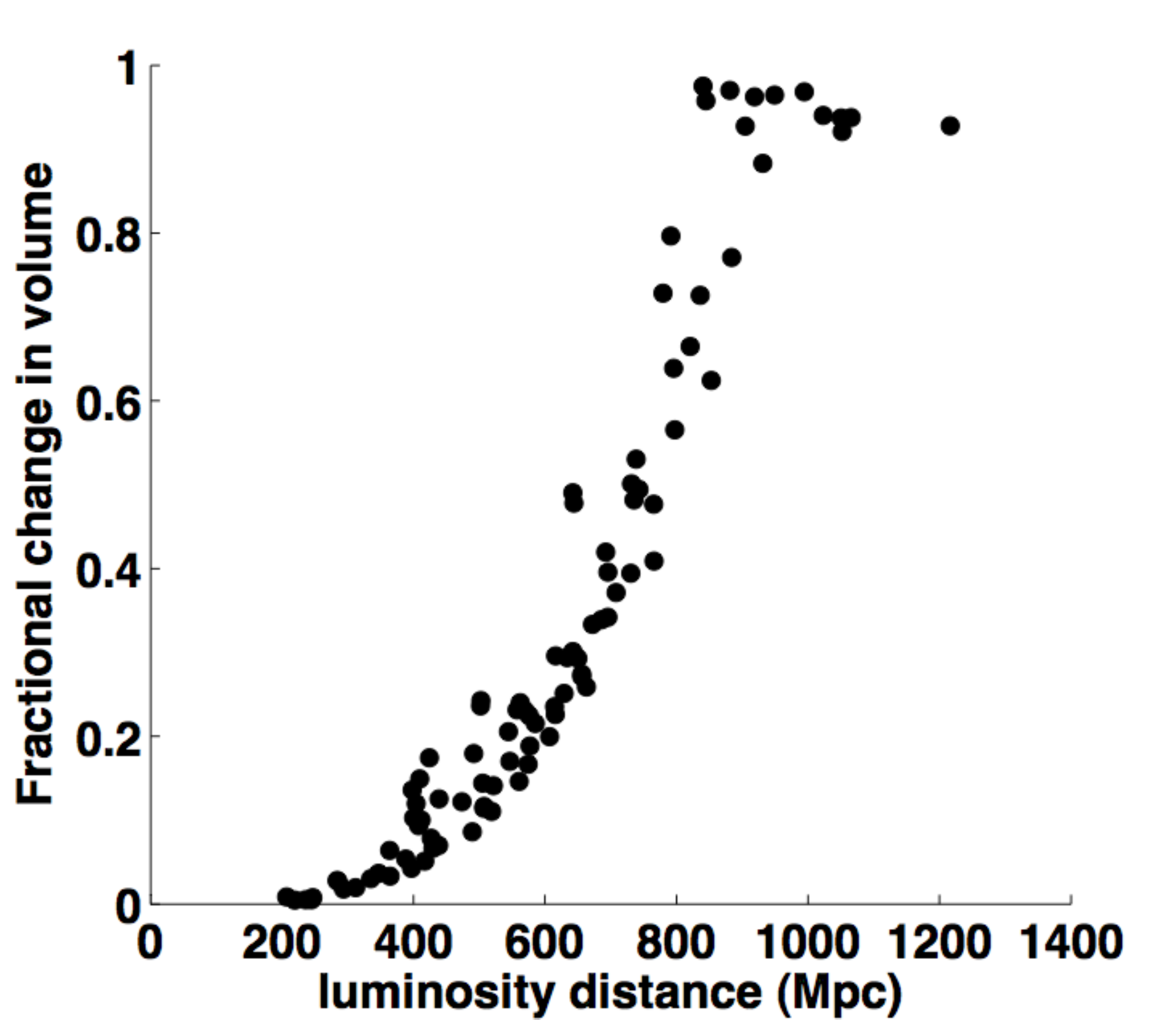}}
\caption{Fractional reductions in volume for a detected sample of
  NS-NS mergers observed by GW Net3a (top panel) and NS-BH mergers
  observed by GW Net5b (bottom panel). This fraction is the ratio of
  the volume encompassed by GW distance measures (Figure 6 in Section 4.5)
  and the volume encompassed within the GW maximum detectable distance. Each circle represents a detected
  NS binary merger. The different colors
represent different case studies of NS-NS mergers examined in
\S\ref{sec:indbin}: green square is Case I (beamed binary), red circle is Case
II (a nearby binary), light blue triangle is Case III (a merger at low Galactic
latitude), purple diamond is Case IV (a merger at high Galactic latitude), and
orange star is Case V (a binary in a dense galaxy cluster
environment).}
\label{fig: volred}
\end{figure}

Following from Figures~\ref{fig:volabs_net1}
and~\ref{fig:volabs_net4}, next, we compare how well we fare when computing
GW volumes either using the event's distance measure or using only the GW
horizon distance. In particular, we compute $r_{\mathrm{vol}}$ the fractional change
in volume:
\begin{equation}
r_{\mathrm{vol}} = \frac{(\min[\mathrm{d}_u, \mathrm{d}_h])^3 - \mathrm{d}_l^3}{\mathrm{d}_h^3},
\label{eqn:fracvol}
\end{equation}
for samples of NS binary mergers detected using different GW networks and
triggering criteria (see Figures~\ref{volred_net1} and
\ref{volred_net4}). 

We find that the fractional change correlates with distance to the binary and the
reduction is higher for nearer binaries. Specifically, for NS-NS mergers, the fractional change in volume is
$<60 \%$ for those events located within 200 Mpc ($\sim
1/3$ of the detected binaries). For NS-BH binaries, the fractional
change in volume is $< 60 \%$ for those events located within 700 Mpc ($\sim
1/2$ of the detected binaries).

\private{
 shows the fractional
change in volume as a function of the mean distance for NS-BH mergers
detected by GW Case II. For NS-NS mergers, the fractional change in volume is
$<60 \%$ for those events located within 200 Mpc ($\sim
1/3$ of the detected binaries). For NS-BH binaries, the fractional
change in volume is $< 60 \%$ for those events located within 700 Mpc ($\sim
1/2$ of the detected binaries).}

\vspace{0.2in}

The implications for search strategies are as follows. In cases where the upper limit of localization
volume is much less than GW maximum detectable distance (c.f. case
studies 2, 3 and 4), the EM counterpart search can be to brighter apparent magnitudes. This
significantly reduces false positive numbers and spectroscopic
follow-up is easier. More frequent are the cases where knowledge of the lower distance limit reduces the total number of host galaxies.

The number density of galaxies is 0.6 per deg$^{2}$ per Mpc at z=0.1
and L$>$0.1\,L$_{\star}$ and scales with z$^2$ \citep{Blanton:2003}.  
For a median localization of 10 deg$^2$ and out to a distance of 200 Mpc, the number of galaxies is $\sim$80. 
The median size of a galaxy is [0.8, 0.4, 0.2] arcmin$^{2}$ at [50,
100, 200] Mpc respectively. The diameter assumed here
is for the surface brightness contour of 25 mag arcsec$^{-2}$. This corresponds to a projected offset (offset$_{\rm kpc}$=size$_{\rm rad}\,\times\,$distance$_{\rm kpc}$) 
of [7.3, 10.8, 14.0]\,kpc at [50, 100, 200]\,Mpc respectively. Thus, the total area occupied by
80 galaxies is 0.004 deg$^{2}$, a factor of 2500 smaller
than the localization. We can easily search 10 times the size of the galaxy to accommodate large kicks of the order of
hundred kiloparsec (e.g., \citealt{FBF:2010, Kasliwal:2012}) in NS-NS mergers and still gain a factor of 25 in terms of reduction of false positives.  
However, if the kicks are over a megaparsec \citep{Kelley:2010}, we cannot use the positions of host galaxies to reduce false positives. 

Efforts are underway to complete the CLU galaxy catalog using four narrow-band filters on the Palomar 48-inch Schmidt telescope. 
This will boost the completeness from 50\% to 85\% of the $B$-band light at 200\,Mpc in the three-quarters of the sky accessible from Mount Palomar.
Note that this distance limit is well-matched for majority of NS-NS binaries detected by GW Net 3a. However, particularly for NS-BH binaries, we
should consider an even larger effort to complete galaxy catalogs out to several hundred Mpc.

\section{Conclusions and Discussion}
\label{sec:conclusion}

Observing compact-object binary mergers in both GW and EM will be challenging.
GW interferometers
will only be able to localize the merger to sky errors
ranging from tens to hundreds deg$^2$.  In addition, the estimated rates of GW-detected compact binary
mergers span several orders of magnitude from zero to hundreds of mergers per year. 
Theoretical predictions of EM signatures 
from the optical to radio is also an active area of research and model estimates vary
significantly. Finally, our characterization and understanding of the
transient sky at different wavelengths, timescales and sensitivity remains
incomplete. 

In this paper, we present the first comprehensive end-to-end simulation of the
detectability and identification of neutron star binary mergers with
GW and EM facilities. Our simulation comprises: the construction of
astrophysically-distributed populations of GW-detectable NS-NS and NS-5
M$_{\odot}$ BH binary
mergers, GW source characterization using different GW detector
networks and triggering criteria, establishing
the detectability of plausible EM counterparts by upcoming or current
telescopes (particularly, optical), and identifying the GW event among the few-to-many astrophysical false-positive
transients in different wavelengths. The extent of our analysis is naturally
dependent on the assumptions made: an underlying population of non-spinning NS-NS or NS-5
M$_{\odot}$ BH binary mergers, GW instrumental Gaussian noise with each
interferometer operating continuously, negligible GW systematic errors
arising from the GW waveform and instrument calibration, a high availability
factor for telescopes to follow-up GW mergers, and idealized optical observing conditions.

Our work is novel in six principal ways. First, we construct GW-detected populations of astrophysically-distributed NS binary
mergers using CLU (a volume-limited local universe galaxy catalog out
to 200 Mpc), instead of at fixed SNR
or distances as in earlier works. Second, we incorporate advanced LIGO
detectors using optical squeezed light into our analysis. Third, we compute explicit
marginalized 3D GW volumes using MCMC. Fourth, we consider how GW volumes can assist follow-up by
optical telescopes and other EM facilities. Fifth, we quantify the tradeoff between depth and area for 
a variety of optical telescopes, including 4-7m class and HSC
telescopes for the first time. Sixth, by examining individual NS
binary mergers, we suggest how to pinpoint the GW event amongst a
possible plethora of astrophysical Galactic and extragalactic false-positives.

By expanding the parameter space of NS binary populations and GW networks, we show that GW detectable distances and sky area errors
may range an order, or several orders of magnitudes
respectively.  From our case studies of GW-detected NS
binary mergers, we find that:

\begin{itemize}

\item Thanks to the GW Malmquist effect, the fraction of NS binary mergers beamed towards
us (with $\theta_{\mathrm j} < 6^{\circ}$) is boosted to large
distances (400 Mpc-1.3 Gpc). However, we have shown empirically that
the fraction is still tiny (1-2.5\%). For this subset of events, the easiest identifiable EM
  counterpart would be a contemporaneous SGRB.  All-sky $\gamma$-ray
  detectors are essential to ensure joint GW and EM observations. Optical squeezing implemented in GW
interferometers will also increase the beamed fraction by a further factor
of nine to ten. 

\item For the small number of golden nearby
binaries ($<$ 100 Mpc), which have small GW localizations given the high SNRs of
GWs, we should intensively follow-up such events at
all wavelengths. Whilst the number of detected golden binaries is
independent of network, the total GW-detected fraction ranges from 2\%-10\% from GW Net3a to GW Net5b. The number of
false-positives in their small localization areas or volumes should be small (\S6.1.2).

\item For the majority of binaries which are neither beamed nor nearby, the challenges in detecting the isotropic optical counterpart are surmountable
by optimizing the depth \emph{versus} area tradeoff. Initially, we
expect small telescopes (especially given their larger number and
wide-field cameras) to play an important role in detecting
counterparts. GW networks with fewer detectors will have poorer
localizations and lower distance sensitivity. Small telescopes have strength in numbers and can be expected to be more flexible for rapid follow-up for GW triggers. 
Later, with increasing detector numbers and instrument sensitivity and
use of coherent triggers, GW networks will have higher distance sensitivity and improved
localizations, larger telescopes will be essential (\S5.2.1).  Given weather and limited sky accessibility of 
optical telescopes (due to sun constraints and altitude constraints),  we advocate for a world-wide network of telescopes 
of different sizes with wide-field cameras spread across different
latitudes and longitudes. Furthermore, we advocate the building of larger FoV infrared cameras.
\end{itemize}

Our simulation and detailed case study analysis motivate us to search for EM counterparts to GW binaries. 
We find that although there are challenges, they are surmountable by timely advance preparation. Hence, we conclude
here with four action items that will better prepare us to securely identify the detected EM counterpart:

\begin{enumerate}

\item We should complete as much as possible host galaxy catalogs out
  to $z \sim 0.1$ to
  increase identifying an EM counterpart through two different means: (i) spatial coincidence with a nearby galaxy can quickly eliminate
false positives for a subset of NS binary mergers by orders of magnitude. This is critical to prioritize
candidates for prompt spectroscopic and panchromatic
follow-up (\S6.1.4, \S6.1.5, \S6.2). (ii) together with low-latency
GW volume errors, in some cases, the number of galaxies can be reduced
to a tractable number for possible targeted follow-up with relatively smaller field-of-view
facilities e.g. radio, infrared, X-ray, large aperture optical
telescopes (\S6.1.2, \S6.1.4, \S6.2). An ongoing effort is the Census of the Local Universe (CLU) using
narrow-band filters (H$\alpha$) on the Palomar 48-inch. Another
planned survey (H I) is WALLABY with ASKAP in the Southern Hemisphere \citep{Kaplan:2012}.

\item We should construct deep ($\sim$ 26
apparent \,mag) all-sky quiescent stellar
source catalogs which would help eliminate foreground false positives,
particularly at low Galactic latitudes (\S6.1.3). We estimate that we will require catalogs that are approximately two magnitudes deeper
than the EM counterpart and span optical and
infrared wavelengths. Current ongoing efforts may not be deep enough, e.g. SDSS, PTF, PanStarrs,
SkyMapper, WISE, VISTA. 

\item We should rehearse the search for transients in large sky localizations. For example, the successful identification and
  follow-up of an optical afterglow of a Fermi/GBM GRB in a three deg.$^{2}$ error circle (c.f. GRB120716A, \citealt{GCN2012})
  is encouraging. Such efforts are a full dress rehearsal for elusive
  EM counterparts in GW constrained large swaths of the sky \citep{Kulkarni:2012}.

\item We should continue to construct a complete inventory of
transients within several hundred Mpc. Just in the past few years, we have uncovered multiple, new classes of optical 
transients, which are fainter, faster and rarer than supernovae
(e.g., \citealt{KasliwalPhD:2011}). We may indeed even be lucky enough to see an EM
counterpart to a NS binary merger prior to hearing the GWs!

\end{enumerate}

In summary, given the diversity of properties and locations of possible EM counterparts and
challenges in their identification, we advocate a \emph{comprehensive, multiwavelength
multi-pronged} approach to observe compact binary
mergers in GWs and EM waves.

\section{Acknowledgements}


M.M.K. acknowledges generous support from the Hubble Fellowship and Carnegie-Princeton Fellowship. 
A.G. thanks the Summer Undergraduate Research Fellowship program at Caltech.

We are very grateful to Jean-Michel D{\'e}sert, Dale Frail, Chris
Hirata, Shri Kulkarni and Setu Mohta for careful reading of the manuscript. We
thank Ernazar Abdikamalov, Paul Groot, Gregg Hallinan, Brian Metzger, Udi Nakar, Sterl Phinney, Tony Piro, Tom Prince,
Jon Sievers, Linqing Wen for useful discussions. SMN thanks the ITC for
hospitality, and useful discussions there with Edo Berger and Josh Grindlay. We thank Anand Sengupta and
Tarun Souradeep for providing LIGO India's
(previously referred to as IndIGO) position and orientation. We thank Haixing
Miao for providing the anticipated advanced LIGO noise curve with
optical squeezing, and Masaki Ando, Larry Price and Stan Whitcomb for
KAGRA and LIGO follow-up references. We thank Neil Gehrels, David Kaplan, Peter Nugent and
Fang Yuan for providing specifications of Lobster-ISS, WIYN, La Silla Quest and Skymapper respectively. Some
of the simulations were performed using the Sunnyvale cluster at
CITA. Part of this work was performed at the Jet Propulsion Laboratory, California Institute of Technology, under contract with the National Aeronautics and Space Administration. Government sponsorship acknowledged. Copyright 2012.

\bibliography{EMGW}

\end{document}